\documentclass[12pt]{amsart}
\usepackage{lmodern}
\usepackage[english]{babel}
\usepackage[utf8]{inputenc}
\usepackage[T1]{fontenc}
\usepackage{amsmath,amssymb,amsthm}
\usepackage{dsfont}
\usepackage[margin=1in, footskip=30pt, a4paper]{geometry}
\usepackage{xcolor}
\usepackage{graphicx}
\usepackage{hyperref}
\usepackage{refcount}
\usepackage{natbib}
\usepackage{tikz}
\usepackage{tikzsymbols}
\usepackage{enumitem}
\usepackage[doublespacing]{setspace}
\usepackage{bm}
\usepackage{url}
\usepackage{subcaption}
\usepackage[bottom]{footmisc}
\usepackage{changepage}
\usepackage{color}
\usepackage{longtable}
\usepackage{grffile}
\usepackage{multicol}
\usepackage{etoolbox}
\usepackage{xfrac}
\usepackage{booktabs}
\usepackage{array}
\usepackage{float}
\usepackage{caption}

\allowdisplaybreaks
\hypersetup{
    pdffitwindow=false,                         
    pdfstartview={XYZ null null 1.00},          
    pdfnewwindow=true,                          
    colorlinks=true,                            
    linkcolor={red!50!black},                   
    citecolor={blue!50!black},                  
    urlcolor={red!50!black},                    
    linkbordercolor={white},                    
}
\newcommand*{\hyref}[2]{\hyperref[#2]{#1~\ref{#2}}}

\usetikzlibrary{arrows.meta,patterns.meta,positioning,calc,backgrounds,decorations.pathreplacing,decorations.pathmorphing,automata,arrows}
\theoremstyle{plain}
\newtheorem{theorem}{Theorem}
\newtheorem{lemma}{Lemma}
\newtheorem{proposition}{Proposition}
\newtheorem{corollary}{Corollary}

\newtheorem{claim}{Claim}
\theoremstyle{definition}

\newtheorem{example}{Example}
\theoremstyle{definition}
\newtheorem{remark}{Remark}

\newcommand{\x}{\textbf{x}}

\newcommand{\p}{\textbf{p}}
\newcommand{\bv}{\textbf{v}}
\newcommand{\bc}{\textbf{c}}

\newcommand{\E}{\mathbb{E}}

\DeclareMathOperator*{\minimize}{\mathrm{minimize}}
\DeclareMathOperator*{\argmax}{arg\,max}
\DeclareMathOperator*{\argmin}{arg\,min}
\newrobustcmd*{\citefirstlastauthor}{\AtNextCite{\DeclareNameAlias{labelname}{given-family}}\citeauthor}

\tikzset{>=latex}
\tikzset{every path/.style={-}}

\pgfdeclarelayer{background}
\pgfdeclarelayer{edgelayer}
\pgfdeclarelayer{nodelayer}
\pgfsetlayers{background,edgelayer,nodelayer,main}

\tikzset{
  none/.style={},
  new edge style 1/.style={draw, thick},
  new edge style 2/.style={draw, thick, dashed},
  new edge style 4/.style={draw, thick, dotted},
  new edge style 5/.style={draw, thick, color=blue},
  new edge style 6/.style={draw, thick, color=blue},
  new edge style 7/.style={draw, thick, fill=none},
  new edge style 8/.style={draw, line width=1.6pt}
}

\definecolor{myblue}{RGB}{0,90,200}

\tikzset{
  areaB/.style={draw=none, fill=myblue!50},
  areaC/.style={draw=none, fill=myblue!30},
  areaA/.style={draw=none, fill=myblue!10},
  fractile/.style={
    draw=blue!70!black,
    line width=0.9pt,
    decorate,
    decoration={random steps, segment length=9pt, amplitude=1.0pt}
  }
}

\tikzset{
    ->,
    >=stealth', 
    node distance=3cm,
    every state/.style={thick}, 
    initial text =,
}

\title{The fiscal limits of efficient trade}
\author{Toan Le \quad \NoCaseChange{\textcircled{\scriptsize r}} \quad Saara H\"{a}m\"{a}l\"{a}inen \quad \NoCaseChange{\textcircled{\scriptsize r}}  \quad C\'{e}sar Barilla}
\date{\today. The symbol \textcircled{\scriptsize r} indicates randomized author order}
\thanks{Toan Le: Helsinki GSE, University of Helsinki; email: \href{mailto.vk.le@helsinki.fi}{toan.vk.le@helsinki.fi}. 
Saara H\"{a}m\"{a}l\"{a}inen: Helsinki GSE, University of Helsinki; email: \href{mailto.hamalainen@helsinki.fi}{saara.hamalainen@helsinki.fi}. 
C\'{e}sar Barilla: University of Oxford, Department of Economics and Nuffield College; email: \href{mailto.barilla@economics.ox.ac.uk}{cesar.barilla@economics.ox.ac.uk}. We thank P\'eter Es\"{o}, Paul Klemperer, Egor Starkov and seminar participants at Oxford and the Nordic Economic Theory Seminar for insightful comments.
}

\begin{document}

\begin{abstract} 
Sustaining efficient trade under private information requires external subsidies. Although a designer cannot avoid deficits on average, they can control \emph{exposure}. We characterize efficient and truthful mechanisms that minimize the worst-case deficit in markets with many buyers and sellers, under interim participation. This problem is dual to maximizing expected trade surplus under correlation ambiguity, with constraints on submarkets with one agent removed. In bilateral trade, the best deficit ceiling is the competitive surplus in a benchmark continuum market, achieved by a ``competitive pivot mechanism.'' In larger markets, optimal mechanisms are characterized in terms of worst-case residual welfare contributions, but solutions exhibit a complex combinatorial structure. We obtain explicit guarantees under natural participation charge restrictions and establish how the best deficit excess decays as market thickens. Our duality characterization is general and extends to many mechanism design environments (public goods, partnership dissolution), and alternative participation constraints.
\vspace{1em}

\\ \textbf{JEL-codes:} D82, D47, D44, C61, H41
\end{abstract}

\maketitle
\pagestyle{plain}

\section{Introduction}\label{sec:introduction}
Private information is a fundamental barrier to efficient trade. When buyers privately know their values for goods and sellers privately know their production costs, truthful and voluntary trading rules that maximize gains from trade must leave information rents to the agents. The lesson of \citet{Vickrey1961} and \citet{MyersonSatterthwaite1983} is that these rents have fiscal consequences. Ensuring efficient trade, incentive compatibility, and voluntary participation is incompatible with budget balance, requiring the market designer to be prepared to finance deficits. That is, efficient trade is fiscally costly.

This impossibility theorem implies that a deficit is unavoidable in expectation. However, it does not say that the designer has no control over its distribution and, therefore, their \emph{exposure} to deficits. Two mechanisms may run the same average deficit, yet one may be safer if it rules out large losses. This is a crucial distinction for policymakers and economic institutions that have a limited budget to cover deficits.\footnote{
    Agencies such as the Congressional Budget Office (US) and the Office for Budget Responsibility (UK) not only produce central forecasts under different economic policies but also assess their exposure under adverse contingencies \citep{CBOIntro2026,OBRWhatWeDo2026}.
    } 
A regulated marketplace, procurement platform, public agency, or clearinghouse may be prepared to bear a cost to facilitate efficient trade, but is typically unable to write an unlimited check.

In this paper, we characterize efficient trading mechanisms that are truthful (in dominant strategies) and minimize worst-case deficits in many-buyer, many-seller markets. Participation is voluntary at the interim stage, when agents only know their own types. From the designer's perspective, solving this problem addresses two practical questions. First, it quantifies the scope of fiscal exposure, by providing the smallest budget sufficient to cover deficits under every type realization. Second, it identifies which mechanisms achieve this guarantee, and how they improve the distribution of deficits.

The class of mechanisms we consider occupies a pragmatic intermediate case between two well-known benchmarks. With truthfulness and stronger ex post participation, the designer is essentially forced to use the Vickrey-Clarke-Groves (VCG) mechanism, leaving no freedom to reshape deficits \citep{Vickrey1961, Clarke1971, Groves1973}. With weaker Bayesian incentive compatibility and interim participation, a well-chosen version of the d'Aspremont-Gérard-Varet (AGV) mechanism can smooth budget imbalances across states and collapse the deficit distribution to its mean \citep{dAspremontGerardVaret1979,Arrow1979,BorgersNorman2009}. The incentive properties we focus on (truthfulness with interim participation) deliver the same \emph{expected} deficit as both VCG and AGV,\footnote{
    This follows from the revenue equivalence theorem \citep[see, e.g.,][]{Borgers2015}.
    } 
but leave room to reshape its \emph{distribution}. Our question of deficit control, therefore, turns on how close tail deficit can be brought to the mean. In our optimal mechanisms, truthful revelation is prior free and robust \citep[in the sense of][]{BergemannMorris2005}, while participation requires only first-order beliefs and is thus simple \citep[in the sense of][]{BorgersLi2019}.

Our main methodological innovation is to connect the \emph{ex post} deficit guarantee problem to an equivalent problem of maximization of \emph{expected} gains from trade under correlation ambiguity in the spirit of, e.g., \citet{Carroll2017}, via a duality approach. In the dual problem, Nature chooses any joint distribution of agents' types such that, whenever any one agent is removed, the remaining agents' types follow the original independent product distribution. Higher-order dependence among all agents is allowed, as long as this dependence disappears in ``leave-one-out'' submarkets. This welfare maximization problem is a multistochastic Monge-Kantorovich problem, which is a generalization of optimal transport recently introduced by \citet{GKZ2019,GKZ2022}.\footnote{
    The multistochastic Monge-Kantorovich problem is a generalization of a classical optimal transport problem, where marginal constraints are replaced by constraints over all dimension $k$ marginal distributions in a space of dimension $d$. Our case with constraints over ``leave-one-out'' couplings, i.e. joint distributions of types of all but one agents, is the $k=d-1$ problem. Little is known about such problems in general, see \citet{GKZ2019,GKZ2022}, which introduced the problem, for details.
    }
Nature's objective is \emph{adversarial} because the social surplus captures information rents, hence the budget of implementing efficient trade. The problem identifies the largest expected deficit that VCG induces with ambiguous correlation. This value coincides with the best type-by-type deficit guarantee despite there being no ambiguity in the original problem, revealing an economic duality between seemingly unrelated problems. Intuitively, the structure of Nature's choice set mirrors participation constraints: the ambiguity set corresponds to joint distributions that leave the expectation of every \emph{participation charge} unchanged.\footnote{
    The duality structure relies on the reduction of admissible mechanisms to the design of participation charges, each of which depends on the reports of all agents except for the agent being charged. Such mechanisms are known as Groves schemes \citep{Groves1973}, and are without loss under efficiency and truthfulness. This turns the deficit control problem into an infinite dimensional linear program.
    }

Bilateral trade has particularly transparent economics. Nature's problem reduces to a standard optimal transport problem that couples high-value buyers with low-cost sellers, concentrating gains from trade in the states that generate the largest informational rents. Even though the actual market contains only two traders, the smallest achievable deficit ceiling equals the competitive surplus in a benchmark market with a unit mass of buyers and a unit mass of sellers, under the exogenously given type distributions. The optimal mechanism uses the benchmark competitive price as a pivot that governs the distribution of information rents, while its team component implements efficient trade \textit{{\`a} la} \cite{Groves1973}: the buyer is charged according to the seller's producer surplus, and the seller is charged according to the buyer's consumer surplus. This solution is an instance of a more generally applicable mechanism that we term the \emph{competitive pivot mechanism}. Efficient and truthful bilateral trade is thus preserved while sharply reducing fiscal exposure. For example, with uniformly distributed types, our competitive pivot mechanism reduces the worst-case deficit from \(1\) under VCG to only \(\sfrac{1}{4}\).

With multiple buyers and sellers, the structure and duality extend, but interactions are richer.  The informational content of participation charges, which can depend on realized types of all \emph{other} agents, disciplines the scope of adversarial dual configurations. From the perspective of each agent, the participation component corresponds to a charge for \emph{joining} a (sub)market of counterparts. The key difference with the bilateral case is that this submarket is now a meaningful trading environment. The designer may only partially extract the expected benefits of entry, as they need to incentivize all types to participate; charges must optimize for the worst-case type, taking each agent as the marginal entrant. This is paralleled in the dual representation, where Nature can no longer simply match every high buyer with every low seller, but must instead satisfy ``leave-one-out'' restrictions: the ambiguity set is such that types in all submarkets obtained by removing one agent are independent. This prevents Nature from concentrating all realizations in the most rent-intensive states. Potentials for the adversarial problem enable the construction and interpretation of solutions. The potentials themselves can be interpreted as worst-case individual contributions to social welfare, taking each agent as marginal. Resulting optimal transfers are VCG transfers (agents pay the externality they impose on others by joining), adjusted by a mean-zero component which captures the ``quality'' of the market they joined, relative to expectations: agents pay to join a ``good'' market and are compensated for joining a ``bad'' one.

The problem has an explicit solution in the case of three agents (e.g., one seller and two buyers) with uniform types. The intricate form of the optimal adversarial distribution illustrates the strength of the leave-one-out constraints, which force Nature into complex manipulations. Its mirror image in the revenue guarantee problem underlines the added power of participation charges in a larger market. The solution relies on a reformulation of the problem as minimizing an expected minimum of three random variables under pairwise independence, which is solved in \citet{Mallows2003}. Optimal potentials exhibit a recursive, self-similar structure. This echoes results in \citet{GKZ2019,GKZ2022} that multistochastic problems can lead to fractal optimal couplings with complex combinatorial structure, even in well-behaved specifications. This suggests that closed-form optimal mechanisms in the general problem are challenging beyond special cases.

A natural route for more practical recommendations is to look for guarantees among constrained classes of mechanisms. We show that several well-chosen restrictions provide economically intuitive and explicit solutions that achieve very good improvements. If participation charges are required to be additively separable, the best guarantee is achieved by the general competitive pivot mechanism. In this case, we recover the intuition from the bilateral trade case and significant, if not maximal, improvement over VCG. Requiring only partial separability (\emph{projection-core mechanism}) or  restricting participation charges to depend only on opposite-side reports (\emph{opposite-side mechanism}) yields further improvements. Lastly, a decomposition of welfare into \emph{price-by-price} components yields a tractable reduction to a collection of binary tests of whether types lie above or below a candidate price. Collating price-by-price solutions yields the \emph{price-decomposable mechanism} and a sharp bound on deficit guarantees. Surprisingly, when all type distributions share a common median and both buyer and seller counts are odd, the bound is exact and the price-decomposable mechanism is truly optimal: it achieves the lowest deficit ceiling. This price-by price decomposition further allows us to investigate the effect of market thickness. We establish that the gap between the optimal deficit ceiling and the mean deficit shrinks exponentially as the number of participants grows.

The approach we develop for trade environments is fully generalizable to other mechanism design problems, and other choices of constraints and objectives. We provide applications to the canonical environments of public good provision and partnership dissolution. The core of our duality method extends verbatim; the specifics of the analysis and the characterization of optimal mechanisms need only be adapted to the social surplus function generated by the environment. We also consider \emph{ex~ante} participation, which is a minor modification of our problem. The same fiscal exposure arises, but the designer can collect larger participation charges; efficient trade can therefore sometimes be implemented without ever running a deficit. We give \emph{necessary and sufficient conditions} under which this is the case; our results nest and generalize the sufficient median condition for achieving strong budget balance from \citet{JagadeesanTordjman2025}.

\subsection*{Related Literature}
The classical benchmark is \citet{MyersonSatterthwaite1983}, who show, in bilateral trade, that efficiency, incentive compatibility, individual rationality, and budget balance are generally incompatible. Subsequent work studies how mechanisms trade off these constraints in three canonical environments: \emph{bilateral, or two-sided, trade} \citep{Chatterjee1983,Williams1987,Williams1999AMechanisms,HagertyRogerson1987,Gresik1989,Satterthwaite1989,McAfee1991continuousquantity,McAffee1992TradeSacrifice,Loertcher2015longway,Delacretaz2019decomposability,LoertscherMezzetti2019,LoertcherMarx2022AER}; \emph{partnership dissolution} \citep*{CramtonGibbonsKlemperer1987,LoertscherWasser2019}; and \emph{public-good provision} \citep{GuthHellwig1986,MailathPostlewaite1990}. A closely related contribution is \citet{JagadeesanTordjman2025}, who show that efficient mechanisms can be deficit-free under ex ante participation. Our paper revisits these tradeoffs from a robust mechanism design perspective.

Our paper relates to VCG redistribution mechanisms. In many allocation problems, particularly multi-unit auctions, VCG raises substantial revenue while preserving efficiency, truthfulness, and \emph{ex post} participation, which can make the mechanism appear extractive \citep{Vickrey1961,Clarke1971,Groves1973}. The redistribution literature asks how much of the VCG revenue can be rebated without sacrificing these incentive properties \citep{Bailey1997,Cavallo2006,GershkovMoldovanu2010}. \citet{Moulin2009} and \citet{GuoConitzer2009} design rebate rules that minimize the worst-case revenue left undistributed. We study a fundamentally mirror problem: when VCG creates deficits, we redesign participation charges to minimize the worst-case deficit.

The paper contributes to robust information and mechanism design. The Wilson doctrine favors rules that rely less on fine details of agents' beliefs \citep{Wilson1987}, and robust implementation asks when outcomes are insensitive to high-order beliefs \citep{BergemannMorris2005, BorgersLi2019}. Two papers are close. \citet{Carroll2017} studies screening under correlation ambiguity, while \citet*{BergemannBrooksMorris2019} compare auction rules by (expected) revenue guarantees across information structures. Our adversary is different. It selects the coupling that keeps all leave-one-out marginals while maximizing fiscal liability. Duality is the key: the greatest VCG ex ante deficit under correlation ambiguity equals the smallest ex post deficit ceiling.

Methodologically, we connect robust fiscal mechanism design to the broad literature on optimal transport \citep{Villani2009} and related problems, in particular its multimarginal \citep{Pass2015} and multistochastic \citep{GKZ2019,GKZ2022} versions. Optimal transport has become a useful tool in economic theory, notably in matching \citep{Ekeland2010,Galichon2016}, multidimensional screening \citep{Rochet1998,Daskalakis2017ecma}, and Bayesian persuasion \citep{DworczakKolotilin2024,KolotilinCorraoWolitzky2025}. Our contribution is to show that a similar structure governs the maxmin budget problem. Our dual is a multistochastic problem over \emph{leave-one-out couplings}: joint distributions whose every one-agent deletion equals the independent prior. This connection is a novel motivation for the (so far little explored) applications of multistochastic transport; our specific problem and solution techniques contribute to the development of that literature.

\section{Model}\label{sec:model}
\subsection{Environment}
We consider a market consisting of a set $N$ of $n$ buyers, each with unit demand, and a set $M$ of $m$ sellers, each with unit supply. All market participants have quasilinear preferences. Each buyer $i \in N$ has a private value $v_i$ independently drawn from a distribution $F_i$ with density $f_i$, yielding a payoff of $u_i^B = v_i q_i^B - t_i^B$ when consuming $q_i^B \in [0,1]$ unit for a payment of $t_i^B$. Symmetrically, each seller $j \in M$ has a private cost $c_j$ independently drawn from a distribution $G_j$ with density $g_j$, yielding a payoff of $u_j^S = - c_j q_j^S -t_j^S$ when producing $q_j^S \in [0,1]$ unit for a receipt of $-t_j^S$. Type distributions share a common support, normalized to $[0,1]$. Goods are perfectly divisible; however, to study indivisible goods, $q_i^B$ and $q_j^S$ can equivalently be interpreted as the probabilities of purchasing and producing the unit. 

By the revelation principle, we restrict our attention to direct mechanisms, which are defined as a pair $({\bf Q}, {\bf T})$ consisting of allocation rule $\mathbf{Q} = (\mathbf{Q}^B, \mathbf{Q}^S)$ and a transfer rule $\mathbf{T} = (\mathbf{T}^B, \mathbf{T}^S)$ for the buyers and sellers, respectively. For any profile of reported values and costs $(\hat{\mathbf{v}}, \hat{\mathbf{c}}) \in [0, 1]^n \times [0, 1]^m$, the mechanism assigns each buyer $i$ a probability of receiving a unit, $Q^B_i(\hat{\mathbf{v}}, \hat{\mathbf{c}}) \in [0,1]$, and each seller $j$ a probability of producing their unit, $Q^S_j(\hat{\mathbf{v}}, \hat{\mathbf{c}}) \in [0,1]$. The mechanism additionally extracts a payment $T^B_i(\hat{\mathbf{v}}, \hat{\mathbf{c}})$ from a buyer $i$ and $T^S_j(\hat{\mathbf{v}}, \hat{\mathbf{c}})$ from a seller $j$, where a negative value implies a transfer from the mechanism to the agent. (Physical) feasibility dictates that the total number of units allocated to buyers cannot exceed the total number of units sourced from sellers, requiring $\sum_{i=1}^n Q^B_i(\hat{\mathbf{v}}, \hat{\mathbf{c}}) \le \sum_{j=1}^m Q^S_j(\hat{\mathbf{v}}, \hat{\mathbf{c}})$ for all $(\hat{\mathbf{v}}, \hat{\mathbf{c}})$. 

For buyer $i$ with true value $v_i$ and seller $j$ with true cost $c_j$, the ex post payoffs under the reporting profile $(\hat{\mathbf{v}}, \hat{\mathbf{c}})$ are given by $U^B_i(v_i; \hat{\mathbf{v}}, \hat{\mathbf{c}})= v_i Q^B_i(\hat{\mathbf{v}}, \hat{\mathbf{c}}) - T^B_i(\hat{\mathbf{v}}, \hat{\mathbf{c}})$ and $U^S_j(c_j; \hat{\mathbf{v}}, \hat{\mathbf{c}})= -c_j Q^S_j(\hat{\mathbf{v}}, \hat{\mathbf{c}}) - T^S_j(\hat{\mathbf{v}}, \hat{\mathbf{c}}).$ The values of outside options for all market participants is normalized to zero.

\subsection{Desiderata}
The market designer's objective is to engineer the rules of voluntary exchange such that private incentives align with the maximization of social surplus, and to do so with minimal worst-case budget deficit. Given realized types $(\mathbf{v},\mathbf{c})$, a mechanism $\mathcal{M}=(\mathbf{Q},\mathbf{T})$ induces ex post deficit:\footnote{
    A negative deficit simply means positive revenue for the designer.
    }
\[
     \mathrm{Deficit} (v,c;\mathcal{M}):= - \sum_{i=1}^n T_i^B(\mathbf{v},\mathbf{c}) - \sum_{j=1}^m T_j^S(\mathbf{v},\mathbf{c}).
\]
The worst-case deficit produced by the mechanism is obtained by taking the largest possible deficit across all possible types:
\[
    \mathrm{Deficit}^\star(\mathcal{M}):= \sup_{(\mathbf{v},\mathbf{c}) \in [0,1]^{n+m}} \; \mathrm{Deficit}(\mathbf{v},\mathbf{c};\mathcal{M}).
\]
In our main specification, the designer's chosen mechanism will be required to verify:
\begin{enumerate}
   \item[\hypertarget{cond:EF}{(EF)}] \textbf{Efficiency.} The allocation rule $(\mathbf{Q}^B,\mathbf{Q}^S)$ maximizes (reported) surplus from trade. Let $v_{(1)} \ge \dots \ge v_{(n)}$ and $c_{[1]} \le \dots \le c_{[m]}$ be the ordered reported types. Trade occurs for exactly $k^*(\mathbf{v},\mathbf{c})$ pairs where $v_{(k^*)} \geq c_{[k^*]}$ and $v_{(k^*+1)} < c_{[k^*+1]}$, allocating $Q^B_i=1$ to the top $k^*$ buyers and $Q^S_j=1$ to the bottom $k^*$ sellers.\footnote{
       In the boundary case where all possible matches generate surplus (i.e, $k^* = \min\{n,m\}$), the market clears entirely. The second inequality is vacuously satisfied by adopting the standard convention for empty order statistics: $v_{(n+1)} = -\infty$ and $c_{[m+1]} = +\infty$.
       }
  \item[\hypertarget{cond:DIC}{(DIC)}] \textbf{Truthfulness.} Truthtelling is a (weakly) dominant strategy for all buyers and sellers. That is, for all buyers $i \in N$ and sellers $j \in M$:
  \begin{eqnarray*}
      v_i &\in& \argmax_{\hat{v}_i\in [0,1]} U_i^B(v_i; \hat{\bf v},\hat{\bf c}), \quad \forall (v_i,\hat{\bf v}_{-i}, \hat{\bf c}) \in [0,1]^{n+m},
      \\
      c_j &\in& \argmax_{\hat{c}_j\in [0,1]} U_j^S(c_j; \hat{\bf v},\hat{\bf c}), \quad \forall (\hat{\bf v}, c_j,\hat{\bf c}_{-j}) \in [0,1]^{n+m}.
  \end{eqnarray*}
  
  \item[\hypertarget{cond:IIR}{(IIR)}] \textbf{Interim Participation}. Agents are willing to participate after observing their own types. For all buyer $i$ and seller $j$ with value and cost $v_i, c_j \in [0,1]$: $\E_{\mathbf{v}_{-i}, \mathbf{c}}[U^B_i(v_i; \mathbf{v}, \mathbf{c})] \ge 0$ and $\E_{\mathbf{v}, \mathbf{c}_{-j}}[U^S_j(c_j; \mathbf{v}, \mathbf{c})] \ge 0.$

\end{enumerate}
The designer's problem is:
\begin{align*}
    & \minimize_{\mathcal{M}} \quad \mathrm{Deficit}^\star(\mathcal{M})
    \qquad \text{ subject to  \hyperlink{cond:EF}{(EF)}, \hyperlink{cond:DIC}{(DIC)}, \hyperlink{cond:IIR}{(IIR)} }
\end{align*}

\subsection{Reformulation with participation charges}
The efficiency constraint pins down the allocation rule; together with incentive compatibility, it partially determines the transfers---up to \emph{participation charges}. This reduces the designer's problem to a choice over these charges, with individual rationality taking the form of an upper bound constraint. We next set up this simpler problem, which forms the basis of our results.

First, define the gains from trade, obtained by maximizing aggregate welfare over all feasible allocations, as a function of realized (or reported) types. This takes the explicit form of the surplus from the efficiently matched pairs:
\[
    W(\mathbf{v},\mathbf{c})
   := \max_{\mathbf{Q} \text{ feasible}} \biggl\{ \sum_{i=1}^n v_i Q_i^B(\mathbf{v},\mathbf{c}) - \sum_{j=1}^m c_j Q_j^S(\mathbf{v},\mathbf{c}) \biggr\}
    = \sum_{k=1}^{\min\{n,m\}} (v_{(k)} - c_{[k]})^{+},
\]
where $x^+:= x \mathds{1}_{x \geq 0}$ denotes the positive part: for any $x \in \mathbb{R}$. To verify the interim participation constraints, we often consider the welfare of the market when specific agents are removed. Let $W_{-i}(\mathbf{v}_{-i},\mathbf{c})$ denote the maximum gains from trade achievable in the sub-economy that excludes buyer $i$. Similarly, let $W_{-j}(\mathbf{v},\mathbf{c}_{-j})$ denote the maximum gains from trade achievable in the sub-economy that excludes seller $j$.

By the Green--Laffont--Holmstr\"om characterization \citep{GreenLaffont1977, Holmstrom1979}, any truthful efficient mechanism with interval type spaces belongs to the family of Groves schemes, with the following transfer rules:
\begin{eqnarray*}
    T^B_i(\mathbf{v}, \mathbf{c}) &=& h^B_i(\mathbf{v}_{-i}, \mathbf{c}) -  \big[W(\mathbf{v}, \mathbf{c})-v_i Q^B_i(\mathbf{v}, \mathbf{c})\big] \label{eqn:BuyerTransfer}
    \\ 
    T^S_j(\mathbf{v}, \mathbf{c}) &=& h^S_j(\mathbf{v}, \mathbf{c}_{-j})-\big[W(\mathbf{v}, \mathbf{c}) + c_j Q^S_j(\mathbf{v}, \mathbf{c})\big], \label{eqn:SellerTransfer} 
\end{eqnarray*}
where $h^B_i: [0,1]^{n+m-1} \to \mathbb{R}$ and $h^S_j: [0,1]^{n+m-1} \to \mathbb{R}$ are \emph{arbitrary} participation charges for each buyer $i\in N$ and each seller $j\in M$. 

The last two terms in these transfer equations constitute the \emph{team component}. This component aligns each agent's incentives with the maximization of social welfare. The remaining components $h^B_i(\mathbf{v}_{-i}, \mathbf{c})$ and $h^S_j(\mathbf{v}, \mathbf{c}_{-j})$ are the \emph{participation charges}. As these charges only depend on the reported type of the \textit{other} participants, they automatically preserve truthful reporting incentives. The following lemma simplifies our design problem by expressing the deficit in terms of participation charges and translating voluntary participation into bounds on average charges.

\begin{lemma}\label{lm:ParticipationCharges} Fix an efficient Groves mechanism $\mathcal{M}= (\mathbf{Q},\mathbf{T})$ as defined above.
\begin{enumerate}
    \item The deficit induced by the mechanism $\mathcal{M}$ when type profile $(\mathbf{v},\mathbf{c}) \in [0,1]^{n+m}$ is realized can be decomposed into:
    \[
        \mathrm{Deficit}(v,c;\mathcal{M}) = (n+m-1) W(\mathbf{v}, \mathbf{c})- \sum_{i=1}^n h^B_i(\mathbf{v}_{-i}, \mathbf{c}) - \sum_{j=1}^m h^S_j(\mathbf{v}, \mathbf{c}_{-j}).
    \]
    
    \item Let $\overline{W}_{-i} = \mathbb{E}[W_{-i}(\mathbf{v}_{-i}, \mathbf{c})]$ and $\overline{W}_{-j} = \mathbb{E}[W_{-j}(\mathbf{v}, \mathbf{c}_{-j})]$ be ex ante welfare in the sub-economies without buyer $i$ and seller $j$. $\mathcal{M}$ satisfies \hyperlink{cond:IIR}{(IIR)} if and only if:
    \[
        \mathbb{E}[h^B_i(\mathbf{v}_{-i}, \mathbf{c})] \le \overline{W}_{-i} \,\,\, \forall i \in N\,\,\,\text{ and  }\,\,\,  
        \mathbb{E}[h^S_j(\mathbf{v}, \mathbf{c}_{-j})] \le \overline{W}_{-j} \,\,\, \forall j \in M.
    \]
\end{enumerate}
\end{lemma}
\hyref{Lemma}{lm:ParticipationCharges} is a standard result in mechanism design; in the following sections, we use it to reduce our deficit control problem to an optimal choice over participation charges. We can immediately see that larger participation charges generate a smaller deficit.
Standard revenue equivalence implies the expected deficit of such mechanisms remains fixed under \hyperlink{cond:EF}{(EF)}, \hyperlink{cond:DIC}{(DIC)}, and \hyperlink{cond:IIR}{(IIR)}.\footnote{
    Here we apply revenue equivalence by assuming that IIR binds at the worst types: buyers with $v_i=0$ and sellers with $c_j=1$ are indifferent between participating and staying out.
    } 
However, by engineering the participation charges, the designer can drastically reduce tail risk.

\subsection{Remarks} 
While we focus on a trade environment, this specific application understates the generality of our framework. The machinery driving our results extends almost verbatim to a wide class of mechanism design environments, as well as different constraints and objectives. We develop these extensions in \hyref{Section}{sec:extensions}. Our analysis carries over with only minimal changes in any environment where the Green--Laffont--Holmstr\"om payment formula holds; alternative applications simply induce different efficient allocation rules and welfare function, or type-dependent outside options. We also consider a variant of our problem under \textit{ex ante individual rationality} (\hyref{Section}{sec:extension-AIR}): agents commit to participating in a mechanism before learning their types. Our analysis extends naturally to this case, requiring only a minor modification of the upper bound on the expected participation charges.

\section{Bilateral trade benchmark}\label{sec:1B1S}
We begin with bilateral trade.
In this setting, where a single buyer with value $v$ faces a single seller with cost $c$, a mechanism is a triplet $(Q, T^B, T^S)$, representing the probability of trade and the individual transfer functions. This benchmark most cleanly illustrates the central economic duality behind the general result. The approach is particularly tractable in this case and admits a closed-form solution; we obtain an optimal mechanism which takes a particularly interesting form: the ``competitive pivot mechanism.''

\subsection{Motivating example.}
Consider bilateral trade with buyer value \(v\) and seller cost \(c\), both independently uniform on \([0,1]\). Efficiency dictates that trade should take place whenever \(v\ge c\). Under VCG, the buyer pays the seller's cost \(c\) and the seller receives the buyer's value \(v\), so the realized VCG deficit is the trade surplus \((v-c)^+.\) More generally,
\[
    \mathbb E[\mathrm{Deficit}^{\mathrm{VCG}}(v,c)] = \int_0^1\int_0^v (v-c)\,\mathrm{d}c\,\mathrm{d}v = \frac16,
    \qquad \max_{v,c\in[0,1]}\mathrm{Deficit}^{\mathrm{VCG}}(v,c)=1.
\]
Thus, even though the expected deficit is modest, the worst-case fiscal exposure is large.

Now anchor the Groves transfers at the competitive price \(p^\star=\sfrac{1}{2}\) that clears the thick benchmark market with demand $1-F(p)=1-p$ and supply $G(p)=p$, by setting:
\[
    T^{B,\mathrm{CP}}(v,c) = c\,\mathbf 1\{v\ge c\}
    + \left(\frac12-c\right)^+
    - \frac18,
    \quad 
    T^{S,\mathrm{CP}}(v,c)
    = -v\,\mathbf 1\{v\ge c\} + \left(v-\frac12\right)^+
    - \frac18.
\]
The added terms, relative to VCG, depend only on the other agent's report, so truthful reporting remains a dominant strategy and efficiency is preserved. The constant $\sfrac{1}{8}$ ensures that participation is voluntary at the interim stage. The resulting deficit is
\[
    \mathrm{Deficit}^{\mathrm{CP}}(v,c) =(v-c)^+ -
    \left(\frac12-c\right)^+ - \left(v-\frac12\right)^+
    + \frac14.
\]
It maintains the expected deficit, \(\mathbb E[\mathrm{Deficit}^{\mathrm{CP}}]=\frac16,\) but limits the upper deficit support:
\[
    \mathrm{Deficit}^{\mathrm{CP}}(v,c)\le \frac14
    \qquad \text{for all }(v,c)\in[0,1]^2.
\]
Hence, the competitive pivot mechanism reduces worst-case deficit exposure from \(1\) to \(1/4\), a 75 percent reduction, without changing the mean deficit. \hyref{Figure}{fig:deficit-distributions} displays the corresponding deficit distributions, including their atoms.

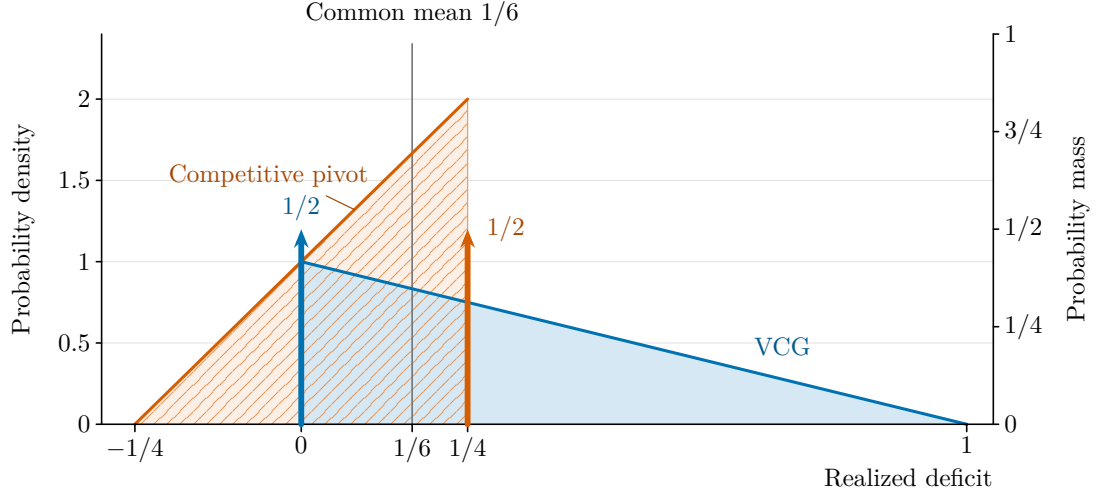
\begin{figure}[t]
    \centering
\begin{tikzpicture}[
    x=8.8cm, y=2.15cm,
    every node/.style={font=\fontsize{9.5}{11.4}\selectfont},
    line cap=round, line join=round,
    density curve/.style={line width=1.15pt},
    fine line/.style={draw=black!60, line width=0.45pt},
    axis/.style={draw=black, line width=0.6pt},
    annotation/.style={draw=CPtext, line width=0.45pt},
    mass arrow/.style={draw=black, line width=2.2pt,
        -{Stealth[length=7pt,width=5.5pt]}}
]
\definecolor{VCGblue}{HTML}{0072B2}
\definecolor{CPorange}{HTML}{D55E00}
\definecolor{VCGtext}{HTML}{005A8D}
\definecolor{CPtext}{HTML}{A74700}
\def\xmin{-0.30}
\def\xmax{1.04}
\def\ymax{2.4}
\pgfmathsetmacro{\meanDeficit}{1/6}
\foreach \h in {0.5,1,1.5,2} {
    \draw[black!12,line width=0.3pt] (\xmin,\h) -- (\xmax,\h);
}
\path[fill=CPorange!10,draw=none]
    (-0.25,0) -- (0.25,2) -- (0.25,0) -- cycle;
\path[fill=VCGblue!16,draw=none]
    (0,0) -- (0,1) -- (1,0) -- cycle;
\path[pattern={Lines[angle=45,distance=3.5pt,line width=0.35pt]},
      pattern color=CPorange!65]
    (-0.25,0) -- (0.25,2) -- (0.25,0) -- cycle;
\draw[fine line,draw=VCGblue!60] (0,0) -- (0,1);
\draw[fine line,draw=CPorange!60] (0.25,0) -- (0.25,2);
\draw[black!55,line width=0.6pt] (\meanDeficit,0) -- (\meanDeficit,2.34);
\node[anchor=south,inner sep=1pt] at (\meanDeficit,2.43)
    {Common mean $1/6$};
\draw[density curve,draw=CPorange] (-0.25,0) -- (0.25,2);
\draw[density curve,draw=VCGblue] (0,1) -- (1,0);
\node[anchor=west,inner sep=0pt,text=CPtext] (cpname) at (-0.20,1.53)
    {Competitive pivot};
\draw[annotation] (0.025,1.43) -- (0.08,1.32);
\node[anchor=south west,inner sep=0pt,text=VCGtext] at (0.68,0.42) {VCG};
\draw[axis] (\xmin,0) -- (\xmax,0);
\draw[axis] (\xmin,0) -- (\xmin,\ymax);
\foreach \h/\lab in {0/0,0.5/0.5,1/1,1.5/1.5,2/2} {
    \draw[axis] (\xmin,\h) -- ({\xmin-0.008},\h);
    \node[anchor=east,inner sep=2pt] at ({\xmin-0.008},\h) {$\lab$};
}
\foreach \x/\lab in {-0.25/{-1/4},0/0,0.1666667/{1/6},0.25/{1/4},1/1} {
    \draw[axis] (\x,0) -- (\x,-0.045);
    \node[anchor=north,inner sep=2pt] at (\x,-0.045) {$\lab$};
}
\node[rotate=90,anchor=south,inner sep=0pt]
    at (-0.40,1.2) {Probability density};
\node[anchor=north east,inner sep=0pt]
    at (\xmax,-0.27) {Realized deficit};
\draw[axis] (\xmax,0) -- (\xmax,\ymax);
\foreach \mass/\lab in {0/0,0.25/{1/4},0.5/{1/2},0.75/{3/4},1/1} {
    \pgfmathsetmacro{\massheight}{\ymax*\mass}
    \draw[axis] (\xmax,\massheight) -- ({\xmax+0.008},\massheight);
    \node[anchor=west,inner sep=2pt]
        at ({\xmax+0.008},\massheight) {$\lab$};
}
\node[rotate=90,anchor=north,inner sep=0pt]
    at (1.15,1.2) {Probability mass};
\draw[mass arrow,draw=VCGblue] (0,0) -- (0,1.2);
\draw[mass arrow,draw=CPorange] (0.25,0) -- (0.25,1.2);
\node[anchor=south,inner sep=2pt,fill=white,text=VCGtext] at (0,1.22) {$1/2$};
\node[anchor=west,inner sep=2pt,fill=white,text=CPtext] at (0.27,1.2) {$1/2$};
\end{tikzpicture}
    \caption{Same mean, different fiscal exposure.}
    \label{fig:deficit-distributions}
    \begin{minipage}{0.9\linewidth}
        \footnotesize\raggedright
        \textit{Notes:} Buyer value and seller cost are independent $U[0,1]$. Densities use the left axis; thick arrows use the probability-mass scale on the right. Each arrow represents mass $1/2$: at deficit $0$ for VCG and $1/4$ for the competitive pivot. Both distributions have mean $1/6$.
    \end{minipage}
\end{figure}

\subsection{Deficit exposure minimization problem}
With one buyer and one seller, there are gains from trade if and only if the value of the buyer exceeds the cost of the seller, hence the social welfare function takes the simple form \(W(v,c)=(v-c)^+\).
As a result, when setting participation charges \(h^B,h^S\) for the buyer and seller in a Groves scheme, the ex post deficit is given by:
\[
    (v-c)^+ - h^B(c) - h^S(v).
\]
The designer's \emph{worst-case deficit exposure} is the maximal ex post deficit across all conceivable type realizations.
A key observation is that we can introduce an auxiliary deficit guarantee variable $\delta$ and replace the objective in the designer's problem by a pointwise constraint that $\delta$ is a deficit ceiling, i.e., everywhere above the realized deficit.
This enables rewriting the problem as a \emph{linear program}.
\begin{align*}
    \textbf{(DG)}\qquad
    \delta^\star := \min_{h^B,h^S,\delta}\;\;  &\delta 
    \\
    \text{s.t.}\qquad
    & (v-c)^+-h^B(c) - h^S(v)   \le \delta \qquad \forall v,c \in [0,1], \label{const:deficit_ceiling} \tag{Deficit ceiling} 
    \\
    & \int_0^1 h^B(c)\,\mathrm{d}G(c) \le 0, \tag{Buyer IIR} 
    \\
    & \int_0^1 h^S(v)\,\mathrm{d}F(v) \le 0. \tag{Seller IIR}
\end{align*}
\textbf{(DG)} is an infinite-dimensional linear program, with a deficit ceiling $\delta$ and participation charges $h^B$ and $h^S$ as choice variables, subject to both pointwise and integral constraints. This formulation of the deficit guarantee problem will be our workhorse throughout the paper; most importantly, it reveals a deep connection to a seemingly unrelated problem.

\subsection{Adversarial surplus maximization problem}
The central idea of our approach is that the designer's problem admits an equivalent (dual) formulation. Consider the following problem, economically meaningful but a priori unrelated: maximizing expected gains from trade under correlation ambiguity. Call the fictional decision-maker in this decision problem Nature. Nature chooses a joint distribution $\lambda$ of the buyer value and seller cost, which must be consistent with their exogenously specified individual distributions $F$ and $G$. The objective is the expectation of the gains from trade $(v-c)^+$ under this joint distribution. The problem is therefore:
\[
   \textbf{(AS)}\qquad \max_{\lambda \in \Lambda(F,G)} \int (v-c)^+\,\mathrm{d}\lambda,
\]
where $\Lambda(F,G)$ denotes the set of couplings with marginals $F$ and $G$, that is, the set of all joint distributions of buyer values and seller costs consistent with the priors. We call this problem adversarial because, from the perspective of the designer, this maximizes expected information rents for the agents. Another way to interpret this problem is as maximizing the expected VCG deficit under correlation ambiguity; see, e.g., \citet{Carroll2017}. Formally, this is a classical optimal transport problem.

\subsection{Duality}
The key claim is that the two problems we introduced are dual to each other, in the sense that the revenue guarantee from optimal deficit exposure is equal to the value in the adversarial surplus maximization problem:
\begin{align*}
    \delta^\star = \max_{\lambda \in \Lambda(F,G)}
    \int (v-c)^+\,\mathrm{d}\lambda.
\end{align*}
This draws a surprising link between two forms of robustness: on the one hand, a \emph{pointwise} objective, where the worst-case is over type \emph{realizations}, on the other hand, an \emph{expected} objective, where the worst-case is over type \emph{correlation}. It is worth emphasizing that this formal equivalence holds despite the fact that there is no actual ambiguity in our environment (types are independent and their distributions are known). Rather, ambiguity is present only in the dual problem, in a sense, as an ``as if'' representation.

This equivalence will be proven more generally in \hyref{Theorem}{thm:duality_nm}; nonetheless, it can be derived directly here. To see where the result comes from, heuristically, observe that the pointwise deficit-ceiling constraint gives rise to a nonnegative dual multiplier $\lambda(v,c)$ satisfying $\int \lambda(v,c)\,\mathrm{d}v\,\mathrm{d}c=1$. Because this constraint must hold for each type realization $(v,c)$, $\lambda$ can be interpreted as a probability measure over buyer values and seller costs. The participation constraints imply that the marginals of $\lambda$ must agree with the priors.

\subsection{Optimal coupling} 
Since the welfare function $W(v,c) = (v-c)^+$ is submodular, classical results from optimal transport theory imply that the expectation is maximized by perfectly negatively correlating the random variables (\emph{``countermonotone coupling''}). Such a coupling is constructed by linking the quantiles of the marginal distributions in reverse order. Let $Z$ be a random variable \emph{uniformly distributed} on the interval $[0,1]$. The countermonotone coupling $\lambda^\star$ is defined by the exact type assignment:
\[
    v = F^{-1}(Z), \qquad c = G^{-1}(1-Z).
\]
The expected gains from trade can thus be expressed as
\[
    \E_{\lambda^\star}[(v-c)^{+}]=\int_0^1\big(F^{-1}(z)-G^{-1}(1-z)\big)^+\,\mathrm{d}z=\int_0^1 \min\{ 1-F(p), G(p)\} \,\mathrm{d}p.
\]

Nature seeks to maximize the average surplus under ambiguous correlation. To do so, intuitively, it correlates the highest-valuation buyers (the top percentiles of $F$) exclusively with the lowest-cost sellers (the bottom percentiles of $G$). This negative correlation concentrates the maximum possible probability mass into states where gains from trade exist ($v > c$) and aggressively stretches trade surplus $v-c$ for those matched pairs, thus maximizing the expected information rent $(v-c)^+$ that the designer is forced to finance.

\begin{figure}[t]
    \centering
\begin{tikzpicture}[
    x=8.3cm, y=6.7cm,
    every node/.style={font=\fontsize{9.5}{11.4}\selectfont},
    line cap=round, line join=round,
    declare function={
        demand(\p)=1-\p;
        supply(\p)=\p+0.6*\p*(1-\p);
        independent(\p)=demand(\p)*supply(\p);
    },
    demand curve/.style={draw=black, line width=1.2pt},
    supply curve/.style={draw=black, line width=1.2pt},
    mean curve/.style={draw=black, line width=0.65pt},
    guide/.style={draw=black!60, line width=0.35pt},
    axis/.style={draw=black, line width=0.6pt,
                 -{Latex[length=1.8mm,width=1.2mm]}}
]
\definecolor{MeanFill}{HTML}{BDD7E7}
\definecolor{CeilingFill}{HTML}{F4E6C7}
\pgfmathsetmacro{\pclear}{(2.6-sqrt(4.36))/1.2}
\pgfmathsetmacro{\qclear}{1-\pclear}

\path[fill=CeilingFill, draw=none]
    (0,0)
    -- plot[domain=0:\pclear, samples=120, variable=\t]
       ({supply(\t)},\t)
    -- plot[domain=\pclear:1, samples=120, variable=\t]
       ({demand(\t)},\t)
    -- cycle;

\path[pattern={Dots[distance=4pt,radius=0.55pt]},
      pattern color=black!70, draw=none]
    (0,0)
    -- plot[domain=0:\pclear, samples=120, variable=\t]
       ({supply(\t)},\t)
    -- plot[domain=\pclear:1, samples=120, variable=\t]
       ({demand(\t)},\t)
    -- cycle;

\path[fill=MeanFill, draw=none]
    (0,0)
    -- plot[domain=0:1, samples=180, variable=\t]
       ({independent(\t)},\t)
    -- cycle;

\draw[guide] (0,\pclear) -- (\qclear,\pclear);
\draw[guide] (\qclear,0) -- (\qclear,\pclear);

\draw[black!30,line width=0.35pt] (0,1) -- (1,1) -- (1,0);

\draw[demand curve] (0,1) -- (1,0);
\draw[supply curve]
    plot[domain=0:1, samples=180, variable=\t]
    ({supply(\t)},\t);
\draw[mean curve]
    plot[domain=0:1, samples=180, variable=\t]
    ({independent(\t)},\t);

\draw[axis] (0,0) -- (1.045,0);
\draw[axis] (0,0) -- (0,1.045);
\draw[line width=0.5pt] (1,0) -- (1,-0.012);
\draw[line width=0.5pt] (0,1) -- (-0.012,1);
\draw[line width=0.5pt] (\qclear,0) -- (\qclear,-0.012);
\draw[line width=0.5pt] (0,\pclear) -- (-0.012,\pclear);
\node[anchor=north east, inner sep=1pt] at (-0.012,-0.012) {$0$};
\node[anchor=north, inner sep=2pt] at (1,-0.015) {$1$};
\node[anchor=east, inner sep=2pt] at (-0.015,1) {$1$};
\node[anchor=north, inner sep=2pt] at (\qclear,-0.015) {$q^\star$};
\node[anchor=east, inner sep=2pt] at (-0.015,\pclear) {$p^\star$};
\node[anchor=west, inner sep=0pt] at (0,1.085) {Price $p$};
\node[anchor=east, inner sep=0pt] at (1.045,-0.12) {Quantity $q$};

\node[anchor=east, inner sep=0pt] at (0.91,0.88) {Supply $G(p)$};
\node[anchor=west, inner sep=0pt] at (0.60,0.07) {Demand $1-F(p)$};
\fill[black] (\qclear,\pclear) circle[radius=1.25pt];
\node[anchor=west, inner sep=0pt] at ({\qclear+0.05},\pclear) {$E$};

\node[align=center, inner sep=1pt] at (0.15,0.58) {mean\\deficit};
\node[align=center, inner sep=1pt] at (0.61,0.65)
    {best\\deficit excess};
\draw[black, line width=0.45pt] (0.49,0.60) -- (0.36,0.59);
\fill[black] (0.36,0.59) circle[radius=0.7pt];

\node[anchor=west, inner sep=1pt] (productlabel) at (0.33,0.80)
    {$(1-F(p))\times G(p)$};
\draw[black, line width=0.45pt,
      -{Latex[length=1.3mm,width=0.9mm]}]
    (productlabel.west) -- ({independent(0.70)},0.70);
\end{tikzpicture}
    \caption{Optimal budget exposure in bilateral trade.}
    \begin{minipage}{0.9\linewidth}
        \footnotesize\raggedright
        \textit{Notes:} The blue shaded area is the average deficit; the dotted amber area is the gap to the optimal deficit ceiling. The combined shaded area equals the competitive surplus.
    \end{minipage}
    \label{fig:1B1S}
\end{figure}
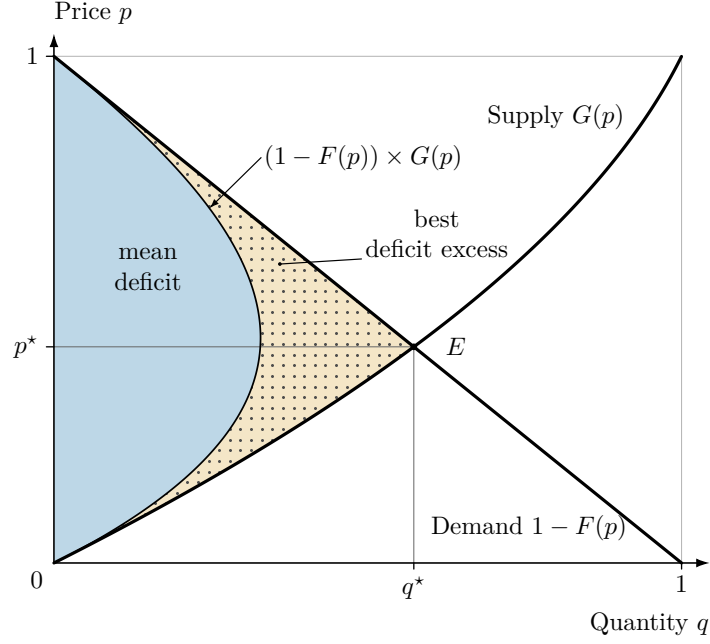

\subsection{The limits of efficient trade}
The explicit solution of the adversarial dual enables a precise characterization of the optimal deficit guarantee.
\begin{proposition} 
\label{prop:duality_1B1S} 
The minimum deficit ceiling for bilateral trade is:
\[
    \delta^{\star} = \int_0^1 \min\{1-F(p), G(p)\} \,\mathrm{d}p.
\]
\end{proposition}

The proposition also admits a competitive-market interpretation, illustrated in \hyref{Figure}{fig:1B1S}. Consider a thick market with a unit mass of buyers and a unit mass of sellers, whose types follow $F$ and $G$. Demand is $1-F(p)$ and supply is $G(p)$, so the competitive price $p^\star$ satisfies \( 1-F(p^\star)=G(p^\star).\)
Buyers with values above $p^\star$ trade with sellers whose costs lie below it. Social surplus is the sum of consumer and producer surpluses:
\[
    \int_{p^\star}^{1}(v-p^\star)\,\mathrm{d}F(v)
    +\int_{0}^{p^\star}(p^\star-c)\,\mathrm{d}G(c)
    =\int_0^1\min\{1-F(p),G(p)\}\,\mathrm{d}p.
\]

\noindent The competitive price decentralizes this surplus-maximizing allocation.\footnote{
    This yields \emph{another} optimal coupling. Let $q^\star=1-F(p^\star)=G(p^\star)$. With probability $q^\star$, draw $v$ and $c$ independently from their respective distributions conditional on $v>p^\star$ and $c<p^\star$. With probability $1-q^\star$, draw them independently conditional on $v<p^\star$ and $c>p^\star$.
    } 
Duality identifies this planner's maximal surplus with the designer's minimal deficit ceiling.\footnote{
    \citet{LoertscherMuir2025} quantify matching gains in growing markets relative to efficient bilateral trade, using a thick competitive market as the benchmark. Our duality identifies this benchmark's surplus with the minimum deficit ceiling in efficient bilateral trade.
    }

The mean deficit also has a simple expression. Under independent types, with participation constraints binding
at the worst types, it equals
\(
    \int_0^1(1-F(p))G(p)\,\mathrm{d}p.
\)\footnote{
    For example, see \citet[p.~279]{MyersonSatterthwaite1983}.
    }
This expression permits an intuitive graphical comparison with the optimal deficit ceiling in \hyref{Figure}{fig:1B1S}. The full shaded area is the competitive surplus and hence the optimal deficit ceiling. The solidly shaded region, bounded by the curve $(1-F(p))\,G(p)$, is the average deficit. The dotted remainder is therefore the best deficit exposure above this average.
\subsection{The competitive pivot mechanism}
In practice, how can the designer cut the worst-case deficit down from the trivial upper bound of $1$ to the competitive market benchmark surplus? To operationalize the dual bound and construct the optimal mechanism, we exploit a geometric property of the gains from trade, which we term ``wedge domination.''  For any arbitrary reference price $p \in [0,1]$ and any realized types $(v,c)$, the gains from trade are universally bounded by the sum of the buyer's and seller's respective surpluses relative to $p$:
\[
    (v-c)^{+} \;\le\;    (v-p)^{+} + (p-c)^{+} .
\]
Taking expectations over any adversarial coupling \(\lambda \in \Lambda(F,G)\), the left side evaluates to the expected adversarial surplus, while the right side depends \emph{only} on the fixed marginals $F$ and $G$. This establishes that \( \E_{\lambda}[(v-c)^{+}] \le \E_{G}[(p-c)^{+}] + \E_{F}[(v-p)^{+}] \) universally.

This upper bound is minimized at the competitive price $p^{\star}$ where \( G(p^{\star}) = 1-F(p^{\star}) \). The sharpest upper bound is precisely the competitive social surplus in a benchmark thick economy.  This gives an explicit solution for the Kantorovich dual potentials of the transport problem: \(\phi^B(c):= (p^{\star}-c)^{+}\) and \(\phi^S(v):= (v-p^{\star})^{+}\). We can then use these to construct optimal transfer rules in our original guarantee problem.

\begin{proposition}[Competitive Pivot Mechanism]\label{prop:OptimalTransfers_1B1S}
The minimum deficit ceiling $\delta^\star$ for bilateral trade is attained by the following participation charges:
\begin{align*}
    h^{B, \mathrm{CP}}(c) &= (p^{\star}-c)^{+} - \int_0^{p^{\star}} G(p)\,\mathrm{d}p, 
    \\
    h^{S, \mathrm{CP}}(v) &= (v-p^{\star})^{+} - \int_{p^{\star}}^1 (1-F(p))\,\mathrm{d}p.
\end{align*}
\end{proposition}

To see why this specific construction is optimal, we can verify the interim participation constraints, and that it guarantees the deficit $\delta^\star$ from \hyref{Proposition}{prop:duality_1B1S}.\footnote{
    \citet{JagadeesanTordjman2025} use a variant of competitive pivot to achieve efficient, truthful, deficit-free bilateral trade under ex ante IR when the buyer's median is at least the seller's.
}
First, we verify voluntary participation. The expected participation charge extracted from the buyer, evaluating over the seller's prior distribution, is exactly zero:
\[
    \E[h^{B,\mathrm{CP}}(c)] = \E[(p^{\star}-c)^{+}] - \int_0^{p^{\star}} G(p)\,\mathrm{d}p =  0.
\]
Symmetrically, \(\E[h^{S, \mathrm{CP}}(v)] = 0\) for the seller. Thus, the interim participation constraint \hyperlink{cond:IIR}{(IIR)} is satisfied. Second, we evaluate the ex post deficit generated by the mechanism: \((v-c)^+- h^B(c) - h^S(v)\). Substituting the optimal transfer rules yields:
\[
    \mathrm{Deficit}^\mathrm{CP}(v,c) =  \int_0^{p^{\star}} G(p)\,\mathrm{d}p + \int_{p^{\star}}^1 (1-F(p))\,\mathrm{d}p -\Bigl[ (p^{\star}-c)^{+} + (v-p^{\star})^{+} - (v-c)^+ \Bigr].
\]
By the wedge domination inequality, the bracketed term is bounded below by zero for all possible value and cost realizations. Therefore, the global maximum occurs exactly when the bracketed term binds at zero, which confirms that the competitive pivot mechanism bounds the worst-case deficit to the competitive surplus.

It is instructive to construct the competitive pivot from VCG. There are no charges under VCG: whenever trade occurs, the buyer pays the cost and the seller receives the value, leaving the designer to subsidize the difference. The competitive pivot additionally charges each agent the other side's realized surplus at $p^\star$ and refunds its expected value. The net adjustment is therefore a ``surplus surprise'': each agent pays extra when the market is better than expected and is compensated when it is worse. Low cost is good news for the buyer, while high value is good news for the seller. Before refund, at least one side of every trade settles at $p^\star$. The charges also apply without trade, ensuring that the designer's subsidy needs never exceed the competitive total surplus.

\section{General two-sided trade environment}\label{sec:nBmS}
We now turn to the multilateral trade environment with many buyers and sellers. The economic duality between deficit guarantee and expected social welfare generalizes, but the larger market reveals that the underlying correlation ambiguity is more complex than the bilateral trade case suggests. This points to a structural interpretation of optimal mechanisms via a conservative welfare contribution formulation.

\subsection{Deficit exposure minimization problem}
As in the bilateral trade case, the first step is to formulate the designer's objective as a pointwise constraint: to minimize the worst-case deficit, we introduce the guarantee as an auxiliary choice variable $\delta$ subject to the pointwise constraints that $\delta$ uniformly bounds the ex post deficit from above. 
Using the inequality expressions for \hyperlink{cond:IIR}{(IIR)} derived in \hyref{Lemma}{lm:ParticipationCharges}, the designer seeks to solve:
\[
    \begin{aligned}
    \textbf{(DG)}\qquad
    \delta^\star = & \inf_{\mathbf{h}^B, \mathbf{h}^S, \delta}\;\; \delta 
    \\
    \text{s.t.}\quad
    & (n+m-1)W(\mathbf{v}, \mathbf{c}) - \sum_{i=1}^n h^B_i(\mathbf{v}_{-i}, \mathbf{c}) - \sum_{j=1}^m h^S_j(\mathbf{v}, \mathbf{c}_{-j}) \leq \delta \qquad \forall (\mathbf{v}, \mathbf{c}), 
    \\
    & \mathbb{E}[h^B_i(\mathbf{v}_{-i}, \mathbf{c})] \le \overline{W}_{-i} \qquad \forall i \in N, 
    \\
    & \mathbb{E}[h^S_j(\mathbf{v}, \mathbf{c}_{-j})] \le \overline{W}_{-j} \qquad \forall j \in M.
    \end{aligned}
\]
As before, this is an infinite-dimensional \emph{linear program} with functions as choice variables, and both pointwise and integral constraints. In the bilateral trade case, having explicit solutions enabled us to gloss over the appropriate choice for the class of functions that participation charges must belong to. This is a more delicate problem in the general case, which we will discuss in \hyref{Section}{sec:existence-of-potentials-nm}. Regardless of attainment, continuous participation charges provide a convenient class over which to formulate the primal problem and establish duality. Since they are dense in the larger space of integrable functions---the natural minimal class to ensure that the integral constraints are well defined---the value $\delta^\star$ is the same across both classes. 
Hence we omit the functional class specification from the problem at this stage, and specify the problem as an infimum rather than a minimum.

\subsection{Adversarial surplus maximization problem}\label{sec:adversarial-problem-nm}
Analogously to the bilateral trade case, we introduce the problem of an adversarial nature, which can manipulate the \emph{correlation} of types to maximize the \emph{expected information rents} for the agents (equivalently: social surplus). The main difference is that, with many buyers and many sellers, we consider a more restrictive class of correlational manipulations. Let \(\Lambda(\mathbf{F},\mathbf{G})\) be the set of joint distributions over types which preserve what we term ``leave-one-out independence'' under the prescribed value and cost distributions \(\mathbf{F}=(F_i)_{i=1,..,n}\) and \(\mathbf{G}=(G_j)_{j=1,...,m}\). This means that for any $\lambda \in \Lambda(\mathbf{F},\mathbf{G})$, the induced distribution over the types of all but one agent (for any choice of the agent being left out), is the product measure. The marginals excluding any buyer $i$ and seller $j$ must satisfy, respectively,
\[
    \lambda_{-i} = \prod_{k\in N\setminus\{i\}}F_k \otimes\prod_{\ell\in M}G_\ell,
    \qquad
    \lambda_{-j} = \prod_{k\in N}F_k \otimes\prod_{\ell\in M\setminus\{j\}}G_\ell.
\]
This leads to the following surplus maximization problem:
\[
    \textbf{(AS)} \qquad \max_{\lambda\in\Lambda(\mathbf{F},\mathbf{G})} \int W\,\mathrm{d}\lambda.
\]
Notice that in the case of one buyer and one seller, the ``leave-one-out'' independence constraints have no bite beyond the prescription of marginals, so this problem collapses back to a standard optimal transport problem. However, with multilateral trade, this is no longer an optimal transport problem but instead a \emph{multistochastic Monge-Kantorovich} problem---a generalization recently introduced by \citet{GKZ2019,GKZ2022}.

As before, and perhaps even more so, this problem may look a priori unrelated to our guarantee problem, but our duality approach reveals a deep connection. The independence requirements on the ambiguity set will prove to be tightly related to the informational content of the participation charges, which can condition on everyone else's type but the agent being charged.

\subsection{Duality}\label{sec:duality-nm}
The following duality theorem relates the two problems. It establishes that the economic duality between pointwise guarantee and ex ante guarantee with correlation ambiguity is general, and it is our main technical tool. Its proof, which can be found in \hyref{Appendix}{app:omitted-proofs}, follows from convex duality techniques.
\begin{theorem}[Strong Duality]\label{thm:duality_nm}
The optimal worst-case deficit for multilateral trade is:
\[
    \delta^{\star} = (n+m-1) \max_{\lambda \in \Lambda(\mathbf{F}, \mathbf{G})} \mathbb{E}_{\lambda}\bigl[W(\mathbf{v}, \mathbf{c})\bigr] - \sum_{i=1}^n \overline{W}_{-i}  - \sum_{j=1}^m \overline{W}_{-j}.
\]
\end{theorem}
Several observations are in order.  Two new terms appear relative to the case of bilateral trade: a market size scaling ($n+m-1$) for the adversarial surplus value, and the sum of expected welfare in leave-one-out submarkets. One can easily verify that this subsumes \hyref{Proposition}{prop:duality_1B1S}: with only one buyer ($n=1)$ and one seller ($m=1$), the adversarial problem collapses to standard optimal transport and the constants are zero (expected welfare in a submarket where either the seller or buyer withdraws is null). The non-zero welfare terms capture that, unlike with one buyer and one seller, the designer is able to reappropriate part of the social welfare via the participation charges (a point that will be echoed in the comparative statics of \hyref{Section}{sec:CS-market-thickness}), because agents are no longer necessarily pivotal for the existence of profitable trade.

To relate this back to the interpretation in terms of VCG deficit (which is no longer \emph{exactly} the social welfare) observe that the problem can be written in the following equivalent form by including the market size scaling and constant terms.
\begin{align*}
    \delta^\star 
    & = - \min_{\lambda \in \Lambda(\mathbf{F}, \mathbf{G})} \int W \,\mathrm{d}\lambda - \sum_i \biggl( \int W \,\mathrm{d}\lambda - \overline{W}_{-i}  \biggr) -\sum_j \biggl(  \int W \,\mathrm{d}\lambda - \overline{W}_{-j} \biggr)
    \\
    & =  - \min_{\lambda \in \Lambda(\mathbf{F}, \mathbf{G})} \int \biggl( W - \sum_i \bigl( W - W_{-i} \bigr)  -\sum_j \bigl(  W  - W_{-j} \bigr) \biggr) \,\mathrm{d}\lambda.
\end{align*}
The first equality simply rearranges terms, and the second uses the leave-one-constraint which implies that $\overline{W}_{-i} = \mathbb{E}[W_{-i}(\mathbf{v}_{-i},\mathbf{c})] = \int W_{-i} \,\mathrm{d}\lambda$ for all admissible $\lambda$. The integrand is exactly the VCG revenue, which can also be interpreted as follows: the designer is the ex post residual claimant to social welfare $W$ but each agent must retain a rent which is their marginal contribution to social welfare ($W-W_{-i}$). Nature's problem, and the optimal deficit guarantee, is a worst-case in the expectation of this object, over joint distributions of type. Given the structure of the mechanism only discriminates on the leave-one-out marginals, individual contributions to welfare are left free in the extra correlational dimension; the optimal $\lambda^\star$ generates the most adversarial distribution by concentration mass towards states where trade is unusually efficient in a way that cannot be detected by participation charges.

\subsection{Conservative welfare contributions}\label{sec:dual-potentials-nm}
The duality result in \hyref{Theorem}{thm:duality_nm} provides a way to compute the value of the optimal deficit guarantee, but does not yet provide a mechanism that achieves it.

Mirroring the approach for the bilateral trade case, we can construct optimal participation charges using the potentials of the adversarial problem. These potentials are obtained by taking another dual in Nature's surplus maximization problem, yielding a welfare imputation problem. General explicit solutions for the adversarial coupling and its potentials remain challenging in the multistochastic case (we discuss the extent to which we can obtain characterizations in \hyref{Sections}{sec:2B1S}, \ref{sec:geometry-of-optimal-couplings}, and \ref{sec:price-by-price-decomposition}). However, the definition of optimal potentials and the construction of participation charges suggest an economic interpretation of the optimal mechanism in terms of a conservative imputation of individual contributions to welfare.

The idea underlying the imputation problem is to assign each agent a potential contribution to welfare associated with entering the market (``individual imputation''). This imputation is conservative in the sense that each participant's contribution depends only on the types of the other agents, and the sum of individual contributions must cover welfare for every realized type vector. The precise problem is to find one such imputation which minimizes total expected welfare contributions.
\begin{align*}
    \textbf{(WI)} \qquad \inf_{(\phi^B,\phi^S)} & \sum_{i=1}^n \int \phi_i^B \,\mathrm{d}F_{-i} \,\mathrm{d}G + \sum_{j=1}^m \int \phi^S_j \,\mathrm{d}F \,\mathrm{d}G_{-j}
    \\
    \text{ subject to }& \sum_{i=1}^n \phi_i^B(\mathbf{v}_{-i},\mathbf{c}) + \sum_{j=1}^m \phi_j^S(\mathbf{v},\mathbf{c}_{-j}) \geq W(\mathbf{v},\mathbf{c}) \; \; \forall \mathbf{v},\mathbf{c}
\end{align*}
This problem is itself dual to the adversarial surplus problem \textbf{(AS)}, in the sense that it has the same value. It also satisfies the following primal dual relation: for any optimal $\bm \phi^B = (\phi^B_i)_{i\in N}$ and $\bm\phi^S = (\phi^S_j)_{j\in M}$ and any $(\mathbf{v},\mathbf{c})$ in the support of the optimal $\lambda^\star$, we must have:
\[
    W(\mathbf{v},\mathbf{c}) = \sum_{i=1}^n \phi^B(\mathbf{v}_{-i},\mathbf{c}) + \sum_{j=1}^m \phi_j^S(\mathbf{v},\mathbf{c}_{-j}).
\]
A clearer way to see that $\phi_i^B(\mathbf{v}_{-i},\mathbf{c})$ captures a self-consistent notion of agent $i$'s \emph{minimal} contribution from entering the market is to observe that for any optimal $(\bm{\phi}^B,\bm{\phi}^S)$ in the above problem (almost surely):
\[
    \phi^B_i(\mathbf{v}_{-i},\mathbf{c}) = \inf_{v_i \in [0,1]} \bigg\{  W(\mathbf{v},\mathbf{c}) - \sum_{k \neq i} \phi_k^B(\mathbf{v}_{-k},\mathbf{c}) - \sum_j \phi_j^S(\mathbf{v},\mathbf{c}_{-j}) \bigg \}.
\]
In other words, if $(\bm{\phi}^B_{-i},\bm{\phi}^S)$ quantifies the individual contributions of \emph{other} agents to welfare as a function of the market they face, then $\phi^B_i(\mathbf{v}_{-i},\mathbf{c})$ is a worst-case over the residual welfare not attributed to other agents, taken over $i$'s type.

\subsection{Optimal mechanisms}\label{sec:optimal-mechanisms-nm}
We can use the individual conservative welfare contributions (i.e. the dual potentials in the adversarial problem) to construct optimal participation charges in the guarantee problem.
\begin{theorem}[Deficit-guarantee Mechanism]\label{thm:optimal-mechanism-from-potentials}
Assume that $(\bm{\phi}^B,\bm{\phi}^S)$ solve the welfare imputation problem.
Then, the optimal deficit guarantee $\delta^\star$ is attained by a Groves transfers with the following participation charges:
\begin{align*}
    h_i^{B}(\mathbf{v}_{-i},\mathbf{c}) &= \overline{W}_{-i} + (n+m-1) \bigl(  \phi_i^B(\mathbf{v}_{-i},\mathbf{c}) - \mathbb{E}[\phi_i^B] \bigr), 
    \\
    h_j^{S}(\mathbf{v},\mathbf{c}_{-j}) &= \overline{W}_{-j} + (n+m-1) \bigl(  \phi_j^S(\mathbf{v},\mathbf{c}_{-j}) - \mathbb{E}[\phi_j^S] \bigr).
\end{align*}
\end{theorem}

This result suggests an interpretation of optimal mechanisms in terms of the \emph{worst-case trade opportunities} created by an agent joining, which is directly related to the \emph{quality} of the market they face. The optimal participation charge combines the ex ante expected welfare $\overline{W}_{-i}$ in the submarket without $i$ with an adjustment reflecting the \emph{worst-case trade opportunities} in the market they face. If $\phi_i^B(\mathbf{v}_{-i},\mathbf{c})$ is large relative to its expectation $\mathbb{E}[\phi_i^B]$, the conservative welfare contribution imputed to agent $i$ is large \emph{even} in the worst case over their type. Thus, even if their type is ``bad'', the market supports a relatively high welfare contribution, suggesting that they face ``good'' trading partners.
This adjustment is scaled by the size of the submarket, $(n+m-1)$, to make it commensurable with social welfare.

With this interpretation,  the resulting transfer rule is directly analogous to VCG, except for the notion of counterfactual welfare. Recall that VCG transfers are given by:
\[
    T^{B,\mathrm{VCG}}_i(\mathbf{v},\mathbf{c}) = W_{-i}(\mathbf{v}_{-i},\mathbf{c}) - \bigl( W(\mathbf{v},\mathbf{c}) - v_i Q_i^B(\mathbf{v},\mathbf{c}) \bigr).
\]
This is the celebrated ``pay-the-externality'' formulation: $i$ pays the difference between the counterfactual welfare in the submarket without $i$, and the welfare in the market including $i$ net of $i$'s private value.
By contrast, optimal transfers in the deficit guarantee problem are given by:
\[
     T_i^B(\mathbf{v},\mathbf{c}) = \overline{W}_{-i} + (n+m-1) \bigl(\phi_i^B(\mathbf{v}_{-i},\mathbf{c}) - \mathbb{E}[\phi_i^B] \bigr) - \bigl( W(\mathbf{v},\mathbf{c}) - v_i Q_i^B(\mathbf{v},\mathbf{c}) \bigr).
\]
These optimal transfers have the same structure as VCG transfers, but replace the welfare of the submarket without $i$ with an ex ante benchmark adjusted by a conservative welfare imputation.
An alternative rewriting in terms of the VCG transfer plus a correction term is insightful:
\[
    T_i^B(\mathbf{v},\mathbf{c}) = T^{B,\mathrm{VCG}}_i(\mathbf{v},\mathbf{c}) +
    \bigl((n+m-1) \phi_i^B - W_{-i} \bigr) (\mathbf{v}_{-i},\mathbf{c}) - \mathbb{E}\bigl[(n+m-1)\phi_i^B - W_{-i} \bigr].
\]
The correction term measures the deviation of $(n+m-1)\phi_i^B-W_{-i}$ from its mean, capturing the \emph{quality} of the market that $i$ gets to join under conservative assumptions about their type.
In a ``good'' market, this difference is high relative to its ex ante expectation, and $i$ is \emph{charged more} than under VCG. In a ``bad'' market, the difference is low, and $i$ is \emph{charged less} than under VCG (or even possibly compensated). Intuitively, the optimal mechanism adjusts VCG transfers to charge agents more when they face favorable trading opportunities under worst-case type dependence, and less when such opportunities are unfavorable.

\subsection{Existence of optimal participation charges}
\label{sec:existence-of-potentials-nm}
Attainment of the optimal deficit guarantee by regular (e.g. continuous) participation charges is surprisingly challenging.\footnote{
    Existence of regular dual potentials for the multistochastic problem \textbf{(AS)}, is known to be difficult. Known failures exist for the multistochastic Monge-Kantorovich problem \citep[see][for an example]{GKZ2022}, even with a very regular objective; analogue failures are found in multimarginal transport and martingale transport. The result also amounts to regularization of solutions in an infinite-dimensional linear program with a highly degenerate constraint structure. Concretely, the fact that dual potentials for this problem depend on overlapping coordinates means irregularities may propagate and compensate, despite the regularity of $W$. This in particular prevents the adaptation of proof techniques from standard optimal transport. Theorem 5.19 in \citet{GKZ2022} is the best available result.
    }
Existence can, nonetheless, be guaranteed in the class of integrable functions $L^1([0,1]^{n+m-1})$. This yields participation charges that attain the optimal deficit guarantee almost surely.\footnote{
    Technically, up to a zero thickness set, which is strictly stronger. See \hyref{Appendix}{app:proof-existence-prop}.
    }
We can improve on optimal $L^1$ potentials and show existence of solutions such that the guarantee holds for \emph{every} type profile, but at the cost of allowing unbounded payments in measure-zero sets.
\begin{proposition}
\label{prop:existence-L1-bounded-below-potentials}
    The optimal value in the deficit guarantee problem is attained within $L^1([0,1]^{n+m-1})$ by (i) participation charges which verify the revenue guarantee almost surely, and (ii) participation charges with extended values in $(-\infty,+\infty]$ which are bounded below, only possibly take the value $+\infty$ on a measure-zero set, and satisfy the guarantee pointwise over all possible type realizations.
\end{proposition}

Going beyond this result requires establishing an a priori \emph{upper bound} on participation charges, including on measure zero sets. \hyref{Appendix}{app:proof-existence-prop} contains the proof of the proposition and a discussion of the obstacles towards a stronger existence result. While exact attainment with bounded charges remains open, \hyref{Appendix}{app:bounded-charges} establishes the existence of approximately optimal bounded charges, whose deficit guarantee converges to $\delta^\star$ as the exogenously imposed limited-liability bound grows large.
Existence of \emph{continuous} solutions is even more demanding; although we conjecture it to be plausible under our specific $W$,\footnote{
    This conjecture rests only on a very specific combination of the leave-one-out constraints with combinatorial properties of $W$. For instance, continuity, convexity, or even piecewise linearity of $W$ alone are insufficient to guarantee existence of continuous potentials. The difficulty may be dimension-dependent (in particular, the techniques in \citet{GKZ2022} for the $N=3$ case do not straightforwardly generalize), which suggests the answer may not be uniform in N, and that the ``weaker'' result is the best that can be obtained in full generality, unless very specific properties of $W$ are exploited.
    }
it requires a priori estimates on minimizing sequences which, so far, appear to be out of reach of the standard toolkit. 

\begin{figure}[t]
    \centering
    \input{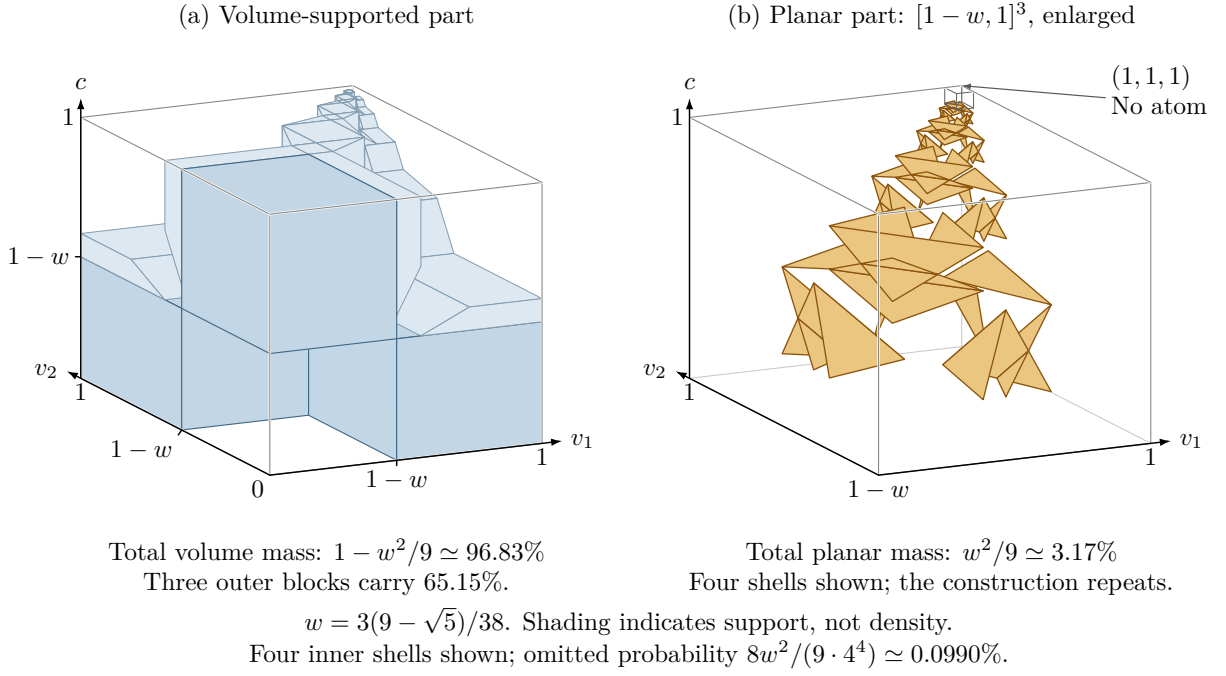}
    \caption{Geometry of the optimal adversarial coupling.}
    \label{fig:2b1s-coupling}
    \vspace{2mm}
    \begin{minipage}{0.95\linewidth}
        \footnotesize\raggedright
        \textit{Notes:} Panel~(a) shows the volume-supported components (in blue); panel~(b) enlarges the upper corner and shows the planar components (in amber). Four dyadic shells are displayed; smaller copies continue toward $(1,1,1)$, which is an accumulation point, not an atom. Shading indicates support, not density. The construction follows \citet{Mallows2003}, Section~4.
    \end{minipage}
\end{figure}

\subsection{An explicit example: 2 buyers, 1 seller.}
\label{sec:2B1S}
Consider the case with two buyers and one seller, with all types uniformly distributed on $[0,1]$ (henceforth ``2B1S-U''). This is the simplest case that goes beyond the special nature of bilateral trade, in that removing one of the buyers still leaves a meaningful submarket. Uniform types provide a natural benchmark that highlights the intrinsic complexity of the problem.

A key observation that contributes to making this case approachable is the following reduction of the welfare function to a \emph{maximum over all types}:
\begin{align*}
    W(v_1,v_2,c) = \max\{v_1,v_2,c\} - c
\end{align*}
Since the distribution of $c$ is fixed, it can be ignored in the welfare maximization problem. As a result, Nature's adversarial objective reduces exactly to maximizing the maximum of three uniformly distributed random variables, under a pairwise independence constraint. This surprisingly complex problem is solved in \citet{Mallows2003}. The following proposition adapts his solution to construct optimal participation charges, with the detailed connection supplied in \hyref{Appendix}{app:2b1s-mallows-details}.
\begin{claim}\label{claim:2b1s-explicit-mallows-solution}
    The following charges solve the 2B1S-U deficit guarantee problem:
    \[
        h_i^{B}(v_{-i},c)=\frac23+\frac{2\kappa}{3}-c-2\varphi(1-v_{-i},1-c), \qquad h^{S}(v_1,v_2)=\frac{2\kappa}{3}-2\varphi(1-v_1,1-v_2),
    \]
     where buyer charges are symmetric for $i=1,2$, $\kappa=\frac{297+5\sqrt5}{1444}$ and
    \[
        \varphi(x,y)=\min\left\{\frac{x}{2}-\eta(y),\eta(x)+\eta(y), \frac{y}{2}-\eta(x)\right\}.
    \]
    with $\eta(0)=0$, $\eta(x)=w/6$ for $x\ge w = \frac{3(9-\sqrt5)}{38}$ and for $0<x<w$, put $\eta(x)=\max\{a(x)/12,\ x/2-a(x)/3\}$, with $a(x)=w 2^{-\left\lfloor\log_2(w/x)\right\rfloor}$.
    Their optimal deficit ceiling is
    \[
        \delta^\star=\frac23-2\kappa =\frac{553-15\sqrt5}{2166}\approx 0.2398.
    \]
\end{claim}

\hyref{Figure}{fig:2b1s-coupling} shows the optimal adversarial coupling. At first glance, we observe that the coupling concentrates about 65\% of the mass in three regions where exactly one type is high (the three ``blocks'' in the figure); intuitively, a second high type is redundant for the maximum. At the same time, the optimal coupling rules out ``three-way bad markets'': it puts zero mass in the region where \emph{every} type is below a cutoff ($1-w \approx 0.466$).

Strikingly, the rest of the mass is in ``fractally'' located self-similar planes in the top right quadrant, where all types are high. The rationale is that the max operator is homogeneous of degree one, and uniform marginal constraints are scale-invariant. Pairwise independence forces some mass onto states where two agents have high valuations.  Nature then pushes that mass into the top corner,  where the subproblem is a rescaled copy of the original. 

Furthermore, on the self-similar planar components, the seller's cost is a deterministic affine function of the buyers' values. That is, given $c$, the two buyers move one-for-one in opposite directions along a segment. On these segments,  the buyers' values are conditionally countermonotone,  as in bilateral trade. However, each agent pair remains independent, as prescribed, and the negative dependence appears only conditionally on the third agent.

\begin{figure}[htbp]
    \centering
    \input{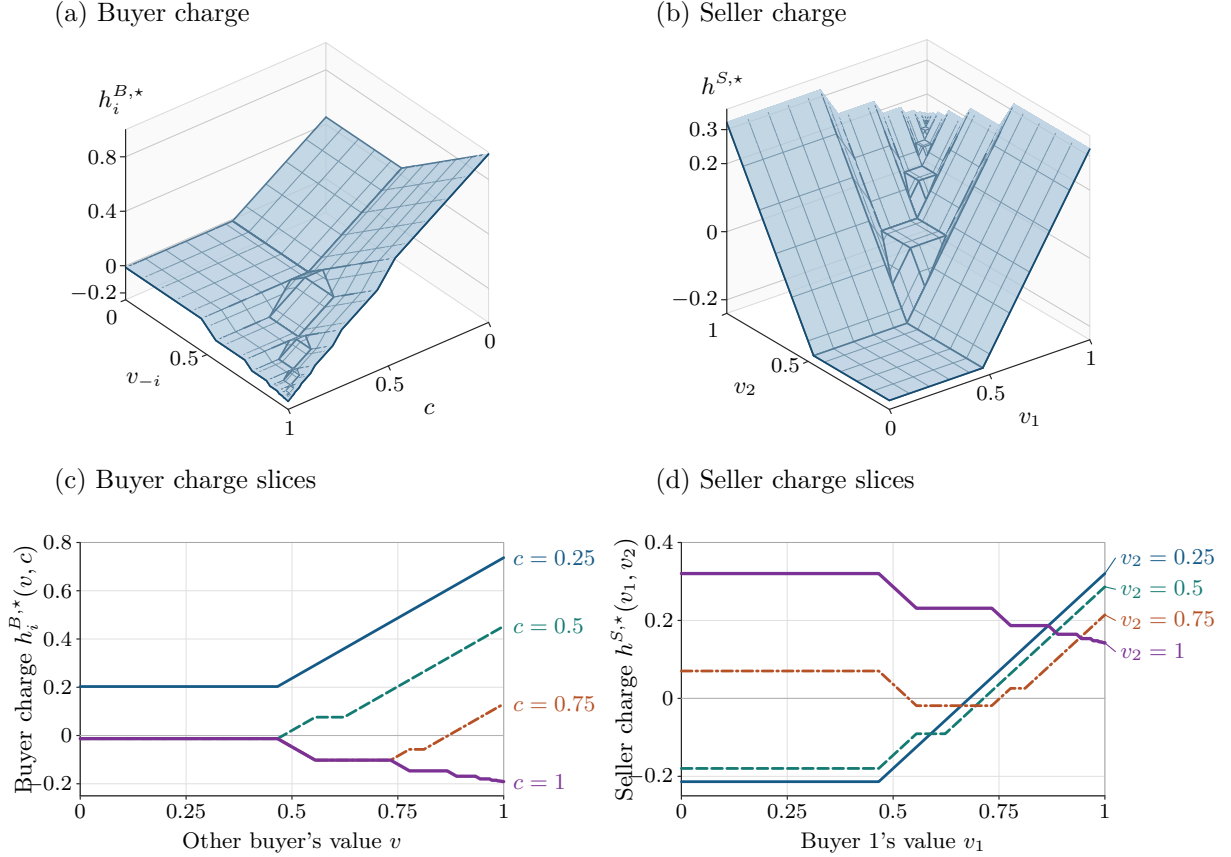}
    \caption{Optimal participation charges in the 2B1S-U case.}
    \label{fig:2b1s-optimal-charges}
    \vspace{2mm}
    \begin{minipage}{0.95\linewidth}
        \footnotesize\raggedright
        \textit{Notes:} The upper panels show the buyer and seller charges. The lower panels show charges at fixed seller costs $c=0.25,0.5,0.75,1$ and fixed buyer values $v_2=0.25,0.5,0.75,1$, respectively.
    \end{minipage}
\end{figure}

Panels~(a) and~(b) of \hyref{Figure}{fig:2b1s-optimal-charges} provide three-dimensional plots of the optimal buyer and seller charges from \hyref{Claim}{claim:2b1s-explicit-mallows-solution}.  Although they are much more regular than the coupling (i.e., continuous and Lipschitz) they still inherit a self-similar structure and non-monotonicities. Panel~(c) shows the buyer charge as a function of the other buyer's value, for four possible seller costs. Note also that the buyer charge is nonincreasing in seller cost; yet, its response to the other buyer's value can be nonmonotone. Panel~(d) shows the seller charge as one buyer's type varies, fixing the other buyer's type for four possible values. Along the slice $v_2=1$, slopes alternate between zero and $-1$ and corners accumulate at $1$, but their heights shrink; this illustrates the self-similarity.

\subsection{On the complexity and desirability of exactly optimal guarantees.}\label{sec:geometry-of-optimal-couplings}
The regularity of the solution to the 2B1S-U case gives an exact case of continuous dual attainment despite a singular coupling. This anecdotally supports our conjecture on the continuous potentials from \hyref{Section}{sec:existence-of-potentials-nm}. Yet, the complexity of the solution, even in this simplest possible extension of bilateral trade, suggests that general analytical characterizations are unlikely to be accessible; indeed, fractal couplings naturally occur in multistochastic problems \citep[see][]{GKZ2022}.\footnote{
    \citet{Mallows2003} concludes: ``Is it not remarkable that going from one-dimensional marginal constraints to two-dimensional ones leads to so much difficulty? I have not even tried to think about any problems with three-dimensional marginals.'' This is echoed by the fact that all arguments towards any kind of solution property in \citet{GKZ2022} are specific to the three-dimensional case.
    }

At the same time, we may ask whether achieving exact optimality is even \emph{economically desirable}. From a market-design perspective, the optimal charges in the 2B1S-U case appear, at face value, quite difficult for any actual institution to implement. Moreover, their intricate dependence on the prior sits somewhat uneasily with our robustness motivation. Finally, how much is actually gained by achieving the exact optimal guarantee? These questions motivate the next section, where we study simpler classes of mechanisms and bound their performance relative to the optimum.

\section{Simple mechanisms and value bounds}
\label{sec:restricted-mechanisms}
A recurring theme of the economic duality between deficit guarantee and surplus maximization is that the informational content of the participation charges is the mirror image of restrictions in the ambiguity set that Nature can exploit. Restricting the designer's toolkit to simpler mechanisms is therefore reflected in Nature having a broader set of distributions to choose from. Parsimonious charges on one side pass through the duality structure and give tractability on the other. An alternative approach is to focus on the \emph{objective} rather than mechanisms, to bound performance. Both routes are complementary. We focus on two main objects of interest: getting concrete bounds on the guarantee and deriving explicit simple mechanisms that perform well, if not optimally.

\subsection{Competitive pivot mechanism}\label{sec:generalized-competitive-pivot}
We first consider mechanisms with \emph{additively separable} participation charges. This is the most restrictive class of mechanisms we study, but also the simplest to implement. Each agent's participation charge is a sum of functions of individual reports by other market participants, so that the contribution of each report can be evaluated independently of the others. Formally, for each agent $a$, require the participation charge to take the form
\[
    h_a(\boldsymbol{\theta}_{-a})
    =\sum_{\ell\ne a}h_a^\ell(\theta_\ell),
\]
where $\boldsymbol{\theta}=(\mathbf v,\mathbf c)$ denotes the
profile of buyers' values and sellers' costs.

The restriction has a direct counterpart in the adversarial problem. 
Recall that unrestricted participation charges require Nature to preserve the joint distribution of every group of all but one agent. 
Under additive separability, however, the expectation of each charge depends only on the individual type distributions. 
Nature is therefore constrained only to preserve these one-dimensional marginals, and may otherwise choose any dependence structure across agents. 
The resulting dual problem is a standard multimarginal optimal transport problem \citep[see][for a review]{Pass2015}:
\[
    \label{eq:additive-transport}
    \max_{\pi\in\Pi(\mathbf{F},\mathbf{G})}
    \mathbb \int W d\pi
\]
where $\Pi(\mathbf{F},\mathbf{G})$ denotes the set of joint distributions with individual marginals prescribed by $F_1,...F_n$ and $G_1,...,G_m$. This new ambiguity set grants Nature considerably more freedom to correlate types than in \textbf{(AS)}.

This problem admits a simple solution, and the optimal mechanism turns out to be the generalization of the competitive pivot from bilateral trade. Formally, define the generalized \textbf{competitive pivot mechanism} by the following participation charges:
{\small
\begin{align*}
    h_i^{B,\mathrm{CP}}(\mathbf v_{-i},\mathbf c) 
    = & \overline{W}_{-i} 
    + \sum_{k\ne i} \left( (v_k-p^\star)^+ -\mathbb E\bigl[(v_k-p^\star)^+\bigr] \right)
    + \sum_{j=1}^{m} \left( (p^\star-c_j)^+ -\mathbb E\bigl[(p^\star-c_j)^+\bigr] \right),
    \\
    h_j^{S,\mathrm{CP}}(\mathbf v,\mathbf c_{-j})
    = & \overline{W}_{-j}
    + \sum_{i=1}^{n} \left[(v_i-p^\star)^+ -\mathbb E \bigl[(v_i-p^\star)^+\bigr] \right] 
    +\sum_{k\ne j} \left( (p^\star-c_k)^+ -\mathbb E\bigl[(p^\star-c_k)^+\bigr] \right),
\end{align*}}
where, as before, the thick market's competitive price $p^\star$ serves as the common pivot for the participation charges and is determined by
\(
    \sum_{i=1}^{n}[1-F_i(p^\star)] =\sum_{j=1}^{m}G_j(p^\star).
\)

\begin{claim}\label{claim:competitive-pivot-nm}
The competitive pivot mechanism minimizes the deficit ceiling among efficient, truthful and interim individually rational mechanisms with additively separable participation charges.
It achieves the minimum deficit ceiling of
\begin{equation*}
    \delta^{\mathrm{CP}} = (n+m-1) \; \mathbb{E} \biggl[ \sum_{i=1}^{n}(v_i-p^\star)^+ + \sum_{j=1}^{m}(p^\star-c_j)^+ \biggr] -\sum_{i=1}^{n}\overline{W}_{-i} -\sum_{j=1}^{m}\overline{W}_{-j}.
\end{equation*}
\end{claim}

The proposition extends the bilateral trade result to arbitrary numbers of buyers and sellers. Observe in particular, that the first expectation term in $\delta^\mathrm{CP}$ is the competitive surplus in the thick benchmark market; in the case $n=m=1$, both the deficit guarantee and participation charges collapse back to the expressions in \hyref{Propositions}{prop:duality_1B1S} and \ref{prop:OptimalTransfers_1B1S}.
In both cases, the competitive price enters the optimal participation charges, even though the mechanism implements efficient trade rather than the competitive allocation; with multiple agents, however, the competitive pivot mechanism is only optimal when restricting the designer's ability to exploit interactions among other agents' reports, but need not attain the unrestricted fiscal limit.

The proof intuition also instructively parallels the bilateral trade case, and lays the foundation for further interpreting the result. For any candidate trading price $p\in[0,1]$, define the ex post welfare bound:
\[
    \label{eq:aggregate-competitive-surplus}
    W(\mathbf v,\mathbf c; p) = \sum_{i=1}^{n}(v_i-p)^+ + \sum_{j=1}^{m}(p-c_j)^+,
\]
and let $\overline{W}(p)=\mathbb E[W(\mathbf v,\mathbf c;p)]$ be the corresponding ex ante welfare bound. Efficient gains from trade satisfy $W(\mathbf v,\mathbf c)\le W(\mathbf v,\mathbf c; p)$ for every price $p$. Since the expectation of $W(\mathbf v,\mathbf c;p)$ depends only on the prescribed individual marginals, $\overline{W}(p)$ provides an upper bound on Nature's objective.

The best bound comes from choosing $p=p^\star$, the competitive price in a thick market with the same composition and type distributions. 
It turns out to be sharp: Nature can construct a coupling under which the number of buyers willing to trade at $p^\star$ equals the number of willing sellers in every realization, while preserving the individual marginal distributions. Consequently, $\max_{\pi\in\Pi} \mathbb E_\pi[W(\mathbf v,\mathbf c)] = \overline{W}(p^\star)$. The dual potentials of this transport problem are given by the individual competitive surplus functions evaluated at $p^\star$. Adapting our duality result under the restriction implies that these potentials identify the optimal participation charges.
Summing the participation charges yields the expression for $\delta^{\mathrm{CP}}$ in \hyref{Claim}{claim:competitive-pivot-nm}. 

A rearranging of the resulting participation charges, paralleling the one in \hyref{Section}{sec:nBmS}, gives a compelling intuition for the mechanism. Define $W_{-i}(\mathbf{v}_{-i},\mathbf{c};p) = \sum_{k \neq i} (v_k-p)^+ + \sum_j(p-c_j)^+,$ the ex post competitive surplus under price $p$, excluding buyer $i$. Buyer $i$'s transfer rule under the competitive pivot mechanism can be rewritten as:
\begin{align*}
    T^{B,\mathrm{CP}}_i = & T_i^{B,\mathrm{VCG}} +   \underbrace{\bigl(W_{-i}(p^\star)-W_{-i}\bigr)(\mathbf{v}_{-i},\mathbf{c}) - \mathbb{E}\bigl[W_{-i}(p^\star)-W_{-i}\bigr]}_{\mathrm{market\,\,quality\,\,index}} .
\end{align*}
Buyer $i$ pays their VCG transfer, adjusted by a market quality index constructed from the difference between surplus at the competitive price and true welfare, both evaluated in the submarket without $i$ and relative to expectations.
In other words, the competitive market price $p^\star$ is used as a benchmark to evaluate the relative quality of the finite market; intuitively in a thick market the wedge $\overline{W}_{-i}(p^\star)-W_{-i}$ is null, so this term is only meaningful in the finite market.

\subsection{Projection-core mechanism}\label{sec:projection-core-mechanisms}
A natural step forward is to consider intermediate cases between additive separability and unrestricted dependency. The market with two buyers and one seller (2B1S) provides a natural illustration. Consider for instance retaining additive separability only for the seller's participation charge, while allowing each buyer's charge to depend jointly on the other buyer's value and the seller's cost:
\[
    h_1^B(v_2,c),\qquad
    h_2^B(v_1,c),\qquad
    h^S(v_1,v_2)=h^{S,1}(v_1)+h^{S,2}(v_2).
\]

This class contains the additively separable mechanisms studied above, but allows the designer to exploit interactions between buyer and seller reports.
As expected, it gives a correspondingly intermediate restriction in Nature's ability to correlate types: Because each buyer's charge depends jointly on the other buyer's value and the seller's cost, Nature must preserve both buyer--seller joint distributions.  The seller's additive charge imposes only individual marginal constraints; in particular, Nature may correlate the two buyers' values, even though each buyer must remain independent of the seller. This is a more constrained adversarial problem than under the competitive pivot mechanism, and hence yields a weakly improved deficit guarantee.

To state the solution, first define the \textbf{projection-core mechanism} by the following participation charges:
\begin{align*}
    h_i^{B,\mathrm{PC}}(v_{i'},c) &=(v_{i'}-c)^+
      +(p^\dagger-\max\{v_{i'},c\})^+
      -(\min\{v_{i'},c\}-p^\dagger)^+\\
    &\quad-\int_0^{p^\dagger}F_{i'}(p)G(p)\,\mathrm{d}p
      +\int_{p^\dagger}^{1}
        [1-F_{i'}(p)][1-G(p)]\,\mathrm{d}p
      \qquad \mathrm{for\,\,} i'\neq i,\\
    h^{S,\mathrm{PC}}(v_1,v_2)
    &=(v_1-p^\dagger)^+ + (v_2-p^\dagger)^+
      -\int_{p^\dagger}^{1}
        [1-F_1(p)]\,\mathrm{d}p-\int_{p^\dagger}^{1}
        [1-F_2(p)]\,\mathrm{d}p,
\end{align*} 
where $p^\dagger$ is the solution to \(F_1(p^\dagger)+F_2(p^\dagger)=1\), the median of the equal-weight mixture of the buyers' distributions. 

\begin{claim}
\label{claim:projection-core}
Consider trade among two buyers and one seller.
The projection-core mechanism minimizes the deficit ceiling among efficient, truthful and interim individually rational mechanisms with an additively separable seller charge. It achieves the minimum deficit ceiling of
\begin{equation*}
    \delta^{\mathrm{PC}} = \int_0^{p^\dagger} G(p)[F_1(p)+F_2(p)]\,\mathrm{d}p + \int_{p^\dagger}^{1} G(p)[2-F_1(p)-F_2(p)]\,\mathrm{d}p.
\end{equation*}
\end{claim}

The projection-core charges are anchored at price $p^\dagger$. Unlike the competitive price, this price depends only on the buyers' distributions.
To see why this price arises, consider Nature's optimal coupling:
\[
    v_1=F_1^{-1}(Z),
    \qquad
    v_2=F_2^{-1}(1-Z),
\]
where $Z$ is uniformly distributed on $[0,1]$ and seller cost is drawn from $G$ independently of $Z$. This preserves both buyer--seller joint distributions. To see why the coupling is adverse, recall that welfare is
\(
    W=(\max\{v_1,v_2\}-c)^+.
\)
Since only one buyer can trade, pairing two high values together leaves the lower one contributing nothing to welfare. Pairing high values with low values instead spreads strong trading opportunities across more realizations. For every seller cost, it therefore maximizes expected gains from trade and, equivalently, the expected VCG deficit.

The projection-core mechanism illustrates the benefit of relaxing additive separability and the gradual increase of complexity. While the competitive pivot mechanism uses the individual competitive surpluses of other agents, the projection-core mechanism allows each buyer's charge to reflect the interaction between the other buyer's value and the seller's cost. The underlying restriction can be extended to markets with more buyers. However, the explicit projection construction does not generalize directly, since having three or more buyers reintroduces higher-dimensional leave-one-out constraints, making the countermonotone coupling used in \hyref{Claim}{claim:projection-core} infeasible.

\subsection{Opposite-side mechanism}\label{sec:opposite-side-mechanisms}
An alternative restriction to additive separability which still admits a tractable solution for arbitrary numbers of buyers and sellers is requiring each agent's participation charge depends only on the reports of agents on the other side of the market:
\[
h_i^B=h_i^B(\mathbf c),
\qquad
h_j^S=h_j^S(\mathbf v).
\]
We term this class \emph{opposite-side participation charges}.
Unlike additive separability, this restriction allows a charge to depend jointly on multiple reports, but only when they come from the opposite side. It is therefore neither a relaxation nor a tightening of the preceding restrictions in general.

Because sellers' charges can depend jointly on all buyers' reports, Nature must preserve the joint distribution of buyers' values (similarly for sellers' costs). Nature is free to correlate the two sides, but cannot introduce dependence among agents within either side. The adversarial problem thus becomes an optimal transport problem between the two sides of the market, where each side is treated as a multidimensional type.

This problem admits a useful characterization in terms of a \emph{quantity-price menu}. Let $\boldsymbol{\rho}=(\rho_1,\ldots,\rho_K)$, where $K=\min\{m,n\}$, $\rho_0=0$, and $\rho_k$ is the total price for trading $k$ units. Define the corresponding buyer and seller surplus functions by
\begin{align*}
    \Psi_{\boldsymbol{\rho}}^B(\mathbf v)
    &= \max_{0\le k\le K} \left\{ \sum_{\ell=1}^k v_{(\ell)}-\rho_k \right\},
    \qquad
    \Psi_{\boldsymbol{\rho}}^S(\mathbf c)=\max_{0\le k\le K}  \left\{\rho_k-\sum_{\ell=1}^k c_{[\ell]}\right\}.
\end{align*}
For every type profile and every menu,
\(
    W(\mathbf v,\mathbf c)
    \le
    \Psi_{\boldsymbol{\rho}}^B(\mathbf v)
    +\Psi_{\boldsymbol{\rho}}^S(\mathbf c).
\)
These functions therefore provide a separable upper bound on efficient welfare, and their expectations depend only on the fixed distributions of the two sides.
The \textbf{opposite-side mechanism} is defined by setting the corresponding participation charges to:
\begin{align*}
    h_i^{B,\mathrm{OS}}(\mathbf c)
    &=\overline{W}_{-i} + \frac{n+m-1}{n}\left(
    \Psi_{\boldsymbol{\rho}^\star}^S(\mathbf c)
    -\mathbb E[\Psi_{\boldsymbol{\rho}^\star}^S]\right),\\
    h_j^{S,\mathrm{OS}}(\mathbf v) &=\overline{W}_{-j}
    +\frac{n+m-1}{m}\left(\Psi_{\boldsymbol{\rho}^\star}^B(\mathbf v)
    -\mathbb E[\Psi_{\boldsymbol{\rho}^\star}^B]\right).
\end{align*}
for any selection $\boldsymbol{\rho}^\star \in \argmin_{\substack{0\le\rho_k\le k\\ k=1,\ldots,K}} \left\{\mathbb E[\Psi_{\boldsymbol{\rho}}^B] +\mathbb E[\Psi_{\boldsymbol{\rho}}^S] \right\}$.

\begin{claim}
\label{claim:opposite-side}
The opposite-side mechanism minimizes the deficit ceiling among efficient, truthful and interim individually rational mechanisms with participation charges that depend only on opposite side reports.
It achieves the minimum deficit ceiling of
\[
    \delta^{\mathrm{OS}} =(n+m-1)\min_{\substack{0\le\rho_k\le k\\ k=1,\ldots,K}}
    \left\{\mathbb E[\Psi_{\boldsymbol{\rho}}^B]
    +\mathbb E[\Psi_{\boldsymbol{\rho}}^S] \right\}
    -\sum_{i=1}^{n}\overline{W}_{-i} - \sum_{j=1}^{m}\overline{W}_{-j}.
\]
\end{claim}

The quantity-price characterization parallels the single-price logic underlying the competitive pivot mechanism. Instead of equating expected demand and supply at one price, the optimal menu equates the entire distributions of quantities chosen by the two sides. These prices are auxiliary instruments for constructing participation charges, rather
than transaction prices: the mechanism continues to implement efficient trade through Groves transfers. As with the competitive pivot and the theoretical optimum, we can interpret the charges as stemming from a different notion of counterfactual welfare, or relative quality of the submarkets with one agent removed, via the same decompositions and rearrangements. These mechanisms also have a computational advantage: regardless of the number of type profiles, finding the optimal opposite-side mechanism requires optimizing only $\min\{n,m\}$ quantity prices.

\subsection{Price-by-price decomposition}\label{sec:price-by-price-decomposition}
Our final relaxation takes a different approach and starts from an intuitive decomposition of the gains from trade in terms of the number of agents willing to trade at each price. Splitting the problem \emph{price-by-price} and piecing solutions together yields clean deficit bounds which, surprisingly, turn out to be sharp and attain exact unrestricted optimality in the special case where buyers and sellers' distribution share the same median, and there is an odd number of traders on each side of the market.

For any candidate price $p\in[0,1]$, define the trade indicators:
\[
    y_i^B(p;v_i)=\mathds{1}\{v_i\ge p\},
    \qquad
    y_j^S(p;c_j)=\mathds{1}\{c_j\le p\}.
\]
The binary profile
\(
    \mathbf{y}(p;\mathbf{v},\mathbf{c})
    =  \bigl((y_i^B(p;v_i))_{i=1}^n,\,(y_j^S(p;c_j))_{j=1}^m\bigr)
    \in\{0,1\}^{n+m}
\)
records each agent's willingness to trade at $p$. Demand and supply counts are obtained by summation: 
\(D(p;\mathbf{v})=\sum_{i=1}^n y_i^B(p;v_i)\)
and
\(S(p;\mathbf{c})=\sum_{j=1}^m y_j^S(p;c_j).\)
Importantly, once a tentative price $p$ is fixed, this depends only on which buyers and sellers are willing to trade, and not their exact private types. Welfare can be expressed as an integral over prices:
\[
    W(\mathbf{v},\mathbf{c})=\int_0^1\min\big\{D(p ; \mathbf{v}),S(p; \mathbf{c})\big\}\,\mathrm{d}p
\]
This is what we term the ``price-by-price decomposition''.
At each price, the integrand counts how many buyers and sellers are willing to trade. We can also view it as the efficient surplus in a binary-type market with buyer values $y_i^B$ and seller costs \(1-y_j^S\): each profitable trade generates one unit of surplus.

We exploit this decomposition to bound the dual objective from above through a collection of simple binary type problems, one for each price. For a binary-type profile \(\mathbf{y}=(\mathbf{y}^B,\mathbf{y}^S)\in\{0,1\}^{n+m}\), intuitively representing individual trading decisions, define the feasible number of trades:
\(
    \mathcal{W}_{n,m}(\mathbf{y})
    = \min\left\{  \sum_{i=1}^n y_i^B,\, \sum_{j=1}^m y_j^S \right\}.
\)
For any $\lambda\in\Lambda(\mathbf{F},\mathbf{G})$, let $\lambda_p$ be the distribution of $\mathbf{y}(p;\mathbf{v},\mathbf{c})$ induced by $\lambda$. The dual objective can then be rewritten as
\[
    \mathbb{E}_{\lambda}[W] =
    \mathbb{E}_{\lambda}\left[
        \int_0^1 \mathcal{W}_{n,m}\bigl(\mathbf{y}(p;\mathbf{v},\mathbf{c})\bigr)\,\mathrm{d}p \right]=\int_0^1 \mathbb{E}_{\lambda_p} [\mathcal{W}_{n,m}]\,\mathrm{d}p.
\]
The key idea is to optimize directly and separately over the joint distributions $\lambda_p$ of the trading indicators, subject to the leave-one-out marginal constraints, without requiring that they arise from a single joint distribution over continuous types. This is a relaxation of the original problem because, in general, the binary trading indicators at different prices cannot be chosen independently; each buyer's willingness to trade is decreasing in price, while each seller's is increasing.

Formally, for any buyer $i$, the indicator $y_i^B(p;v_i)$ is a Bernoulli random variable with success probability $1-F_i(p)$ (identifying the distribution $F_i$ with its CDF). Similarly, for any seller $j$, the indicator $y_j^S(p; c_j)$ is a Bernoulli random variable with success probability $G_j(p)$. Denote their product law by $\mu_p$ (i.e. the product of the previous Bernoulli distributions). With a slight abuse of notation, let $\Lambda(\mu_p)$ be the set of distributions on $\{0,1\}^{n+m}$ that preserve every leave-one-out marginal of $\mu_p$. These are the same constraints as before, now imposed on binary types. Allowing Nature the freedom to optimize separately at each price gives the following upper bound on the adversarial value:
\[
   \max_{\lambda \in \Lambda(\mathbf{F},\mathbf{G})} \mathbb{E}_\lambda[W] \leq \int_0^1
    \max_{\lambda_p\in\Lambda(\mu_p)}
    \mathbb{E}_{\lambda_p}[\mathcal{W}_{n,m}] \,\mathrm{d}p.
\]

It is worth noting that the integrand is the value of an adversarial surplus maximization problem with binary types. This relaxation is tractable: we can solve each such problem in closed form, because for each price, the associated couplings satisfying the leave-one-out constraints admit a simple characterization. To describe these couplings, denote the total number of trading agents by $|\mathbf{y}|=\sum_{a=1}^{n+m}y_a$.

\begin{lemma}[Binary Type Couplings]
\label{lem:binary-leave-one-out}
A joint distribution $\lambda_p \in \Delta\bigl( \{0,1\}^{n+m} \bigr)$ is in $\Lambda(\mu_p)$ if and only if
\(
    \lambda_p(\mathbf{y}) = \mu_p(\mathbf{y})+\epsilon_p(-1)^{|\mathbf{y}|} 
\)
for every $\mathbf{y}\in\{0,1\}^{n+m}$ and for some scalar $\epsilon_p\in[\underline{\epsilon}_p,\overline{\epsilon}_p]$, where
\[
    \underline{\epsilon}_p = -\min_{|\mathbf{y}|\ \mathrm{even}}\mu_p(\mathbf{y}),
    \qquad \overline{\epsilon}_p = \min_{|\mathbf{y}|\ \mathrm{odd}}\mu_p(\mathbf{y}).
\]
\end{lemma}
To see this, consider an arbitrary $\lambda_p \in \Lambda(\mu_p)$ and let $d(\mathbf{y})=\lambda_p(\mathbf{y})-\mu_p(\mathbf{y})$. The leave-one-out restrictions require $d(0,\mathbf{y}_{-a})+d(1,\mathbf{y}_{-a})=0$ for every coordinate $a$ and every fixed $\mathbf{y}_{-a}$: any flip from $0$ to $1$ in one entry must switch the sign of the difference. Hence, this gives $d(\mathbf y)=\epsilon_p(-1)^{|\mathbf{y}|}$ with $\epsilon_p=d(\mathbf{0})$. Conversely, this adjustment cancels when any coordinate is summed out. The bounds $\underline{\epsilon}_p$ and $\overline{\epsilon}_p$ simply ensure that all of the resulting probabilities are nonnegative.

The lemma leaves Nature just one degree of freedom in each fixed-price problem. Its marginal effect on the fixed-price welfare $\E_{\lambda_p}[\mathcal{W}]$ is governed by the coefficient
\[
    C_{n,m}
    :=\sum_{\mathbf y\in\{0,1\}^{n+m}}(-1)^{|\mathbf            y|}\mathcal{W}_{n,m}(\mathbf y)
    =\sum_{k=1}^{\min\{n,m\}}\binom{n-1}{k-1}\binom{m-1}{k-1}
    =\binom{n+m-2}{n-1}>0.
\]
Nature therefore chooses the largest feasible adjustment, $\epsilon_p=\overline{\epsilon}_p$, raising fixed-price welfare by $C_{n,m}\overline{\epsilon}_p$. Integrating these optimal values yields the following price-by-price deficit guarantee.

\begin{claim}[Price-decomposable mechanisms]
\label{claim:bernoulli-mechanism}
The price-by-price deficit ceiling is
\[
    \delta^{\mathrm{PD}} ={}(n+m-1)\int_0^1
    \left[ \mathbb E_{\mu_p}[\mathcal{W}_{n,m}] +C_{n,m}\overline{\epsilon}_p \right]\,\mathrm{d}p-\sum_{i=1}^n\overline W_{-i} -\sum_{j=1}^m\overline W_{-j} \geq \delta^\star.
\]
It is attained by the following price-decomposable mechanism with participation charges
\[
    h_a^{\mathrm{PD}}(\boldsymbol\theta_{-a})
    =\overline W_{-a}+(n+m-1)\int_0^1
    \left[\varphi_{a,p}\bigl(\mathbf y_{-a}(p;\boldsymbol\theta_{-a})\bigr)-\mathbb E_{\mu_p}[\varphi_{a,p}]
    \right]\,\mathrm{d}p,
\]
for each agent $a$, where $\varphi_{a,p}$ are a measurable selection of optimal dual potentials for the fixed-price $p$ problem.
\end{claim}
\begin{remark}
The price-by-price deficit ceiling also decomposes into expected deficit and excess above the mean:
\[
    \delta^{\mathrm{PD}}
    =\mathbb E_\mu[\mathrm{Deficit}]+(n+m-1)C_{n,m}
    \int_0^1\overline{\epsilon}_p\,\mathrm{d}p.
\]
This decomposition will be useful in the next section for studying the effect of market thickening on controlling the deficit exposure.
\end{remark}

While we started by looking for an upper bound on Nature's adversarial objective, the price-by-price relaxation also allows us to construct an explicit mechanism that attains the deficit ceiling $\delta^{\mathrm{PD}}$. First, each fixed-price problem is a finite linear program, so strong duality guarantees the existence of optimal binary dual potentials. Each potential omits one agent's indicator, and together they cover fixed-price welfare:
\[
    \sum_{a\in N\cup M}\varphi_{a,p}(\mathbf y_{-a})
    \ge \mathcal W_{n,m}(\mathbf y) \quad \forall \mathbf{y} \in \{0,1\}^{n+m}.
\]
We then ``glue'' these potentials together by integrating over prices. Since welfare itself decomposes price by price, the collated potentials cover welfare for the original market. Therefore, the resulting charges satisfy
\[
    \sum_{a\in N \cup M}
    h_a^{\mathrm{PD}}(\boldsymbol\theta_{-a})
    \ge (n+m-1)W(\boldsymbol\theta)-\delta^{\mathrm{PD}}\quad \forall \bm \theta \in [0,1]^{n+m}.
\]
Thus the charges cover all but $\delta^{\mathrm{PD}}$ of the
designer's subsidy needs at all type profiles. A measurable selection of fixed-price potentials gives continuous participation charges; details are later provided in \hyref{Appendix}{app:price-by-price-decomposition}.

Nature's room for complex manipulation has a simple intuition (thanks to \hyref{Lemma}{lem:binary-leave-one-out}). At each price, it can only add the same probability mass to every even configuration and subtract it from every odd configuration. It chooses the largest feasible adjustment, $\overline{\epsilon}_p$, at which a least-likely odd configuration loses all its mass. The price-by-price bound allows this choice to be made separately at every price. The price-decomposable mechanism construction requires no consistency among these separate maximizers.

Unrestricted optimality requires more: Nature's optimal binary type laws must also fit together. Precisely, the bound is exact if and only if one admissible joint distribution of the original types induces, for each price, the optimal coupling of binary trading decisions. For the following large family of cases, we can construct such a coupling.

\begin{claim}[Exactness]
\label{claim:bernoulli-exactness}
If all type distributions share a common median and both $n$ and $m$  are odd, then $\delta^{\mathrm{PD}}=\delta^\star$. The price-decomposable mechanism therefore solves \textbf{(DG)}.
\end{claim}

This is quite remarkable because the relaxation discards consistency across all prices, yet a common median and odd counts on both sides of the market suffice to recover it. The bound is not always exact, however, as the uniform 2B1S case illustrates (see \hyref{Section}{sec:comparing-mechanisms-2B1S-uniform}). In general, \(\delta^{\mathrm{PD}}\) is the attained ceiling of the price-decomposable mechanism, but it may exceed the smallest achievable ceiling.

\begin{table}[t]
    \centering
    \small
    \begin{tabular}{@{}l c c c c >{\raggedright\arraybackslash}p{4.6cm}@{}}
        \toprule
         Mechanism & Deficit & Excess & VCG excess & Gap to & What Nature \\
         & ceiling & over mean & removed & optimum & may correlate \\
        \midrule
        VCG              & $1$             & $\sfrac{5}{6}$  & $0\%$    & $+317\%$  & Anything (fixed charges; worst case over types) \\
        Competitive pivot              & $\sfrac{1}{3}$  & $\sfrac{1}{6}$  & $80\%$   & $+39.0\%$ & All pairs; only 1D marginals fixed \\
        Projection-core    & $\sfrac{1}{4}$  & $\sfrac{1}{12}$ & $90\%$   & $+4.24\%$ & The buyer--buyer pair; buyer--seller pairs fixed \\
        Opposite-side    & $\approx 0.3033$ & $\approx 0.1366$ & $83.6\%$ & $+26.5\%$ & Buyer--seller pairs; the buyers' joint law is fixed \\
        Price-decomposable & $\sfrac{1}{4}$ & $\sfrac{1}{12}$ & $90\%$  & $+4.24\%$ & Leave-one-out at each price, indep. across price \\
        Optimal          & $\approx 0.2398$ & $\approx 0.0732$ & $91.2\%$ & $0$ & Third-order dependence only; all pairs fixed \\
        \midrule
        Mean (AGV) & $\sfrac{1}{6}$  & $0$             & 100\%     & ---       & --- \\
        \bottomrule
    \end{tabular}
    \caption{\footnotesize Two buyers and one seller with $U[0,1]$ types. Deficit ceilings of alternative mechanisms and feasible adversarial couplings. \emph{Excess over mean}: ceiling minus the mean deficit $\sfrac{1}{6}$, common to every Groves mechanism with binding interim participation constraints. \emph{VCG excess removed}: one minus the ratio of excess to VCG's excess $\sfrac{5}{6}$. \emph{Gap to optimum}: $(\delta-\delta^\star)/\delta^\star$.}
    \label{tab:mechanisms-comparison-2b1s}
\end{table}

\subsection{Numerical comparisons with two buyers and one seller}\label{sec:comparing-mechanisms-2B1S-uniform}
How do the restricted mechanisms we have introduced fare, relative to each other and to the two benchmarks of VCG and the true optimal mechanism? As a natural testing ground and illustration, we return to the case of two buyers and one seller with uniformly distributed type, and explicitly compare the performance of alternative mechanisms.
\hyref{Table}{tab:mechanisms-comparison-2b1s} summarizes the results; the derivations for detailed results are in \hyref{Appendix}{app:2B1S-table-derivations}

It is apparent that large improvements can be obtained even under restrictions. The competitive pivot, our simplest mechanism and most drastic constraint, already performs excellently, leaving only 20\% of the worst-case deficit excess over the mean, relative to a little less than 9\% a the optimum. The relatively modest relaxation to either price-decomposable or projection-core mechanisms yields a result extremely close to the true optimum, eliminating almost all the residual that the competitive pivot retained (only a little over 1\% of the excess deficit over the mean remains relative to what the optimal mechanism achieves). This highlights that there is generally a large scope to improve fiscal exposure even without going all the way to the optimal guarantee, and taking into account the designer's practical concerns. A natural next step is to ask how this guarantee depends on market size.

\section{The effect of market thickness}\label{sec:CS-market-thickness}
Previous discussions have suggested, intuitively, that the designer's ability to recover surplus increases in a larger market, through the added informational content of participation charges. How much fiscal risk remains as the market grows thicker? A natural measure of fiscal risk is the excess of the best deficit ceiling over the mean deficit, introduced in the previous section to compare mechanisms. This gap measures the tail risk that remains after the mechanism has optimally compressed the deficit distribution. 

Throughout this section, consider $n$ buyers with i.i.d.\ values from $F$ and $n$ sellers with i.i.d.\ costs from $G$, both supported on $[0,1]$, with continuous densities $f$ and $g$. Assume further that there is a unique interior $p^\star$ such that $G(p^\star)=1-F(p^\star)$, and that $F$ and $G$ are continuously differentiable near $p^\star$. 
We adapt previous notations and add the subscript $n$ to make the market size explicit, for instance, denoting $\mathrm{Deficit}_n(\mathbf{v},\mathbf{c}; \mathcal{M})$ the realized deficit under mechanism $\mathcal{M}$.
Recall revenue equivalence fixes the expected deficit in all mechanisms we consider; denote by $\overline{\delta}_n$ this expected deficit. By construction:
\begin{align*}
    \overline{\delta}_n = \mathbb{E}\bigl[ \mathrm{Deficit}_n(\mathbf{v},\mathbf{c}; \mathcal{M}) \bigr] = \mathbb{E}\bigl[ \mathrm{Deficit}_n(\mathbf{v},\mathbf{c};{\mathrm{VCG}}) \bigr]
\end{align*}
for any mechanism $\mathcal{M}$ that is efficient, truthful, and whose IIR constraints bind at the worst types, including VCG in particular.

Let $\delta_n^\star$ denote the optimal deficit ceiling and define the \emph{deficit excess}
\[
    \mathcal E^\star_n
    :=\delta_n^\star - \bar{\delta}_n.
\]
This raises two distinct questions. Does the deficit excess vanish in large markets, i.e., is the designer able to fully compress the tail to the mean? What does the expected deficit itself converge to as the market becomes large?
The price-by-price decomposition from \hyref{Claim}{claim:bernoulli-mechanism} enables us to give a sharp answer to both questions.

\subsection{The expected deficit in large markets}\label{subsec:limit-deficit}
We first characterize the limit of the expected deficit for our class of mechanisms. 
\begin{proposition}[Convergence of Mean Deficit]\label{prop:limit-deficit}
    Let $q^\star:=1-F(p^\star)=G(p^\star) \in (0,1)$.
    \begin{align*}
        \lim_{n \rightarrow \infty} \bar{\delta}_n = \frac{q^\star}{f(p^\star) + g(p^\star)}
    \end{align*}
\end{proposition}
We provide a new self-contained proof of this result via the price-by-price decomposition, which yields an exact finite-market expression (see \hyref{Appendix}{app:limit-deficit-proof} for a detailed proof).
The result itself, however, is not new: with a common support, the VCG mechanism coincides with the double auction with zero fee studied in \citet{Tatur2005}, who shows that it attains the limit stated in \hyref{Proposition}{prop:limit-deficit}. 
Theorem 2 in \citet{LoertscherMezzetti2019} extends the corresponding lower bound to multi-unit traders and dependent type.

The limiting deficit is the trading volume $q^\star$ in the continuum market, adjusted by a price sensitivity term; indeed, the denominator corresponds to the derivative of excess demand at the equilibrium price.
The limit expression can also be written as the ratio of the competitive price to the sum of supply and demand elasticities at that price:
\begin{align*}
    \frac{q^\star}{f(p^\star) + g(p^\star)} = \frac{p^\star}{\varepsilon_D + \varepsilon_S} \quad \text{ where: } \varepsilon_S:= \frac{p^\star G'(p^\star)}{q^\star}, \; \varepsilon_D:=- \frac{p^\star (1-F)'(p^\star)}{q^\star}
\end{align*}
The intuition is that small payment wedges accumulate across many trades; the deficit \emph{per trade} vanishes, but the \emph{aggregate} deficit does not.
Having more buyers and sellers with values \emph{near the competitive price} reduces the limiting deficit.
It also complements results in \citet{Le2024EfficientTrade}, which establish asymptotic normality of the \emph{realized} VCG deficit, and \citet{Muir2017}, which prove joint asymptotic normality of efficient quantity and trade surplus.

\subsection{The deficit excess in large markets}\label{subsec:limit-deficit-excess}
The price-by-price decomposition also provides a sharp bound on the deficit excess $\mathcal E^\star_n$ and a simple intuition for why it should vanish as $n$ grows. Nature must preserve the independent distribution of every leave-one-out economy. Any remaining dependence therefore involves all agents and disappears when any one is removed. At each price, this leaves Nature only the single degree of freedom identified in \hyref{Lemma}{lem:binary-leave-one-out}. As the market thickens, the feasible interval for this adjustment shrinks exponentially with the number of participants, limiting Nature's ability to increase expected welfare and hence fiscal exposure.

The resulting convergence rate depends on the relative location of buyer and seller medians. When they differ, no price makes both sides' trading indicators simultaneously maximally uncertain, and $\mathcal E^\star_n$ decays exponentially. When the medians coincide, there is a knife-edge price at which all trading indicators are maximally uncertain. The deficit excess still vanishes, but the uniform example below shows that convergence can be much slower. The following main theorem makes this asymptotic compression precise.

\begin{theorem}[Convergence of Deficit Excess]
\label{thm:convergence}
Let $\beta(p)=\min\{F(p),1-F(p)\}$ and $\sigma(p)=\min\{G(p),1-G(p)\}$. Then
\[
    \label{eqn:mainbound}
    0\le \mathcal E^\star_n \le (2n-1)\binom{2n-2}{n-1}
    \int_0^1 [\beta(p)\sigma(p)]^{n-1}
    \min\{\beta(p),\sigma(p)\}\,\mathrm{d}p.
\]
Moreover, the excess decays with rate depending on the relative location of medians:
\begin{enumerate}
    \item[i.] If $F$ and $G$ have distinct medians, then $ \nu=4\max_{p\in[0,1]}\beta(p)\sigma(p)<1$, and
    \[
        \mathcal E^\star_n\le \left(n-\frac12\right)\nu^{n-1}.
    \]

    \item[ii.] If $F$ and $G$ share a median $p_0$, and
    $\alpha=\min\{\inf_p f(p),\inf_p g(p)\}>0$, then
    \[
        \mathcal E^\star_n \le \frac{1}{\alpha4^n} \left(2-\frac1n\right) \binom{2n-2}{n-1} =O(n^{-1/2}).
    \]
\end{enumerate}
\end{theorem}

The main tool of the proof is a simple estimate for Nature's ability to (adversarially) improve welfare. At each price, increasing the parity adjustment requires removing the same probability mass from every odd profile. The least-likely odd profile therefore limits the adjustment everywhere else. To bound its probability, fix $n-1$ buyers and $n-1$ sellers at their less likely realizations, then choose the remaining two indicators to obtain odd parity. This gives
\(
    \overline{\epsilon}_p \le [\beta(p)\sigma(p)]^{n-1} \min\{\beta(p),\sigma(p)\}.
\)
By \hyref{Claim}{claim:bernoulli-mechanism}, $\mathcal E^\star_n\le (2n-1)C_{n,n}\int_0^1\overline{\epsilon}_p\,\mathrm{d}p$.
Thus rare trading configurations limit Nature's welfare advantage, and estimating their probabilities yields the two convergence bounds (see \hyref{Appendix}{app:limit-deficit-ceiling} for a detailed proof).  
The following example shows the rates are sharp.

\begin{example}[Uniform $n$B$n$S]
To show that the inverse-square-root rate is sharp when medians coincide, consider $F=G$ uniform on $[0,1]$, so that
$\beta(p)=\sigma(p)=\min\{p,1-p\}$. Integrating the price-by-price optima gives the price-decomposable deficit excess
\[
    \mathcal{E}_n^{\mathrm{PD}} =\frac{(2n-1)C_{n,n}}{(2n+1)4^n}
    \begin{cases}
        1, & n\text{ odd},\\
        1+1/n, & n\text{ even}.
    \end{cases}
\]
For a lower bound, let $\mu$ be the independent uniform law,
set $\chi(t)=1$ for $t\le1/2$ and $\chi(t)=-1$ otherwise, and define probability measure $\lambda$ by
\[
    \frac{\mathrm{d}\lambda}{\mathrm{d}\mu}
    (\mathbf v,\mathbf c)
    =
    1+(-1)^n
    \prod_{i=1}^n \chi(v_i)
    \prod_{j=1}^n \chi(c_j).
\]
Integrating out any coordinate eliminates the perturbation, preserving every leave-one-out marginal. At each price, this coupling induces the parity adjustment $\min\{p, 1-p\}^{2n}$. Weak duality and Stirling's formula therefore give
\[
    \frac{n}{n+1}\mathcal{E}_n^{\mathrm{PD}}
    \le \mathcal{E}_n^\star
    \le \mathcal{E}_n^{\mathrm{PD}},
    \qquad
    \mathcal{E}^\star_n\sim\frac{1}{4\sqrt{\pi n}}.
\]
The rate is therefore sharp. The price-decomposable mechanism matches the optimal leading constant and is exactly optimal for odd $n$, when $\mathcal{E}_n^\star=\mathcal{E}_n^{\mathrm{PD}}$.
\end{example}

\subsection{Takeaways: the large market worst-case guarantee}
Putting \hyref{Theorem}{thm:convergence} and \hyref{Proposition}{prop:limit-deficit} together implies that the worst-case deficit converges to a constant, which is the limit of the expected deficit, hence has the same natural interpretation:
\begin{align*}
    \delta_n^\star \, \xrightarrow[n \rightarrow \infty]{} \, \frac{q^\star}{f(p^\star) + g(p^\star)}.
\end{align*}

Although a closed-form optimal rule is generally out of reach, our price-decomposable mechanism delivers the rates of excess compression given in \hyref{Theorem}{thm:convergence}. It can therefore be viewed as an AGV-like insurance mechanism with dominant-strategy incentive compatibility and interim individual rationality. AGV completely eliminates budget fluctuations under Bayesian incentive compatibility while the price-decomposable mechanism approaches the same budget smoothing as the market thickens while staying truthful.

This also connects the paper to informationally robust mechanism design \citep*{BergemannBrooksMorris2019}. More recently, \citet*{BrooksDuZhang2026} show that suitably designed market order mechanisms guarantee competitive outcomes as the numbers of buyers and units grow, across information structures and equilibria. Our result identifies a different benefit of market thickness. Efficient trade is implemented at every market size; what improves is the predictability of its fiscal cost.

The original AGV construction targets efficient public-good provision with ex post budget balance, taking participation for granted or decided at the ex ante stage \citep{Arrow1979,dAspremontGerardVaret1979}. Throughout our paper, we actually use ``AGV mechanism'' for its voluntary-participation version, which satisfies \hyperlink{cond:IIR}{(IIR)} with a deterministic budget. This construction connects our result to \citet{BorgersNorman2009}, who show that independent types allow payments to be rearranged from ex ante to ex post budget balance while preserving Bayesian incentives and interim participation. Risk-neutral participants effectively insure the designer. In auctions, \citet{EsoFuto1999} similarly show how risk-neutral bidders can fully insure a risk-averse seller without reducing expected receipts. The budgetary distinction matters to a clearinghouse financing efficient trade or a government providing a public good: the average funding requirement and the reserve against adverse realizations are separate concerns. \hyref{Theorem}{thm:convergence} bounds this reserve and shows how market size makes the necessary subsidy increasingly predictable.

This logic also underlies the robustly collusion-proof mechanisms of \citet[Theorem~1 and Remark~2]{CheKim2006}, which preserve allocations and interim payoffs while making the principal's worst-case payoff equal to its average. In our setting, a tight deficit ceiling partially limits the designer's exposure to collusion. Although truthfulness does not rule out profitable coordinated misreports, the ceiling holds at every report profile. Collusion may therefore change the deficit distribution and increase its mean, but cannot force the designer to finance more than the guaranteed ceiling. This limits the fiscal consequences of collusion, even when there is a scope for profitable joint deviations.

\section{Extensions}\label{sec:extensions}
Our approach is not specific to trade environments, and can be extended beyond interim participation constraints. It applies whenever efficiency and truthfulness lead to Groves transfers, so that the designer's only remaining instrument is the choice of participation charges. In each environment below, the same structure reappears: efficient implementation creates a Groves fiscal liability, participation constraints determine how much can be charged, and the sharp worst-case deficit or revenue guarantee is characterized by a multistochastic transport problem under leave-one-out constraints. We present the economic message here and leave the formal derivations to the appendix.

\subsection{The case of ex ante participation}\label{sec:extension-AIR}
The guarantee problem under ex ante participation is completely analogous to that under \hyperlink{cond:IIR}{(IIR)}. A mechanism satisfies \emph{ex ante individual rationality} if each agent's expected payoff from participating is nonnegative before learning their type. For an efficient Groves scheme, this is equivalent to
\begin{equation}\tag{AIR}\label{cond:AIR}
    \mathbb E\bigl[h_i^B(\mathbf v_{-i},\mathbf c)\bigr]
        \le \overline W \,\,\,\text{for every }i\in N, 
    \quad
    \mathbb E\bigl[h_j^S(\mathbf v,\mathbf c_{-j})\bigr]
        \le \overline W \,\,\,\text{for every }j\in M,
\end{equation}
where $\overline W:=\mathbb E[W]$.
The sole difference to our previous \hyref{Lemma}{lm:ParticipationCharges} is that the upper bounds on the expected participation charges are replaced by the expected gain from trade $\overline{W}$, rather than the expected surpluses under exclusion $\overline{W}_{-i}$ and $\overline{W}_{-j}$. Because this leaves the underlying geometry of the linear program entirely unchanged, our results under interim participation~\hyperlink{cond:IIR}{(IIR)} can be straightforwardly adapted. 

To highlight differences, focus on the case of bilateral trade. Under ex ante participation~\eqref{cond:AIR}, the designer gains the leverage to extract participation charges up to twice the expected gains $2\overline{W}$. Therefore, the mechanism may no longer produce a deficit and instead generate net revenue. The largest worst-case revenue guarantee $r^\star$ (the negative of the deficit guarantee $\delta^\star$) under \eqref{cond:AIR} becomes exactly the difference between this maximum ex ante budget and the competitive surplus of benchmark thick market $w^\infty$:
\[
    r^\star = 2\overline{W} -w^\infty = 2\overline{W} -\int_0^1 \min\{1-F(p),G(p)\}\,\mathrm{d}p.
\]

The optimal participation charges preserve the competitive pivot intuition. Each agent pays the designer $\overline W$ plus the other side's ``surplus surprise'': its realized competitive surplus at $p^\star$ minus its expected value. The expected charge is therefore $\overline W$, and each charge is $\overline W$ higher than under interim participation.

\begin{proposition}\label{cor:optimal_exante}
Among efficient, truthful mechanisms satisfying~\eqref{cond:AIR}, the charges
\begin{align*}
    h^B(c)
    &= \overline W + (p^\star-c)^+ 
       -\int_0^{p^\star}G(p)\,\mathrm{d}p,\\
    h^S(v)
    &= \overline W + (v-p^\star)^+ 
       -\int_{p^\star}^1[1-F(p)]\,\mathrm{d}p
\end{align*}
achieve the largest worst-case revenue guarantee \(r^\star=2\overline W-w^\infty.\)
\end{proposition}
\begin{corollary}
Efficient, truthful trade under ex ante participation is possible without ever running a deficit if and only if $2\overline W\ge w^\infty$.
\end{corollary}

The corollary recovers the sufficient condition of \citet{JagadeesanTordjman2025}: the median buyer is willing to trade with the median seller. Indeed, this ordering implies $q^\star\ge1/2$, and hence $\overline W\ge q^\star w^\infty\ge w^\infty/2$. They establish possibility using the same competitive pivot construction yet its optimality in terms of guaranteed revenue is not shown. Another sufficient condition is that the mean buyer is willing to trade with the mean seller, $\mathbb E[v]\ge\mathbb E[c]$. At any price $p$ between these means, independence and Jensen's inequality bound each side's competitive surplus by $\overline W$. Thus
\(
    w^\infty
    \le \mathbb E[(v-p)^+]+\mathbb E[(p-c)^+]
    \le 2\overline W,
\)
so mean ordering also ensures efficient trade without a deficit.

\subsection{Constrained-efficient allocation rules}\label{sec:extension-constrained-efficiency}
Our framework also extends beyond first-best efficiency to accommodate \textit{constrained} efficient rules like affine maximizers. By assigning asymmetric weights $w_i^B, w_j^S$ to buyers and sellers and penalizing certain allocations by a loss function $L(\cdot)$, affine maximizers allow designers to target distributional goals without sacrificing truthfulness. Since the Groves characterization still holds, we can construct Groves transfers by using the affine welfare function 
\[
    \tilde{W}(\textbf{v},\textbf{c}) = \max_{(\textbf{Q}^B, \textbf{Q}^S)} \left\{\sum_{i\in N}w_i^B v_iQ_i^B-\sum_{j\in M} w_j^S c_j Q_j^S-L(\textbf{Q}^B, \textbf{Q}^S) \right\}
\]
subject to feasibility.

\subsection{Partnership dissolution}\label{sec:extension-partnership-dissolution}
Partnership dissolution considers how a jointly owned asset should be allocated when partners privately know their values. Efficiency requires assigning the asset to the partner with the highest valuation, but voluntary participation requires each partner to be compensated for giving up their initial ownership claim. This creates the same fiscal tension as trade: Groves transfers implement efficiency, while participation charges determine if dissolution can be sustained within a budget.

\hyref{Appendix}{app:partnership} shows that the revenue-guarantee problem again has a similar duality structure, with the dual taking the form of a multistochastic Monge-Kantorovich problem. Nature chooses a joint distribution of values that preserves leave-one-out marginals while maximizing, in this case, the expectation of the \emph{highest valuation}. The resulting formula gives an exact worst-case revenue guarantee. Under ex ante participation and a common value distribution, this approach recovers the possibility theorem of \citet{JagadeesanTordjman2025}: efficient dissolution can be implemented without ever running a deficit for arbitrary ownership shares. 

\subsection{Public good provision}\label{sec:extension-public-good-provision}

Public good provision offers a second, conceptually distinct application. The efficient decision is collective rather than bilateral: the public good should be provided whenever the sum of agents' private values exceeds the cost. 
Groves mechanisms again implement the efficient rule, but the resulting transfers expose the designer to state-by-state fiscal imbalances. The question is how large this exposure must be if the designer insists on truthfulness, efficiency, and voluntary participation.

\hyref{Appendix}{app:publicgood} formulates the corresponding revenue-guarantee problem. Nature now chooses the dependence among agents' valuations to maximize the designer's expected fiscal exposure. Under ex ante participation, we obtain a transparent sufficient condition for deficit-free efficient provision: when the aggregate expected value exceeds the cost, participation charges can finance the worst-case exposure. Thus, fiscal imbalances can be eliminated through participation charges when provision is efficient on average.

\section{Concluding remarks}
In this paper, we develop a robust framework for designing mechanisms that bound the worst-case deficit in two sided trade environment and others (partnership dissolution, public good provision). By optimally engineering participation charges and leveraging duality with an adversarial surplus maximization problem, we can construct mechanisms which control the designer's maximum exposure to catastrophic type realizations.

It is informative to compare the distribution of the ex post deficit under our mechanism with that of the standard VCG mechanism. As established by \cite{Le2024EfficientTrade}, as a market thickens, the VCG deficit displays asymptotic normality. Consequently, while the risk of a catastrophic deficit diminishes in large markets, the large support suggests that the tail risk of fiscal ruin always remains.\footnote{As the economics profession starkly learned during the 2008 financial crisis, designing institutions that rely on asymptotic safety while ignoring rare events is a perilous approach.} Our revenue-guarantee mechanism neutralizes this tail risk: the deficit distribution remains centered at the expected deficit, but we sharply cap the support of the deficit from above using reengineered participation charges. 

Our approach complements the broader literature on robust mechanism design by providing tight ex post revenue guarantee across all type realizations. For example, \cite*{BergemannBrooksMorris2019} study expected revenue guarantees across information structures in simple auctions. A natural and challenging direction for future work is to fuse these two paradigms: designing mechanisms that yield optimal revenue guarantees that are robust simultaneously across all possible type realizations and all admissible information structures.

\bibliographystyle{ecta} 
\bibliography{bibliography}
\appendix
\section{Proofs of Lemmas and Theorems}\label{app:omitted-proofs}
\subsection{Proof of \hyref{Lemma}{lm:ParticipationCharges}} To establish the first condition, we sum the Groves' transfers extracted from all buyers and sellers:
\begin{eqnarray*}
    \mathrm{Deficit}({\bf v}, {\bf c}) &=& -\sum_{i=1}^n \bigl(h^B_i({\bf v}_{-i},{\bf c}) - W(\bv,\bc) +v_i Q^B_i\bigr) - \sum_{j=1}^m \bigl(h^S_j({\bf v}, {\bf c}_{-j})-W(\bv, \bc) -c_j Q^S_j \bigr) \nonumber \\
    &=& (n+m-1) W({\bf v}, {\bf c})- \sum_{i=1}^n h_i^B({\bf v}_{-i},{\bf c}) - \sum_{j=1}^m h_j^S({\bf v}, {\bf c}_{-j}) . \label{eqn:ExpostRevenue}
\end{eqnarray*}
The second line simplification is by definition of ex post efficiency: the sum of the allocated values equals the gains from trade: $\sum_{i=1}^n v_i Q^B_i(\mathbf{v}, \mathbf{c}) - \sum_{j=1}^m c_j Q^S_j(\mathbf{v}, \mathbf{c}) = W(\mathbf{v}, \mathbf{c})$. 

To establish the condition for interim participation, observe buyer $i$'s ex post payoff given true value $v_i$ is:
$$U^B_i(\mathbf{v}, \mathbf{c}) = v_i Q^B_i(\mathbf{v}, \mathbf{c}) - T^B_i(\mathbf{v}, \mathbf{c}) = W(\mathbf{v}, \mathbf{c}) - h^B_i(\mathbf{v}_{-i}, \mathbf{c}).$$
Taking the expectation over all other agents' types yields the interim utility $u^B_i(v_i) = \mathbb{E}_{\mathbf{v}_{-i}, \mathbf{c}}[W(v_i, \mathbf{v}_{-i}, \mathbf{c})] - \mathbb{E}[h^B_i(\mathbf{v}_{-i}, \mathbf{c})]$. Because $W$ is non-decreasing in $v_i$, the minimum interim utility occurs at $v_i = 0$, where buyer $i$ does not trade and the welfare is exactly $W_{-i}(\mathbf{v}_{-i}, \mathbf{c})$. Imposing $u^B_i(0) \ge 0$ requires $\mathbb{E}[h^B_i({\bf v}_{-i},c)] \le \overline{W}_{-i}$. An symmetric argument for seller $j$ yields $\mathbb{E}[h^S_j({\bf v},{\bf c}_{-j})] \le \overline{W}_{-j}$.
\hfill $\square$

\subsection{Proof of \hyref{Proposition}{prop:duality_1B1S} and \hyref{Proposition}{prop:OptimalTransfers_1B1S}}
The duality structure and construction of solutions via potentials in the reduced primal problem in \hyref{Proposition}{prop:duality_1B1S} and \hyref{Proposition}{prop:OptimalTransfers_1B1S} follow directly from the more general results in \hyref{Theorem}{thm:duality_nm} and \hyref{Theorem}{thm:optimal-mechanism-from-potentials}. 
In the case with one buyer and one seller, the reduced dual is exactly a standard one-dimensional optimal transport problem, with a ground cost verifying the twist condition so that standard results about monotone rearrangement and explicit values are accessible and deliver the results directly. \hfill $\square$

\subsection{Proof of \hyref{Theorem}{thm:duality_nm}}
We introduce some more compact notation for readability. 
Let $\Theta=[0,1]^{n+m}$ the space of type realizations, and $\Theta_{-1}=[0,1]^{n+m-1}$ that same space when removing one arbitrary agent (dimension).
Let $C:=\mathcal{C}^0(\Theta)$, and $C_{-1}:=\mathcal{C}^0(\Theta_{-1})$, each equipped with the sup norm.
The constraint operator $\oplus: C_{-1}^n \times C_{-1}^m \longrightarrow C$ is defined by $[{\bf h}^B \oplus {\bf h}^S](\bv,\bc) := \sum_i h_i^B(\bv_{-i},\bc) + \sum_j h_j^S(\bv,\bc_{-j})$.
This maps the collection of individual participation charges to the total participation charge as a function on $\Theta$, and is a linear continuous operator.
To write a formal duality result, we follow the general perturbation approach \citep[see e.g.][Chapter 3]{EkelandTemam1999}.
Define the perturbed objective functional:
\begin{align*}
    \Xi: (\mathbb{R} \times C_{-1}^n \times C_{-1}^m) \times (C \times \mathbb{R}^n \times \mathbb{R}^m) \; \longrightarrow & \; \overline{\mathbb{R}} 
    \\
    \bigl( (\delta,{\bf h}^B,{\bf h}^S) ; (\gamma, {\bf z}^B, {\bf z}^S) \bigr) \; \longmapsto \; \delta &+ \Gamma \biggl( {\bf h}^B \oplus {\bf h}^S - (n+m-1) W + \delta - \gamma \biggr) 
    \\
    & + \sum_i \chi \bigl(\overline{W}_{-i}- \mathbb{E} [h_i^B] - z_i^B \bigr) 
    \\
    & + \sum_j \chi \bigl(\overline{W}_{-j}- \mathbb{E} [h_j^S]  - z_j^S \bigr)
\end{align*}
where and $\Gamma,\chi$ are the convex indicators:\\
\begin{minipage}{.45\textwidth}
    \begin{align*}
    \Gamma:  C \longrightarrow & \overline{\mathbb{R}}
    \\
    \gamma \longmapsto & \begin{cases} 0  & \text{ if } \gamma \geq 0 \text{ (pointwise) } \\ +\infty & \text{ otherwise } \end{cases}
    \end{align*}
\end{minipage}
\begin{minipage}{.45\textwidth}
    \begin{align*}
    \chi:  \mathbb{R} \longrightarrow & \overline{\mathbb{R}}
    \\
    x \longmapsto & \begin{cases} 0  & \text{ if } x \geq 0 \\ +\infty & \text{ otherwise } \end{cases}
    \end{align*}
\end{minipage}
\\The $\Gamma$ term captures the pointwise revenue constraint (perturbed by the function $\gamma$) and each of the $R$ term captures the corresponding IIR constraints (perturbed by real numbers $z_i^B$,$z_j^S$ respectively). 
To condense notation further and state the duality result in the most standard fashion, define $X:= \mathbb{R} \times C_{-1}^n \times C_{-1}^m$ and $Y=C \times \mathbb{R}^n \times \mathbb{R}^m$ with respective duals (identifiable to): $X^* = \mathbb{R} \times \mathfrak{M}(\Theta_{-1})^n \times \mathfrak{M}(\Theta_{-1})^m$ and $Y^* = \mathfrak{M}(\Theta) \times \mathbb{R}^n \times \mathbb{R}^m$, where $\mathfrak{M}$ denotes the space of Radon measures.

By construction our starting primal is the unperturbed problem:
\begin{align*}
 \inf_{ x \in X } \Xi(x,0).
\end{align*}
$\Xi$ is a proper convex lsc function (since both $\Gamma$ and $\chi$ are proper convex lsc, and they are composed with continuous linear operators), furthermore is is direct to show that the value in the primal is finite. 
By Proposition 2.3. in \citet{EkelandTemam1999} it suffices to show (qualification condition) that there exists $\bar{x} \in X$ such that $y \mapsto \Psi(\bar{x},y)$ is continuous at $y=0$ to establish that strong duality holds in the sense that:
\begin{align*}
    \inf_{ x \in X } \Xi(x,0) = \max_{ q \in Y^* } - \Xi^*(0,q)
\end{align*}
where $\Xi^* : X^* \times Y^* \rightarrow \overline{\mathbb{R}}$ denotes the Legendre transform of $\Xi$.
In this problem, the qualification condition can be shown immediately by taking $\bar{x} = (\bar{\delta},\bar{h}^B,\bar{h}^S)$ with $\bar{h}^B_i \equiv \bar{h}_j^S \equiv -1$ for all $i,j$, $\bar{\delta} = (n+m-1)\sup_{\bv,\bc} W(\bv,\bc) + \eta$ for $\eta >0$, since within a neighborhood of $y=0$ we have $\Xi(\bar{x},y)=\delta$ (we are in the interior of all constraints), which then directly implies continuity.
This establishes that strong duality holds, so we are only left with deriving an explicit expression of the dual problem.
Write explicitly:
\begin{align*}
    \Xi^* \bigl( (0,0,0) ; (\lambda,\bm\alpha,\bm\beta) \bigr) &= \sup_{\substack{(\delta,{\bf h}^B,{\bf h}^S) \in X \\ (\gamma,{\bf z}^B,{\bf z}^S) \in Y}} \,\, \int \gamma \,\mathrm{d}\lambda + \sum_i \alpha_i z_i^B + \sum_j \beta_j z_j^S
    \\
    & \hspace{4em} + \delta - \Gamma \biggl( {\bf h}^B \oplus {\bf h}^S - (n+m-1) W + \delta - \gamma \biggr)
    \\
    & \hspace{4em} - \sum_i \chi \bigl( \overline{W}_{-i}- \mathbb{E} h^i_B  - z_i^B \bigr) - \sum_j \chi \bigl( \overline{W}_{-j}- \mathbb{E} h^j_S  - z_j^S \bigr)
\end{align*}
Rearrange and perform the usual change of variable to rewrite each of the inner supremum terms in terms of the Legendre transform of $\Gamma$ and $\chi$. 
Putting it all together:
\begin{align*}
    \Xi^* \bigl( (0,0,0) ; (\lambda,\bm\alpha,\bm\beta) \bigr) = - \inf_{(\delta,h^B,h^S)} 
         & \int \bigl( \delta + (n+m-1) W - {\bf h}^B \oplus {\bf h}^S  \bigr) \,\mathrm{d}\lambda + \Gamma^\star(-\lambda)
        \\
        & \quad - \sum_i \alpha_i (\overline{W}_{-i}- \mathbb{E} h^i_B) + \chi^*(-\alpha_i)
        \\
        & \quad - \sum_j \beta_j (\overline{W}_{-j}- \mathbb{E} h^j_S) + \chi^*(-\beta_i)
        \\
        & \quad - \delta
\end{align*}
The term inside the supremum is the Lagrangian, as usually recovered from the perturbation. $\Gamma^*$ and $\chi^*$ encode negativity constraints (over $\mathfrak{M}(\Theta)$ and $\mathbb{R}$ respectively), so the Legendre transform terms taken together encode the constraints that $\lambda$ is a positive measure and $\alpha_i,\beta_j \geq 0$ for all $i,j$.
Further rearranging the term gives:
\begin{align*}
    -\Xi^* \bigl( (0,0,0) ; (\lambda,\bm\alpha,\bm\beta) \bigr) = &  (n+m-1) \int W \,\mathrm{d}\lambda - \sum_i \alpha_i \overline{W}_{-i} - \sum_j \beta_j \overline{W}_{-j} 
    \\
    & \quad + \inf_{\delta} \delta \biggl( \int \,\mathrm{d}\lambda - 1\biggr)
    \\
    & \quad - \sum_i \biggl( \inf_{h_i^B} \alpha_i \mathbb{E} h_i^B -\int h_i^B \,\mathrm{d}\lambda_{-i} \biggr)
    \\
    & \quad - \sum_j \biggl( \inf_{h_j^S} \beta_j \mathbb{E} h_j^S - \int h_j^S \,\mathrm{d}\lambda_{-j} \biggr)
    \\
    & \quad + \Gamma^\star(-\lambda) + \sum_i \chi^*(-\alpha_i) + \sum_j \chi^*(-\beta_i).
\end{align*}
The first line is the dual objective. 
The second line will encode the constraint that $\lambda(\Theta)=1$ i.e. $\lambda$ is a probability measure.
The third and fourth lines encode marginal constraints; indeed taking any $i$ to illustrate:
\begin{align*}
\inf_{h_i^B} \alpha_i \mathbb{E} h_i^B -  \int h_i^B \,\mathrm{d}\lambda_{-i}  = \inf_{h_i^B} \int h_i^B(\bv_{-i},\bc) \bigl( \alpha_i \,\mathrm{d}F_{-i}(\bv_{-i}) \,\mathrm{d}G(\bc) -  \,\mathrm{d}\lambda_{-i}(\bv_{-i},\bc) \bigr)
\end{align*}
where $\lambda_{-i}$ denotes the projection of $\lambda$ on the space of type vectors excluding $i$ (the ``marginal'' on the $-i$ space). 
Hence this term imposes that $\lambda_{-i} = \alpha_i (F_{-i} \otimes G)$ (understanding CDFs as measures).
Putting everything together, we can write the dual problem as:
\begin{align*}\tag{D}\label{eq:main-dual}
\sup_{\substack{\lambda \in \Delta(\Theta) \\ \bm\alpha \in \mathbb{R}^n_+,\bm\beta \in \mathbb{R}^m_+}}& (n+m-1) \int W \,\mathrm{d}\lambda - \sum_i \alpha_i \overline{W}_{-i}- \sum_j \beta_j \overline{W}_{-j}
\\
\text{ s.t. } & \lambda_{-i} = \alpha_i (F_{-i} \otimes G) \quad \forall i\in N,
\\
& \lambda_{-j} = \beta_j (F \otimes G_{-j}) \quad \forall j\in M.
\end{align*}
where we have incorporated the positivity and probability constraints directly as a choice of ambient space in the problem, and isolated the marginal constraints.
Integrating the constraints yields that necessarily $\alpha_i=\beta_j=1$ for all $i,j$ so we get the simplified dual, removing the constant terms:
\begin{align*}\tag{D'}\label{eq:reduced-dual}
    \sup_{\lambda \in \Delta(\Theta)}& \int W  \,\mathrm{d}\lambda
    \\
    \text{ s.t. } & \lambda_{-i} =  F_{-i} \otimes G \quad \forall i\in N,
    \\
    & \lambda_{-j} = F \otimes G_{-j} \quad \forall j\in M.
\end{align*}
Furthermore, the supremum in both \eqref{eq:main-dual} and \eqref{eq:reduced-dual} is attained by strong duality. \hfill $\square$

\begin{remark}
    Assuming continuous participation charges is the natural way to rigorously prove the duality result, because of the topological duality between the spaces of continuous functions and Radon measures. However, this implies no restriction on the value. 
    At minimum, an admissible class of function for participation charges requires integrability, so that the IIR constraints are well defined. 
    It can be shown directly by a density argument that the \emph{value} in the deficit guarantee problem is unchanged when considering either continuous charges or $L^1$ charges.
    Strong duality only gives existence in the problem over couplings (which is, in any case, easy to obtain by a direct argument). 
    Existence of solutions to the original problems may only be available in the relaxed class; we discuss this below in \hyref{Section}{app:proof-existence-prop}.
\end{remark}

\subsection{Proof of \hyref{Theorem}{thm:optimal-mechanism-from-potentials}}
Starting back from the reduced dual problem, we can perform one more dualisation step to get another standard ``Kantorovich-style'' problem, which we will call the \emph{reduced primal}.
The reduced dual \eqref{eq:reduced-dual} above is itself the dual problem of:
  \begin{align*}\tag{P'}\label{eq:reduced-primal}
    \inf_{(\bm\phi^B,\bm\phi^S) \in C_{-1}^n \times C_{-1}^m}&\,\,\,  \sum_i \int \phi^B \,\mathrm{d}F_{-i} \,\mathrm{d}G + \sum_j \int \phi^S_j \,\mathrm{d}F \,\mathrm{d}G_{-j}
    \\
    \text{ s.t. } \,\,\,\,& \bm\phi^B \oplus \bm\phi^S \geq W
  \end{align*}
where recall that we have defined the operator $\bm\phi^B \oplus \bm\phi^S:=\sum_i \phi_i^B + \sum_j \phi_j^S$ understood as a function over $\Theta$. Once we establish that strong duality holds i.e. \eqref{eq:reduced-primal}$=$\eqref{eq:reduced-dual} and there exists a solution to \eqref{eq:reduced-primal}, we can combine primal dual relationships above to get a characterisation of the participation charges in terms of the optimal potentials in \eqref{eq:reduced-primal}.

Strong duality between \eqref{eq:reduced-primal} and \eqref{eq:reduced-dual} follows from a similar argument as previously. We provide a concise version of the perturbation argument for completeness; note however that this duality step also follows from the results in \citet{GKZ2022}.
Define the perturbed functional
{\small
\begin{align*}
  \Xi \big((\bm\phi^B,\bm\phi^S),\gamma\big) :=
  \sum_{i=1}^n \int \phi_i^B\,\mathrm{d}F_{-i} \,\mathrm{d}G
  +\sum_{j=1}^m \int \phi_j^S\,\mathrm{d}F \,\mathrm{d}G_{-j}
  +\Gamma\!\big(\bm\phi^B \oplus \bm\phi^S-W-\gamma\big),
  \,\, \gamma\in C(\Theta)
\end{align*}}
Then the reduced primal rewrites as:
\begin{align*}
  \inf_{(\bm\phi^B,\bm\phi^S)\in X} \Xi \big((\bm\phi^B,\bm\phi^S),0\big).
\end{align*}
Since $W\in C(\Theta)$ and $\Theta$ is compact, $W$ is bounded. Choose constants
$\bar\phi_i^B,\bar\phi_j^S$ such that
\[
  \sum_{i=1}^n \bar\phi_i^B+\sum_{j=1}^m \bar\phi_j^S > \sup_{\Theta} W.
\]
Then $\bar{\bm\phi}^B \oplus \bar{\bm\phi}^S-W$ is strictly positive, hence
$\gamma\mapsto \Xi(\bar{\bm\phi}^B,\bar{\bm\phi}^S,\gamma)$ is finite and continuous at $\gamma=0$.
Therefore the qualification condition is satisfied and strong duality holds:
\begin{align*}
  \inf_{(\bm\phi^B,\bm\phi^S)\in X} \; \Xi \big((\bm\phi^B,\bm\phi^S),0\big) = \max_{\lambda \in Y^*} \; \Xi^*(0,\lambda)
\end{align*}
A straightforward explicit derivation of $\Psi^*$ yields the expected primal-dual relationship:
\[
    \max_{\lambda\in\Lambda(\mathbf{F},\mathbf{G})} \int_\Theta W\,\mathrm{d}\lambda = \inf_{\substack{(\bm\phi^B,\bm\phi^S) \in C_{-1}^n \times C_{-1}^m \\ \bm\phi^B \oplus \bm\phi^S \geq W}} \sum_i \int \phi_i^B \,\mathrm{d}F_{-i} \,\mathrm{d}G + \sum_j \int \phi^S_j \,\mathrm{d}F \,\mathrm{d}G_{-j}.
\]

To construct a solution in the original problem, assume $(\phi_i^B)_{i=1}^n$ and $(\phi_j^S)_{j=1}^m$ solve the reduced primal.
Define:
\[
  h_i^B
  :=
  (n+m-1)\Big(\phi_i^B-\mathbb E[\phi_i^B]\Big)+\mathbb E[W_{-i}],
  \qquad
  h_j^S
  :=
  (n+m-1)\Big(\phi_j^S-\mathbb E[\phi_j^S]\Big)+\mathbb E[W_{-j}],
\]
and set
\[
    \delta^\star
    = (n+m-1)\left(\sum_{i=1}^n \mathbb E[\phi_i^B]+\sum_{j=1}^m \mathbb E[\phi_j^S]
    \right) - \sum_{i=1}^n \mathbb E[W_{-i}]- \sum_{j=1}^m \mathbb E[W_{-j}].
\]

Then, by construction,
\[
    \mathbb E[h_i^B]=\mathbb E[W_{-i}],
    \qquad
    \mathbb E[h_j^S]=\mathbb E[W_{-j}],
\]
so the interim participation constraints bind. Moreover, using the reduced feasibility constraint, for all $(\bv,\bc)$
\[
    \sum_{i=1}^n \phi_i^B(\bv_{-i},\bc)+\sum_{j=1}^m \phi_j^S(\bv,\bc_{-j})\ge W(\bv,\bc),
\]
we obtain pointwise
\[
    (n+m-1)W(\bv,\bc)-\sum_{i=1}^n h_i^B(\bv_{-i},\bc)
    -\sum_{j=1}^m h_j^S(\bv,\bc_{-j})\leq \delta^\star.
\]
Hence $(\delta^\star,{\bf h}^B, {\bf h}^S)$ is feasible for the original primal.

If $(\bm\phi^B,\bm\phi^S)$ is optimal for \eqref{eq:reduced-primal}, then by strong duality for the reduced pair:
\[
    \sum_{i=1}^n \mathbb E[\phi_i^B]+\sum_{j=1}^m \mathbb E[\phi_j^S]
    =\max_{\lambda\in\Lambda(\mathbf{F},\mathbf{G})}\int_\Theta W\,\mathrm{d}\lambda,
\]
so
\[
    \delta^\star = (n+m-1)\max_{\lambda\in\Lambda(\mathbf{F},\mathbf{G})}\int_\Theta W\,\mathrm{d}\lambda - \sum_{i=1}^n \mathbb E[W_{-i}]- \sum_{j=1}^m \mathbb E[W_{-j}],
\]
which is exactly the optimal revenue guarantee in the original problem.
Thus an optimizer of the reduced primal yields, by an explicit affine renormalization, an optimizer of the original primal.    \hfill $\square$

\subsection{Proof of \hyref{Proposition}{prop:existence-L1-bounded-below-potentials}.}\label{app:proof-existence-prop}
Say that a set $E$ has zero thickness if:
\begin{align*}
    E \subseteq \bigcup_{k=1}^{n+m} \pi^{-1}_{-k}(Z_{k}) \text{ where } Z_k \subseteq \Theta_{-k} \text{ and } (\mathbf{F} \otimes \mathbf{G})_{-k}(Z_k) = 0,
\end{align*}
where $\pi_{-k}$ denotes the projection operator from $\Theta$ on $\Theta_{-k}$.
Observe that having zero thickness is a stronger requirement than having measure zero in $\Theta$ (all zero thickness sets have measure zero in $\Theta$ but the converse is false).
Consider the following relaxation of the \textbf{(WI)} problem:
\begin{align*}
    \inf_{(\bm\phi^B,\bm\phi^S) \in (L^1(\Theta^{n+m-1}))^{n+m}} & \sum_i \int \phi_i^B \,\mathrm{d}F_{-i} \,\mathrm{d}G + \sum_j \int \phi^S_j \,\mathrm{d}F \,\mathrm{d}G_{-j}
    \\
    \text{ subject to } & \sum_i \phi_i^B(\mathbf{v}_{-i},\mathbf{c}) + \sum_j \phi_j^S(\mathbf{v},\mathbf{c}_{-j}) \geq W(\mathbf{v},\mathbf{c}) 
    \\
    & \; \; \text{ for all } (\mathbf{v},\mathbf{c}) \text{ up to a zero thickness set}
\end{align*}
Existence of solutions follows from Theorem 5.19 in \citet{GKZ2022}.
This furthermore establishes that this relaxed value equals the value in the adversarial surplus maximization problem \textbf{(AS)}.
Constructing participation charges as in \hyref{Theorem}{thm:optimal-mechanism-from-potentials} gives a solution to the $L^1$ relaxed deficit guarantee problem:
\begin{align*}
    \inf_{\substack{({\bf h}^B,{\bf h}^S) \in (L^1(\Theta^{n+m-1}))^{n+m} \\ \delta \in \mathbb{R}}} & \; \; \delta
    \\
    \text{subject to } & (n+m-1) W - {\bf h}^B \oplus {\bf h}^S \leq \delta \; \text{ a.e. }
    \\
    & \int h_i^B \,\mathrm{d}F_{-i} \,\mathrm{d}G \leq \overline{W}_{-i} \; \forall i
    \\
    & \int h_j^S \,\mathrm{d}F \,\mathrm{d}G_{-j} \leq \overline{W}_{-j} \; \forall j
\end{align*}
which by a direct density argument has the same value $\delta^\star$ as our guarantee problem. 
Let $E \subseteq \bigcup_{k=1}^{n+m} \pi^{-1}_{-k}(Z_{k})$ the (zero thickness, hence measure zero) set on which the constraint might fail.
Define for any $k=1,...,n+m$:
\begin{align*}
    \tilde{\phi}_k = \begin{cases}
        \phi_k \text{ on } \Theta_{-k} \setminus Z_k
        \\
        +\infty \text{ on } Z_k
    \end{cases}
\end{align*}
By definition this belongs to $L^1$ with extended values in $(-\infty,+\infty]$ and the value in either problems above is unchanged. Furthermore the constraint is not only still satisfied (so $\tilde{\phi}$ is optimal) but now holds everywhere. \hfill $\square$

\begin{remark}
    The obstruction to bounded potentials is that, to improve on boundedness up to a zero thickness set, one must find a modification of $\phi_k$ which preserves:
    \begin{align*}
        \phi_k(\bm\theta_{-k}) \geq W(\bm\theta) - \sum_{l \neq k} \phi_l(\bm\theta_{-l})
    \end{align*}
    at all profiles $\bm\theta=(\mathbf{v},\mathbf{c})$. Since this requires that the potentials for other agents are bounded uniformly bounded below on the relevant slices of overlapping coordinates, this falls into a circularity problem which is difficult to resolve. 

    In \hyref{Appendix}{app:bounded-charges} we consider a restricted problem which imposes bounded charges via a limited liability assumption, and show that there exists approximately optimal charges in this class. Whether we can find bounded charges with a pointwise guarantee in the original problem hinges on whether such a cap eventually stops binding as the liability limit becomes large. The reason this is not automatic is that, intuively, an adversarial nature concentrates mass on correlated type configurations that are negligible in the whole market yet consistent with every $(N-1)$-dimensional marginal. To guarantee against them, the designer may want to pay one agent ever larger subsidies (very negative $h_k$) on smaller and smaller sets of the others' reports, compensating with high charges for other agents. Boundedness requires that these kinds of subsidies are not strictly beneficial, so that the gain from removing the limited liability bound disappear at a large enough threshold. This is a calmness/stability property of the problem.

    When $n+m=2$ this holds by the classical $c$-transform argument of optimal transport: each potential can be replaced by a transform of the other, and continuous $W$ makes these transforms bounded. For $n+m \geq 3$ the argument breaks down, because the other agents' charges also depend on the reports one would optimize over. Known results are scarce (e.g. in \citet{GKZ2022} for the case $n+m=3$) and rely on additional structure; the general case remains open.
\end{remark}

\subsection{Proof of \hyref{Proposition}{prop:limit-deficit}}\label{app:limit-deficit-proof}
We reuse the notations introduced in \hyref{Section}{sec:price-by-price-decomposition} for the price-by-price decomposition. By revenue equivalence, the expected deficit across all mechanisms is the same as the VCG deficit. Recall that the ex-post VCG deficit, with $n$ buyers and $n$ sellers, is given by $\sum_{i} (W_n - W_{n,-i}) + \sum_j (W_n - W_{n,-j}) - W_n$, where $W_n$ denotes the welfare function indexed by market size, and similarly $W_{n,-i}$ with agent $i$ removed. To simplify the exposition, with a slight abuse of notations, identify the functions $W_n,W_{n,-i},W_{n,-j}$ with the corresponding random variables for the realized welfare. Similarly, denote $D_n(p) = \sum_{i=1}^n y_i^B(p;v_i)$ the random variable for the total quantity demanded at price $p$, where the $v_i$ are $n$ iid realizations from $F$, and $S_n(p) = \sum_{j=1}^n y_j^S(p;c_j)$ the random variable for the total quantity demanded at price $p$, where the $c_j$ are $n$ iid realizations from $G$.
Recall the price-by-price decomposition:
\begin{align*}
    W_n(\mathbf{v},\mathbf{c}) = \int_0^1 \min \bigl\{ D_n(p), S_n(p) \bigr\}\, \mathrm{d}p.
\end{align*}
Fixing a candidate price $p$, deleting buyer $i$ reduces $D_n(p)$ by exactly one when $p \leq v_i$; hence:
\begin{align*}
    W_n - W_{n,-i} &= \int_0^1 \min \{ D_n(p), S_n(p) \} \, \mathrm{d}p  -  \int_0^{v_i} \min \{ D_n(p)- 1, S_n(p) \} \mathrm{d}p\\ &
    -  \int_{v_i}^1 \min \{ D_n(p), S_n(p) \} \, \mathrm{d}p
    = \int_0^{1} \mathds{1}\{ p \leq v_i \} \mathds{1}\{ D_n(p) \leq S_n(p) \} \, \mathrm{d}p
\end{align*}
Where the second line uses that for all $d \geq 1$ and all $s$: $\min\{d,s\}- \min\{d-1,s\}=\mathds{1}\{d \leq s\}$.
Symmetrically:
\begin{align*}
    W_n - W_{n,-j} = \int_0^1 \mathds{1}\{ p \geq c_j \} \mathds{1}\{S_n(p) \leq D_n(p) \} \, \mathrm{d}p
\end{align*}
Summing the welfare differences over $i$ gives:
\begin{align*}
    \sum_{i=1}^n (W_n - W_{n,-j}) &= \sum_{i=1}^n \int_0^{1} \mathds{1}\{ p \leq v_i \} \mathds{1}\{ D_n(p) \leq S_n(p) \} \, \mathrm{d}p
    \\
    & = \int_0^1 \biggl( \mathds{1}\{ D_n(p) \leq S_n(p) \} \sum_{i=1}^n \bigl( \mathds{1}\{ p \leq v_i \} \bigr) \biggr) \, \mathrm{d}p
    \\
    & = \int_0^1 \biggl( \mathds{1}\{ D_n(p) \leq S_n(p) \} D_n(p) \biggr) \, \mathrm{d}p.
\end{align*}
Symmetrically:
\begin{align*}
    \sum_{j=1}^n ( W_n - W_{n,-j} ) = \int_0^1 \mathds{1}\{S_n(p) \leq D_n(p) \} S_n(p) \, \mathrm{d}p.
\end{align*}
So that the VCG deficit rewrites and simplifies into:
\begin{align*}
    \int_0^1 & \biggl( \mathds{1}\{ D_n(p) \leq S_n(p) \} D_n(p) + \mathds{1}\{S_n(p) \leq D_n(p) \} S_n(p) - \min\{D_n(p), S_n(p) \biggr) \, \mathrm{d}p
    \\
    & = \int_0^1 D_n(p) \mathds{1}\{D_n(p)=S_n(p)\}\, \mathrm{d}p.
\end{align*}
Therefore, by revenue equivalence, and interchanging expectation and integral, the expected deficit is:
\begin{align*}
    \overline{\delta}_n = \int_0^1 \mathbb{E} \bigl[ D_n(p) \mathds{1}\{D_n(p)=S_n(p)\} \bigr] \, \mathrm{d}p.
\end{align*}
Rewriting the expectation under the integral explicitly gives:
\begin{align*}
    \overline{\delta}_n = \int_0^1 \biggl( \sum_{k=1}^n k\binom nk^2 \bigl[(1-F)(p)G(p) \bigr]^k \bigl[F(p)(1-G)(p)\bigr]^{n-k} \biggr) \,\mathrm{d}p.
\end{align*}
To evaluate this integral, set
\begin{align*}
    Z(p)=\sqrt{(1-F(p))G(p)} +\sqrt{F(p)(1-G(p))}, 
    \qquad
    \xi(p)=\frac{\sqrt{(1-F(p)) G(p)}}{Z(p)},
\end{align*}
and note that at the market clearing price $p^\star$, $Z(p^\star)=1$ and $\xi(p^\star)=q^\star$.

The integrand in $\overline{\delta}_n$ can be factorized as:
\begin{align*}
    Z(p)^{2n} \sum_{k=0}^n k \biggl[ {\binom nk}\xi(p)^k \bigl( 1-\xi(p) \bigr)^{n-k} \biggr]^2.
\end{align*}

Next, we use binomial asymptotics to approximate the sum uniformly over $\xi(p)$ as $n$ becomes large. 
Fix any $x \in (0,1)$ and define:
\begin{align*}
    T_n :=  \sum_{k=0}^n k \biggl[ {\binom nk}x^k \bigl( 1-x \bigr)^{n-k} \biggr]^2,
\end{align*}
i.e. the sum term in the above integrand, with $\xi(p)$ replaced by $x$.
The key idea is to use complex exponentials as a device to represent matching indices, this allows us to rewrite the sum as an integral, which in turn enables Gaussian approximation.
Recall the following two useful facts. First, for any integer $m \in \mathbb{Z}$, $\frac{1}{2\pi} \int_{-\pi}^{\pi} e^{\mathrm{i}m\theta} \, \mathrm{d}\theta$ is equal to $1$ if $m=0$ and $0$ otherwise (where $\mathrm{i}$ denotes the imaginary unit).
Second, applying the binomial identity to $1-x$ and $xe^{-\mathrm{i}\theta}$, for any $\theta\in (-\pi,\pi)$:
\begin{align*}
    (1-x+xe^{-\mathrm{i}\theta})^n = \sum_{k=0}^n {\binom n k} (1-x)^{n-k} x^k e^{-\mathrm{i}k\theta}.
\end{align*}
and using $n \binom{n-1}{k-1} = k \binom{n}{k}$ gives:
\begin{align*}
     \sum_{k=0}^n k{\binom n k} (1-x)^{n-k} x^k e^{-\mathrm{i}k\theta} = n x e^{\mathrm{i}\theta}(1-x+xe^{-\mathrm{i}\theta})^{n-1}.
\end{align*}
 we can use this to rewrite:
\begin{align*}
    T_n  
    & = \sum_{k=1}^n \sum_{\ell = 0}^n k {\binom n k} {\binom n \ell} x^{k+\ell}(1-x)^{2n-k-\ell} \frac{1}{2\pi} \int_{-\pi}^{\pi} e^{\mathrm{i}(\ell-k)\theta} \, \mathrm{d}\theta
    \\
    & = \frac{n x}{2 \pi} \int_{-\pi}^{\pi} e^{-\mathrm{i}\theta}(1-x+xe^{\mathrm{i}\theta})^{n-1}(1-x+xe^{-\mathrm{i}\theta})^n \, \mathrm{d}\theta
    \\
    & = \frac{n x}{2 \pi} \int_{-\pi}^{\pi} \bigl[x + (1-x)\cos(\theta) \bigr] \bigl[1- 4x(1-x) \sin^2(\sfrac{\theta}{2}) \bigr]^{n-1} \, \mathrm{d}\theta
    \\
    & = \frac{n x}{2 \pi} \int_{-\pi}^{\pi} \bigl[x + (1-x)\cos(\theta) \bigr] \bigl[1- 2x(1-x) (1-\cos(\theta)) \bigr]^{n-1} \, \mathrm{d}\theta
\end{align*}
Where, to get the last equality, observe that the imaginary part integrates to zero. The next step is to use the integral form to obtain a Gaussian approximation; indeed, observe that the power $(n-1)$ in the integrand concentrates the integral near $0$, within a window of width $\sfrac{1}{\sqrt{n}}$.
To make it explicit, substitute $t = \sqrt{n}\theta$ and write:
\begin{align*}
    \frac{T_n}{\sqrt{n}} = \frac{x}{2\pi} \int_{\mathbb{R}} \mathds{1}\{|t|\leq \pi \sqrt{n} \} \biggl[ x + (1-x) \cos \frac{t}{\sqrt{n}} \biggr]  \biggl[ 1 - 2 x(1-x) \biggl( 1 - \cos \frac{t}{\sqrt{n}} \biggr) \biggr]^{n-1} \mathrm{d} t
\end{align*}
For any $\varepsilon\in(0,1)$, the elementary limits
$(1-\cos\theta)/\theta^2\to 1/2$ and
$-\log(1-u)/u\to 1$ imply that, for sufficiently small $\delta>0$
and all $|\theta|\leq\delta$,
\begin{align*}
    1-\varepsilon
    &\leq x+(1-x)\cos\theta \leq 1,\\
    e^{-(1+\varepsilon)x(1-x)\theta^2}
    &\leq 1-2x(1-x)(1-\cos\theta)
    \leq e^{-(1-\varepsilon)x(1-x)\theta^2}.
\end{align*}
Thus, the contribution of $|t|\leq\delta\sqrt n$ to the integral in $T_n/\sqrt n$ is bounded below and above, respectively, by
\begin{align*}
    \frac{x(1-\varepsilon)}{2\pi}
    \int_{-\delta\sqrt n}^{\delta\sqrt n}
    e^{-(1+\varepsilon)x(1-x)(1-1/n)t^2}\,\mathrm{d}t,
    \qquad
    \frac{x}{2\pi}
    \int_{-\delta\sqrt n}^{\delta\sqrt n}
    e^{-(1-\varepsilon)x(1-x)(1-1/n)t^2}\,\mathrm{d}t.
\end{align*}
By a linear change of variables and the Gaussian integral, these converge to
\[
    \frac{x(1-\varepsilon)}{2\sqrt{\pi(1+\varepsilon)x(1-x)}},
    \qquad
    \frac{x}{2\sqrt{\pi(1-\varepsilon)x(1-x)}}.
\]
The remaining contribution has absolute value at most
\[
    x\sqrt n\,\bigl[1-2x(1-x)(1-\cos\delta)\bigr]^{n-1}
    \longrightarrow 0.
\]
Letting $\varepsilon\downarrow0$ therefore gives
\[
    \frac{T_n}{\sqrt n}
    \longrightarrow
    \frac{x}{2\sqrt{\pi x(1-x)}}.
\]
All bounds, and hence the convergence, are uniform for $x$ in any compact subset of $(0,1)$.

Now return to the price-by-price integral. The preceding uniform approximation implies that, in a sufficiently small fixed neighborhood of $p^\star$,
\[
    \overline{\delta}_n
    = \sqrt n \int_{p^\star-\delta}^{p^\star+\delta}
    Z(p)^{2n}
    \left[
        \frac{\xi(p)}{2\sqrt{\pi\xi(p)(1-\xi(p))}}
        +o(1)
    \right]\,\mathrm{d}p +o(1),
\]
where the first $o(1)$ is uniform in $p$. Indeed, $Z(p)\leq1$, with equality exactly when $1-F(p)=G(p)$. By continuity and uniqueness of $p^\star$, $Z$ is bounded strictly below $1$ outside this neighborhood. 
Since $T_n\leq n$, the omitted contribution is exponentially small.
To evaluate the remaining integral, observe that
\[
    1-Z(p)^2
    =
    \left[
        \sqrt{(1-F(p))(1-G(p))}
        -\sqrt{F(p)G(p)}
    \right]^2.
\]
Rationalizing the difference of square roots therefore gives
\[
    \lim_{p\to p^\star}
    \frac{-\log Z(p)^2}{(p-p^\star)^2}
    =
    \frac{[f(p^\star)+g(p^\star)]^2}
         {4q^\star(1-q^\star)}.
\]
Consequently, for any $\varepsilon\in(0,1)$, reducing $\delta$ if
necessary, we have for $|p-p^\star|\leq\delta$,
{\small
\begin{equation*}
    \exp\left\{
        -(1+\varepsilon)n
        \frac{[f(p^\star)+g(p^\star)]^2}
             {4q^\star(1-q^\star)}
        (p-p^\star)^2
    \right\}
    \leq Z(p)^{2n}
    \leq
    \exp\left\{
        -(1-\varepsilon)n
        \frac{[f(p^\star)+g(p^\star)]^2}
             {4q^\star(1-q^\star)}
        (p-p^\star)^2
    \right\}.
\end{equation*}}
Integrating these Gaussian bounds and letting $\varepsilon\downarrow0$
yields
\[
    \sqrt n\int_{p^\star-\delta}^{p^\star+\delta}Z(p)^{2n}\,\mathrm{d}p
    \longrightarrow
    \frac{2\sqrt{\pi q^\star(1-q^\star)}}{f(p^\star)+g(p^\star)}.
\]
Here the limit holds for every sufficiently small fixed $\delta$: any portion bounded away from $p^\star$ is exponentially negligible.
The same observation, together with continuity and $\xi(p^\star)=q^\star$, allows us to replace in the price integral to get:
\[
    \lim_{n\to\infty}\overline{\delta}_n
    =
    \frac{q^\star}{f(p^\star)+g(p^\star)}. 
\]
This concludes the proof.\hfill $\square$

\subsection{Proof of \hyref{Theorem}{thm:convergence}}
\label{app:limit-deficit-ceiling}
Let \(C_{n,n}=\binom{2n-2}{n-1}\). By \hyref{Claim}{claim:bernoulli-mechanism} and the
expected-deficit decomposition in \hyref{Section}{sec:price-by-price-decomposition},
\[
    0\le \mathcal E^\star_n
    \le (2n-1)C_{n,n} \int_0^1\overline{\epsilon}_p\,\mathrm{d}p.
\]
It remains to bound $\overline{\epsilon}_p$, the probability of a least-likely odd profile under $\mu_p$.

Fix $n-1$ buyers and $n-1$ sellers at their less likely realizations. Of the remaining buyer and seller, assign the coordinate with the smaller less-likely probability to its less likely realization, and choose the last indicator to give odd parity. The probability contribution of this last coordinate is at most one. Hence
\[
    \overline{\epsilon}_p
    \le [\beta(p)\sigma(p)]^{n-1}
        \min\{\beta(p),\sigma(p)\}.
\]
Substituting gives the main bound.

\begin{enumerate}[leftmargin=*]
\item[i.]
Now suppose $F$ and $G$ have distinct medians. Then $\beta(p)$ and $\sigma(p)$ cannot both equal $1/2$. Continuity and compactness therefore give $\nu=4\max_{p\in[0,1]}\beta(p)\sigma(p)<1$. Since $\min\{\beta(p),\sigma(p)\}\le1/2$,
\[
    \mathcal E^\star_n
    \le \frac{2n-1}{2}\binom{2n-2}{n-1}
        \left(\frac{\nu}{4}\right)^{n-1}
    \le \left(n-\frac12\right)\nu^{n-1},
\]
where the last inequality uses $\binom{2n-2}{n-1}\le4^{n-1}$.

\item[ii.]
Finally, suppose $F$ and $G$ share a median $p_0$. Since $F(p_0)=G(p_0)=1/2$ and both densities are bounded below by $\alpha$,
\[
    \beta(p)\le\frac12-\alpha|p-p_0|,
    \qquad
    \sigma(p)\le\frac12-\alpha|p-p_0|.
\]
Integrating over the largest interval on which this envelope is nonnegative gives
\[
    \int_0^1[\beta(p)\sigma(p)]^{n-1}\min\{\beta(p),\sigma(p)\}\,\mathrm{d}p
    \le
    2\int_0^{1/(2\alpha)}\left(\frac12-\alpha x\right)^{2n-1}\,\mathrm{d}x
    =\frac{1}{\alpha n4^n}.
\]
Therefore,
\[
    \mathcal E^\star_n
    \le \frac{1}{\alpha4^n}\left(2-\frac1n\right)\binom{2n-2}{n-1}
    =O(n^{-1/2}),
\]
where the rate follows from Stirling's formula
\citep[see, e.g.,][]{feller1968}.
\end{enumerate}
\hfill $\square$

\section{The uniform two-buyer, one-seller case}
\label{app:2b1s-mallows-details}
This appendix proves \hyref{Claim}{claim:2b1s-explicit-mallows-solution} by connecting the solution of \citet{Mallows2003} to our adversarial welfare problem and the resulting participation charges. The connection has two parts: reflecting types turns welfare maximization into Mallows's expected-minimum problem, and reversing his pair-potential inequality gives the welfare cover used to construct our mechanism.

\subsection{From adversarial welfare to the expected minimum}
With two buyers and one seller, efficient welfare is
\[
    W(v_1,v_2,c)
    =(\max\{v_1,v_2\}-c)^+
    =\max\{v_1,v_2,c\}-c.
\]
The leave-one-out constraints preserve the independent uniform law of every pair of types. Define the reflected coordinates
\(
    x_1=1-v_1, \, x_2=1-v_2, \, x_3=1-c.
\)
Reflection preserves these pair marginals, and
\[
    W(v_1,v_2,c)=1-c-\min\{x_1,x_2,x_3\}.
\]
Since the seller's marginal is fixed, every admissible coupling gives $\mathbb E_\lambda[c]=1/2$. Nature's problem is therefore equivalent to
\[
\begin{aligned}
    \max_{\lambda\in\Lambda(\mathbf F,\mathbf G)} \mathbb E_\lambda[W]
    = \frac12-\min_\gamma \mathbb E_\gamma[\min\{x_1,x_2,x_3\}],
\end{aligned}
\]
where $\gamma$ ranges over joint distributions whose three pair marginals are independent uniforms on $[0,1]^2$.
This is precisely the problem solved by \citet{Mallows2003}. His optimal value is
\[
    \kappa=\frac{297+5\sqrt5}{1444},
    \qquad
    \max_{\lambda\in\Lambda(\mathbf F,\mathbf G)} \mathbb E_\lambda[W]=\frac12-\kappa.
\]

There is a normalization to keep in mind when using the construction in Section~4 of \citet{Mallows2003}. His component tables use the cube $[0,L]^3$, where
\[
    L=36+6(3+\sqrt5),
    \qquad w=\frac{36}{L}=\frac{3(9-\sqrt5)}{38}.
\]
Let $\widetilde\lambda$ denote the full measure in this scaled cube, including all three cyclic copies of the listed components. Its pair marginals are Lebesgue measure on $[0,L]^2$, and its total mass is $L^2$, rather than $L^3$. The corresponding adversarial law for trade is defined by
\[
\begin{aligned}
    \int f(v_1,v_2,c)\,\mathrm{d}\lambda^\star
    =\frac1{L^2}\int f\left(1-\frac{X_1}{L}, 1-\frac{X_2}{L}, 1-\frac{X_3}{L}\right)\,\mathrm{d}\widetilde\lambda(\mathbf X)
\end{aligned}
\]
for every bounded measurable function $f$. Changing variables in each pair marginal shows directly that $\lambda^\star$ preserves the required independent uniform pairs. Thus the reflection and normalization apply to the entire measure, including its planar components, without assuming a three-dimensional density.

\subsection{Mallows's pair potential}
Let $A_k=w2^{-k}$ for $k=0,1,\ldots$. Define $\eta(0)=0$, $\eta(x)=w/6$ for $w\le x\le1$, and, on each dyadic interval $[A_k/2,A_k]$, set
\begin{equation}\label{eq:mallows-eta}
    \eta(x)=
    \begin{cases}
        A_k/12,&A_k/2\le x\le5A_k/6,\\
        x/2-A_k/3,&5A_k/6\le x\le A_k.
    \end{cases}
\end{equation}
The pieces agree at their boundaries. This is equivalent to the floor-function definition in \hyref{Claim}{claim:2b1s-explicit-mallows-solution}. Now define
\begin{equation}\label{eq:mallows-pair-potential}
    \varphi(x,y)
    =\min\left\{\frac{x}{2}-\eta(y),\, \eta(x)+\eta(y),\,\frac{y}{2}-\eta(x)\right\}.
\end{equation}
This is a compact expression for Mallows's piecewise pair potential $g^1$ in Section~3, with his function $f$ written as $\eta$.

The relevant dual inequality can be checked directly:
\begin{equation}\label{eq:mallows-pair-minorant}
    \varphi(x_1,x_2)+\varphi(x_1,x_3)+\varphi(x_2,x_3)
    \le\min\{x_1,x_2,x_3\}.
\end{equation}
Indeed, suppose $x_1$ is a smallest coordinate. Bound the first two terms by $x_1/2-\eta(x_2)$ and $x_1/2-\eta(x_3)$, respectively, and the third by $\eta(x_2)+\eta(x_3)$. Adding cancels the $\eta$ terms and gives $x_1$.

The normalization needed for participation is
\begin{equation}\label{eq:mallows-pair-normalization}
    \int_0^1\!\int_0^1\varphi(x,y)\,\mathrm{d}x\,\mathrm{d}y
    =\frac w3-\frac{w^2}{2}+\frac{19w^3}{81}
    =\frac\kappa3.
\end{equation}
For completeness, on $[w,1]^2$ the potential equals $w/3$, while on a mixed rectangle $0\le x\le w\le y\le1$ it equals
$x/2-w/6$. These regions contribute
\[
    \frac w3(1-w)^2
    +2(1-w)\int_0^w\left(\frac x2-\frac w6\right)\,\mathrm{d}x
    =\frac w3-\frac{w^2}{2}+\frac{w^3}{6}.
\]
Inside $[0,w]^2$, self-similarity gives $\varphi(x/2,y/2)=\varphi(x,y)/2$. Write $I_w=\int_0^w\int_0^w\varphi(x,y)\,\mathrm{d}x\,\mathrm{d}y$. Integration of the affine branches over the two mixed rectangles and the upper square of the dyadic annulus gives
\[
\begin{aligned}
    \int_{[0,w]^2\setminus[0,w/2]^2}\varphi(x,y)\,\mathrm{d}x\,\mathrm{d}y
    &=\frac{w^3}{72}+\frac{59w^3}{1296} =\frac{77w^3}{1296}.
\end{aligned}
\]
The remaining half-square contributes $I_w/8$. Thus $I_w=77w^3/1296+I_w/8=11w^3/162$. Adding the outer contribution gives \eqref{eq:mallows-pair-normalization}. Substituting the stated value of $w$ yields
$3\int\int\varphi\,\mathrm{d}x\,\mathrm{d}y=\kappa$.

Under any feasible law for the reflected coordinates, each pair is independently uniform. Hence \eqref{eq:mallows-pair-minorant} implies $\mathbb E[\min_a x_a]\ge3\int\!\int\varphi=\kappa$. Mallows's construction attains equality and is supported where the pair-potential inequality binds. This supplies the matching coupling certificate, rather than merely a lower bound on the expected minimum.

\subsection{Proof of \hyref{Claim}{claim:2b1s-explicit-mallows-solution}}
To translate the pair minorant into our welfare imputation problem, define
\begin{align*}
    \phi_i^B(v_{-i},c)&=\frac{1-c}{2}-\varphi(1-v_{-i},1-c), \qquad i=1,2,
    \\
    \phi^S(v_1,v_2)&=-\varphi(1-v_1,1-v_2).
\end{align*}
Each potential omits the type of the agent to whom it is assigned. By \eqref{eq:mallows-pair-minorant},
\[
    \phi_1^B+\phi_2^B+\phi^S
    =1-c-\sum_{a<b}\varphi(x_a,x_b)
    \ge1-c-\min_a x_a
    =W.
\]
Under the independent prior, their expectations are
\[
    \mathbb E[\phi_i^B]=\frac14-\frac\kappa3,
    \qquad
    \mathbb E[\phi^S]=-\frac\kappa3.
\]
Their expected sum is $1/2-\kappa$, matching the adversarial welfare value. They therefore solve the welfare imputation problem.

Removing either buyer leaves bilateral trade between two independent uniform types, with expected surplus $1/6$. Removing the seller leaves no trade, so the seller's participation budget is zero.
The centered construction in \hyref{Theorem}{thm:optimal-mechanism-from-potentials} consequently gives
\begin{align*}
    h_i^B(v_{-i},c) &=\frac16+2\bigl(\phi_i^B(v_{-i},c)-\mathbb E[\phi_i^B]\bigr)
    =\frac23+\frac{2\kappa}{3}-c-2\varphi(1-v_{-i},1-c),
    \\
    h^S(v_1,v_2)&=2\bigl(\phi^S(v_1,v_2)-\mathbb E[\phi^S]\bigr)
    =\frac{2\kappa}{3}-2\varphi(1-v_1,1-v_2).
\end{align*}
Thus the charges are exactly those in the claim, with $\mathbb E[h_i^B]=1/6$ and $\mathbb E[h^S]=0$. All interim participation budgets bind.

For a direct pointwise guarantee, define the nonnegative slack
\[
    s(\mathbf x)
    =\min_a x_a-\sum_{a<b}\varphi(x_a,x_b).
\]
Substituting the charges into the deficit formula gives
\begin{equation}\label{eq:mallows-deficit-slack}
    \mathrm{Deficit}(v_1,v_2,c;\mathcal M)
    =\frac23-2\kappa-2s(\mathbf x)
    \le\frac23-2\kappa.
\end{equation}
Equality holds, for example, at $(v_1,v_2,c)=(1,1,1)$.

Conversely, integrate any feasible deficit ceiling against the reflected Mallows coupling $\lambda^\star$.
Since every charge depends on a prescribed pair, its expectation under $\lambda^\star$ equals its expectation under the prior. Weak duality therefore gives
\[
    \delta
    \ge2\mathbb E_{\lambda^\star}[W]-\frac13
    =2\left(\frac12-\kappa\right)-\frac13
    =\frac23-2\kappa.
\]
This matches the pointwise guarantee, proving optimality and
\[
    \delta^\star
    =\frac23-2\kappa
    =\frac{553-15\sqrt5}{2166}
    \simeq0.239824.
\]
Finally, the independent prior gives $\overline W=3/4-1/2=1/4$, so the average deficit is $2\overline W-1/3=1/6$. \hfill $\square$

\subsection{The coupling diagram and the regularity of the charges}
The reflection also explains the geometry in \hyref{Figure}{fig:2b1s-coupling}. The three outer blocks are cyclic permutations of $[0,1-w]^2\times[1-w,1]$, with combined probability $3(1-w)^2\simeq0.6515$. The remaining probability is not entirely planar: it includes both volume and singular components. The total planar mass is $w^2/9$. Within the upper corner, the construction repeats at dyadic scales: halving distances from $(1,1,1)$ quarters the mass of the corresponding components. The point $(1,1,1)$ is an accumulation point, not an atom. These statements concern support and component masses; uniform shading in the diagram does not represent uniform density.

Although this coupling has singular components, its pair potential and the resulting charges are continuous. The function $\eta$ has slopes $0$ and $1/2$, and each branch of \eqref{eq:mallows-pair-potential} is $1/2$-Lipschitz in either coordinate. Taking the minimum preserves this bound. Consequently,
\begin{align*}
    |h_i^B(v,c)-h_i^B(v',c')|&\le |v-v'|+2|c-c'|,
    \\
    |h^S(v_1,v_2)-h^S(v_1',v_2')|&\le |v_1-v_1'|+|v_2-v_2'|.
\end{align*}
In particular, the buyer's charge is nonincreasing in seller cost: the explicit term $-c$ falls by the cost increase, while the potential adjustment can offset at most that amount. Along the upper edge,
\[
    h^S(v,1)=\frac{2\kappa}{3}+2\eta(1-v),
    \qquad
    h_i^B(v,1)=h^S(v,1)-\frac13.
\]
Thus both charges inherit the dyadic structure: their slopes alternate between zero and $-1$, with shrinking corners accumulating at type one. This gives a case of continuous optimal charges despite a singular adversarial coupling, without establishing the same conclusion for general type distributions.

\section{Restricted mechanisms}\label{app:reformulations-approximations}

\subsection{Approximately optimal bounded charges}\label{app:bounded-charges}
Let $K=n+m\ge 2$ and index agents (buyers and sellers) by $k\in\mathcal K$. Write $\mathbf x=(\mathbf v,\mathbf c)\in \Theta^K$, $\mathbf x_{-k}\in \Theta ^{K-1}$ for the profile without agent $k$, $\mu=F^{\otimes n}\otimes G^{\otimes m}$, and $\mu_{-k}$ for the marginal of $\mu$ on the coordinates other than $k$. A mechanism is a family of Borel functions $\bm{h}=(h_k)_{k\in\mathcal K}$, $h_k:\Theta^{N-1}\to\mathbb R$, and
\[
  \oplus[\bm{h}](\mathbf x)=\sum_{k\in\mathcal K} h_k(\mathbf x_{-k}). 
\]
Set $W^{\max}=\sup_{\Theta^N} W$.  For $\ell \in[0,\infty]$, let $\mathcal H_\ell$ be the set of mechanisms with
$h_k\ge -\ell$ everywhere for all $k$ (so $\mathcal H_\infty$ is the unrestricted class), and define
\begin{equation}\label{eq:capped-problem}
  \delta^\star(\ell)=\inf_{h\in\mathcal H_\ell,\ \gamma\in\mathbb R}\ \gamma
  \quad\text{s.t.}\quad
  \begin{cases}
   (K-1)W(\mathbf x)-\oplus[\bm{h}](\mathbf x)\le\gamma & \text{for all }\mathbf x\in \Theta^N,\\[2pt]
   \int h_k\,\mathrm{d}\mu_{-k}\le \bar W_{-k} & \text{for all } k\in\mathcal K .
  \end{cases}
\end{equation}
Thus $\delta^\star=\delta^\star(\infty)$, and $B\mapsto\delta^\star(B)$ is non-increasing.

\begin{proposition}[Bounded mechanisms under limited liability]\label{prop:capped-charges}
Let $\ell \ge 0$, Then:
\begin{enumerate}
\item[(i)] The value $\delta^\star(\ell)$ in the limited liability problem in attained by some $h^\ell$, and the deficit guarantee holds at \emph{every} profile $\mathbf x\in \Theta^N$.
\item[(ii)] The solution can be chosen bounded:
\[
 -\ell \le h^\ell_k\le M_\ell:=(K-1)\big(\overline W^{\max}+\ell\big)-\gamma_0
 \quad\text{for all }k, 
\]
where $\gamma_0:=(K-1) \int W\,\mathrm{d}\mu-\sum_{k}\overline W_{-k}$.
\item[(iii)] Suppose that that $F$ and $G$ have densities bounded above and away from zero; then $\delta^\star(\ell)\downarrow\delta^\star$ as $\ell\to\infty$.
\end{enumerate}
\end{proposition}

\begin{proof}
The pair $h_k\equiv -\ell$, $\gamma=(K-1)\overline W^{\max}+N\ell$ is feasible because
$\overline W_{-k}\ge -B$. For any feasible $(h,\gamma)$, integrate the guarantee against
$\mu$ and use Fubini, $\int h_k(\mathbf x_{-k})\,\mathrm{d}\mu=\int h_k\,\mathrm{d}\mu_{-k}$, and the budget
constraints. This gives $\gamma\ge\gamma_0$. Hence $\delta^\star(\ell)\in[\gamma_0,\infty)$.

Define the following truncation operation. Let $(h,\gamma)$ be feasible with $h\in\mathcal H_\ell$, and put $\tilde h_k=\min(h_k,M_\ell)$.
Fix $\mathbf x$. If no coordinate is truncated at $\mathbf x$, then $\oplus[\tilde{\mathbf{h}}](\mathbf x)=\oplus[\mathbf{h}](\mathbf x)$. Otherwise some $k$ has $\tilde h_k(\mathbf x_{-k})=M_\ell$. Since all other terms are greater than $-\ell$,
\[
\oplus[\tilde{\mathbf{h}}](\mathbf x)\ge M_\ell-(K-1)\ell=(N-1)\overline W^{\max}-\gamma_0
\ge (K-1)W(\mathbf x)-\gamma .
\]
In both cases the guarantee holds at $\mathbf x$. Truncation only lowers each $h_k$, so the
budgets still hold. Hence $(\tilde h,\gamma)$ is feasible, and
$-\ell\le\tilde h_k\le M_\ell$.

To prove existence, take feasible $(h^j,\gamma^j)$ with $\gamma^j\downarrow\delta^\star(\ell)$, truncated in the sense just defined. Then $(h^j_k)_k$ is bounded in the Hilbert space $\mathcal L:=\prod_k L^2(\mu_{-k})$, so a
subsequence converges weakly in $\mathcal L$. By Mazur's lemma there are convex combinations
\[
g^J=\sum_{j\ge J}\lambda^J_j h^j \quad\text{(finitely many nonzero }\lambda^J_j\text{)}
\]
that converge strongly in $\mathcal L$. The weights are the same for every $k$, and passing
to a further subsequence gives $\mu_{-k}$-a.e.\ convergence for every $k$. Set
$\bar\gamma^J=\sum_j\lambda^J_j\gamma^j\to\delta^\star(\ell)$.

Because the constraints are linear in $(h,\gamma)$, each $(g^J,\bar\gamma^J)$ is feasible,
pointwise, with $-\ell\le g^J_k\le M_\ell$. Define the Borel functions
\[
h^B_k:=\limsup_{J\to\infty} g^J_k\in[-\ell,M_\ell].
\]
For every $\mathbf x$, since a sum of $\limsup$s dominates the $\limsup$ of the sum,
\[
\oplus[\bm{h}^B](\mathbf x)\ \ge\ \limsup_J T[g^J](\mathbf x)\ \ge\ (N-1)W(\mathbf x)-\lim_J\bar\gamma^J
=(N-1)W(\mathbf x)-\delta^\star(\ell).
\]
Moreover $h^B_k=\lim_J g^J_k$ holds $\mu_{-k}$-a.e., so bounded convergence gives
$\int h^B_k\,\mathrm{d}\mu_{-k}=\lim_J\int g^J_k\,\mathrm{d}\mu_{-k}\le\overline W_{-k}$. Hence
$(h^B,\delta^\star(B))$ is feasible, and therefore optimal. This proves (i) and (ii).

To show convergence, fix $\eta>0$ and a feasible $(h,\gamma)$ in $\mathcal H_\infty$ with
$\gamma\le\delta^\star+\eta$. Extend $W$ and each $h_k$ beyond their domain by coordinatewise reflection.
Convolve each $h_k$ with a product mollifier $\rho_\varepsilon^{\otimes(K-1)}$. Since the
$k$-th factor of $\rho_\varepsilon^{\otimes K}$ integrates to one,
\[
\textstyle\sum_k (h_k*\rho^{\otimes(K-1)}_\varepsilon)(\mathbf x_{-k})
=(\oplus[h]*\rho_\varepsilon^{\otimes K})(\mathbf x)
\ge (K-1)W(\mathbf x)-\gamma-(K-1)\omega_W(c\,\varepsilon)
\]
at every $\mathbf x$. Here $\omega_W$ is the modulus of continuity of $W$, and the constant
$c$ absorbs the reflection.
The density bounds make $\mu_{-k}$ and Lebesgue measure mutually absolutely continuous
with bounded densities. So $h^\varepsilon_k:=h_k*\rho^{\otimes(K-1)}_\varepsilon\to h_k$ in
$L^1(\mu_{-k})$.
Subtract $e^\varepsilon_k:=\big(\int h^\varepsilon_k\,\mathrm{d}\mu_{-k}-\overline W_{-k}\big)^+\to 0$
from $h^\varepsilon_k$ to restore the budgets.

The resulting mechanism is continuous on a compact set, hence bounded below by some
$-\ell_\varepsilon$. It is feasible with
$\gamma+(N-1)\omega_W(c\varepsilon)+\sum_k e^\varepsilon_k$, so
\[
\delta^\star\le\delta^\star(\ell_\varepsilon)\le\delta^\star+\eta+o(1)\quad\text{as }\varepsilon\to 0 .
\]
Monotonicity of $\delta^\star(\cdot)$ and arbitrariness of $\eta$ yield (iii).
\end{proof}

\subsection{Restricted duality}\label{app:restricted-duality}
The following gives a general argument to adapt the duality structure in the baseline problem to cases with restricted charges.
Index buyers and sellers indifferently by $k$, and denote $\mu$ the independent type coupling. Suppose each charge $h_k$ is restricted to sums $\sum_{S\in\mathcal S_k} f_S(\bm\theta_S)$ of arbitrary functions of the coordinates in $S\subseteq -k$. The proof of \hyref{Theorem}{thm:duality_nm} applies verbatim. The only change is that the coefficient of each $f_S$ forces $\lambda_S=\mu_S$ only for $S\in\bigcup_a\mathcal S_a$.

\subsection{Competitive pivot mechanism}
Proof of \hyref{Claim}{claim:competitive-pivot-nm}.
\label{app:proof-competitive-pivot-nm}
The proposed charges are additively separable and bind the IIR budgets.
Summing them gives
\begin{align*}
    \mathrm{Deficit}(\mathbf v,\mathbf c;\mathcal M^{\mathrm{CP}})
    &=\delta^{\mathrm{CP}}
      +(n+m-1)\bigl[
          W(\mathbf v,\mathbf c)-W(\mathbf v,\mathbf c;p^\star)
      \bigr]
    \le\delta^{\mathrm{CP}}.
\end{align*}
This is weak duality. For a dual certificate to prove strong duality, market clearing allows the willingness
indicators at $p^\star$ to be coupled with equal buyer and seller
counts almost surely. Indeed, their probability vector lies in the
integral polytope
$\{\mathbf y\in[0,1]^{n+m}:
\sum_i y_i^B=\sum_j y_j^S\}$.
Drawing each type conditional on its indicator yields
$\pi^\star\in\Pi(\mathbf F,\mathbf G)$ satisfying
$W(\mathbf v,\mathbf c)=W(\mathbf v,\mathbf c;p^\star)$
almost surely. Hence
$\mathbb E_{\pi^\star}[W]=\overline W(p^\star)$. \hfill $\square$

\subsection{Projection-core mechanisms}\label{app:projection-core-mechanisms}

Nature chooses a coupling from
\begin{equation*}
\Gamma^{\mathrm{PCM}}
=
\left\{
\pi:
\pi_{v_1,c}=F_1\otimes G,\quad
\pi_{v_2,c}=F_2\otimes G
\right\}.
\end{equation*}

Efficient gains from trade are
\[
W(v_1,v_2,c)=(\max\{v_1,v_2\}-c)^+,
\]
and the VCG deficit can be expressed as
\begin{equation*}
D^{\mathrm{VCG}}(v_1,v_2,c)
=
\left|(v_1-c)^+-(v_2-c)^+\right|.
\end{equation*}

\textbf{Proof of \hyref{Claim}{claim:projection-core}.}
For any coupling in $\Gamma^{\mathrm{PCM}}$, the VCG deficit admits the threshold representation
\[
D^{\mathrm{VCG}}(v_1,v_2,c)
=
\int_c^1
\mathbf 1_{\{
\mathbf 1_{\{v_1>t\}}
\ne
\mathbf 1_{\{v_2>t\}}
\}}\,\mathrm{d}t.
\]
Conditional on the seller's cost, the marginal distribution of each buyer's value remains fixed.
The probability that the buyers lie on opposite sides of a threshold $t$ is therefore bounded by
\(
\min\{F_1(t)+F_2(t),
      2-F_1(t)-F_2(t)\}.
\)
The countermonotone coupling:
\[
v_1=F_1^{-1}(Z),\qquad
v_2=F_2^{-1}(1-Z),\qquad
Z\sim U[0,1],
\]
attains this bound simultaneously at all thresholds. 
Integrating over seller costs gives the deficit expression. To construct the optimal charges, define
\begin{equation*}
a_p(v,c) = (p-\max\{v,c\})^+ - (\min\{v,c\}-p)^+.
\end{equation*}
This function measures the signed distance between
$p$ and its projection onto the interval with
endpoints $v$ and $c$.
Observe that
\[
a_p(v,c)+(v-p)^+ = \left|(v-c)^+-(p-c)^+\right|.
\]
The triangle inequality then yields
\[
    D^{\mathrm{VCG}}(v_1,v_2,c)
    \le{} a_p(v_1,c)+a_p(v_2,c)+(v_1-p)^++(v_2-p)^+.
\]
At $p=p^\dagger$, the countermonotone coupling attains equality.  Centering the resulting potentials gives the participation charges. These charges satisfy interim individual rationality and attain the dual bound. \hfill $\square$

\subsection{Opposite-side mechanisms}\label{app:opposite-side-mechanisms}
Writing $\mu_B=\bigotimes_i F_i$ and $\mu_S=\bigotimes_j G_j$, the admissible couplings are
\begin{equation*}
    \Gamma^{\mathrm{OS}}
    =
    \left\{
    \pi:
    \pi_{\mathbf v}=\mu_B,\quad
    \pi_{\mathbf c}=\mu_S
    \right\}.
\end{equation*}

\textbf{Proof of \hyref{Claim}{claim:opposite-side}.} The buyer and seller surplus functions provide an upper bound on welfare under every coupling in $\Gamma^{\mathrm{OS}}$. 
Minimizing their expected sum therefore gives an upper bound on Nature's optimal expected welfare.

At an optimal menu, let $\mathcal{Q}^B(\mathbf v)$ and $\mathcal{Q}^S(\mathbf c)$ denote the quantities maximizing the respective surplus functions. The first-order conditions for the convex price problem imply
\(
\Pr(\mathcal{Q}^B=k)=\Pr(\mathcal{Q}^S=k),
\)
for \(k=0,\ldots,K\). Nature can therefore draw a common quantity and then independently draw the buyer and seller profiles conditional on their preferred quantities equaling that quantity. This preserves the joint distribution within each side and ensures
\[
    W(\mathbf v,\mathbf c)
    =
    \Phi^B(\mathbf v)+\Phi^S(\mathbf c)
\]
under the resulting coupling. Hence the welfare bound is sharp.

Finally, the centered participation charges satisfy interim individual rationality and yield the pointwise deficit bound
\[
    \mathrm{Deficit}^{\mathrm{OS}}(\mathbf v,\mathbf c)
    =
    \delta^{\mathrm{OS}}
    -(n+m-1)
    \left[
    \Phi^B(\mathbf v)+\Phi^S(\mathbf c)
    -W(\mathbf v,\mathbf c)
    \right]
    \le\delta^{\mathrm{OS}}.
\]
The optimal coupling establishes that no mechanism in the restricted class can guarantee a smaller deficit. \hfill $\square$

\subsection{Price-by-price decompositions}\label{app:price-by-price-decomposition}
\textbf{Proof of \hyref{Claim}{claim:bernoulli-mechanism}}. For any $\lambda_p\in\Lambda(\mu_p)$ and any binary dual cover $\sum_{a\in N\cup M}\varphi_{a,p}\ge\mathcal W_{n,m}$,
the leave-one-out constraints give weak duality:
\[
    \mathbb E_{\lambda_p}[\mathcal W_{n,m}]
    \le\sum_{a\in N\cup M}
        \mathbb E_{\lambda_p}[\varphi_{a,p}]
    =\sum_{a\in N\cup M}
        \mathbb E_{\mu_p}[\varphi_{a,p}].
\]
For a matching coupling certificate, take
\[
    \lambda_p^\star(\mathbf y)
    =\mu_p(\mathbf y)
     +\overline{\epsilon}_p(-1)^{|\mathbf y|},
\]
which is feasible by \hyref{Lemma}{lem:binary-leave-one-out}. Choose a least-likely odd profile $\mathbf y_p^\star$ and define
\[
    g_p(\mathbf y)
    =\mathcal W_{n,m}(\mathbf y)
     +C_{n,m}\mathds{1}\{\mathbf y=\mathbf y_p^\star\}.
\]
Its full alternating sum vanishes, so its multilinear expansion contains no term involving every coordinate. Consequently, $g_p$ admits a decomposition
$g_p=\sum_{a\in N\cup M}\varphi_{a,p}$. This is a feasible dual cover, and
\[
    \mathbb E_{\lambda_p^\star}[\mathcal W_{n,m}]
    =\mathbb E_{\mu_p}[g_p]
    =\mathbb E_{\mu_p}[\mathcal W_{n,m}]
     +C_{n,m}\overline{\epsilon}_p.
\]
The matching coupling and cover establish strong duality at each price. Fixing a decomposition for each possible $\mathbf y_p^\star$, with measurable tie-breaking, gives a uniformly bounded,
measurable selection of optimal potentials.

Centering the integrated potentials gives
$\mathbb E_\mu[h_a^{\mathrm{PD}}]=\overline W_{-a}$,
so the resulting Groves mechanism satisfies IIR.
Summing its charges and using the welfare decomposition yields
\[
    \mathrm{Deficit}(\mathbf v,\mathbf c;\mathcal M^{\mathrm{PD}})
    =\delta^{\mathrm{PD}}
     -(n+m-1)C_{n,m}\int_0^1
        \mathds{1}\bigl\{
            \mathbf y(p;\mathbf v,\mathbf c)=\mathbf y_p^\star
        \bigr\}\,\mathrm{d}p
    \le\delta^{\mathrm{PD}}.
\]
When all buyer values are zero and all seller costs are one,
$\mathbf y(p;\mathbf v,\mathbf c)=\mathbf0$ almost everywhere. Since $\mathbf y_p^\star$ is odd, the integral vanishes, proving that the mechanism attains the stated deficit ceiling.
\hfill $\square$

\textbf{Proof of \hyref{Claim}{claim:bernoulli-exactness}}.
\label{app:proof-bernoulli-exactness}
Let $p_0$ be the common median, and set $\chi(t)=1$ for $t\le p_0$ and $\chi(t)=-1$ otherwise. Define
\[
    \frac{\mathrm{d}\lambda^\star}{\mathrm{d}\mu}
    (\mathbf v,\mathbf c)
    =1-\prod_{i=1}^n \chi(v_i)\prod_{j=1}^m \chi(c_j).
\]
This density equals two when an odd number of types exceed $p_0$, and zero otherwise. Each sign has mean zero, so integrating out any one type leaves density one. Hence
$\lambda^\star\in\Lambda(\mathbf F,\mathbf G)$. To identify its induced binary laws, observe that
\[
    \int_0^p \chi(t)\,\mathrm{d}F_i(t)
    =\min\{F_i(p),1-F_i(p)\},
\]
and similarly for $G_j$. The corresponding upper-tail integral has the opposite sign. Thus, writing $A_p(\mathbf y)$ for the set of type profiles inducing $\mathbf y$ at price $p$,
independence under $\mu$ and oddness of $m$ give
{\small \begin{align*}
    \lambda_p^\star(\mathbf y)
    =\int_{A_p(\mathbf y)}
        \frac{\mathrm{d}\lambda^\star}{\mathrm{d}\mu}
        \,\mathrm{d}\mu
    =\mu_p(\mathbf y)+(-1)^{|\mathbf y|}
      \prod_{i=1}^n\min\{F_i(p),1-F_i(p)\}
      \prod_{j=1}^m\min\{G_j(p),1-G_j(p)\}.
\end{align*}}
The product is the smallest probability of any binary profile under $\mu_p$. Below $p_0$, a minimizing profile is $(\mathbf0_n,\mathbf1_m)$; above $p_0$, it is
$(\mathbf1_n,\mathbf0_m)$. Both are odd because $m$ and $n$ are odd, respectively.
The product therefore equals $\overline{\epsilon}_p$, giving
\[
    \lambda_p^\star(\mathbf y)
    =\mu_p(\mathbf y)
     +\overline{\epsilon}_p(-1)^{|\mathbf y|}.
\]
Thus one admissible coupling induces the optimal binary law at every price.

The welfare decomposition and weak duality now yield \( \delta^\star=\delta^{\mathrm{PD}}.\) price-decomposable mechanism gives the reverse inequality,
establishing both equality and optimality.
\hfill $\square$

\subsection{Derivations for \hyref{Table}{tab:mechanisms-comparison-2b1s}}\label{app:2B1S-table-derivations}

Throughout, $n=2$, $m=1$, all types are $U[0,1]$, $\mu$ is the product prior, and $W=(\max\{v_1,v_2\}-c)^+=\max\{v_1,v_2,c\}-c$. The participation budgets are $\overline W_{-i}=\mathbb E(v_{j}-c)^+=\sfrac16$ for each buyer and $\overline W_{-S}=0$, and $n+m-1=2$.
Note that $\mathbb E_\mu W=\int_0^1\Pr(\max v_i\ge p)\Pr(c\le p)\,dp=\int_0^1(1-p^2)p\,dp=\sfrac14$. Every Groves mechanism with binding budgets therefore has mean deficit $2\cdot\sfrac14-\sfrac13=\sfrac16$.
\begin{itemize}
    \item \textbf{VCG.}  Take $h_a=W_{-a}$, which gives $h_1=(v_2-c)^+$, $h_2=(v_1-c)^+$ and $h_S=0$. These budgets bind. If $v_1\ge v_2$, the deficit is $2W-(v_1-c)^+-(v_2-c)^+=(v_1-c)^+-(v_2-c)^+\le 1$, with equality at $(v_1,v_2,c)=(1,0,0)$.
    \item \textbf{Competitive-pivot.} Wedge domination (Section~\ref{sec:generalized-competitive-pivot}) at $p^\star=\sfrac23$ gives:
    \[
    W\le \textstyle\sum_i(v_i-p^\star)^++(p^\star-c)^+,
    \qquad\text{so}\qquad
    \mathbb E_\lambda W\le 2\cdot\sfrac1{18}+\sfrac29=\sfrac13 .
    \]
    Equality holds for the coupling with density $\sfrac94$ on each of the three regions where exactly one type exceeds $\sfrac23$. These regions have mass $\sfrac13$ each, and the coupling has uniform one-dimensional marginals. On its support the wedge inequality binds: either one buyer is above $p^\star$ and the seller below, or the seller is above $p^\star$ and both buyers below, so $W=0$. Hence $\delta^{\mathrm{CPM}}=\sfrac23-\sfrac13=\sfrac13$.

    \item \textbf{Projection core.} Write $x=1-\theta$, so that Nature minimizes $\mathbb E\min\{x_1,x_2,x_3\}$. Conditional on $x_3=t$, each of $x_1,x_2$ must be $U[0,1]$, and their dependence is free. Since $\Pr(\min\{x_1,x_2\}\le s)\le 2s$, the countermonotone pair $x_2=1-x_1$ makes $Z=\min\{x_1,x_2\}\sim U[0,\sfrac12]$ stochastically smallest. Using the same conditional coupling for every $t$ is admissible, so the bound is attained:
    \[
    \min\mathbb E\min_i x_i=\int_0^{1/2}(t-t^2)\,dt+\tfrac12\cdot\tfrac14=\tfrac5{24}.
    \]
    Hence $\max\mathbb E_\lambda W=\sfrac{19}{24}-\sfrac12=\sfrac7{24}$ and $\delta^{\mathrm{PCM}}=\sfrac14$.

    \item \textbf{Opposite-side.} $W$ depends on $(v_1,v_2)$ only through $M=\max\{v_1,v_2\}$, whose CDF is $m^2$. Any coupling of $(M,c)$ lifts to an admissible coupling of $((v_1,v_2),c)$ by drawing $(v_1,v_2)$ from its conditional law given $M$. The problem is therefore bilateral, and by \hyref{Claim}{claim:opposite-side} with $K=1$,
    \[
    \max\mathbb E_\lambda W=\int_0^1\min\{1-p^2,p\}\,dp=\tfrac56(1-\pi^\star)=\tfrac{5(3-\sqrt5)}{12},
    \qquad \pi^\star=\tfrac{\sqrt5-1}{2},
    \]
    using $\pi^{\star2}=1-\pi^\star$ and $\pi^{\star3}=2\pi^\star-1$. Hence $\delta^{\mathrm{OS}}=\sfrac{(13-5\sqrt5)}{6}$.

    \item \textbf{Price-by-price decomposition.} At price $p$, the threshold vector $y=(\mathbf 1\{v_1\ge p\},\mathbf 1\{v_2\ge p\},\mathbf 1\{c\le p\})$ has welfare statistic $H(y)=y_3\,\mathbf 1\{y_1+y_2\ge1\}$. By Section~\ref{sec:price-by-price-decomposition}, an admissible $\lambda_p$ differs from $\mu_p$ only by $-\delta_p(-1)^{|y|}$, which changes $\mathbb E H$ by $\delta_p\sum_y H(y)(-1)^{|y|}=\delta_p$. Nonnegativity on the odd cells gives $\delta_p\le\min\{p^3,\,p(1-p)^2\}=p\min\{p,1-p\}^2$, so
    \[
    \mathbb E_\lambda W\le \tfrac14+\int_0^1p\min\{p,1-p\}^2dp=\tfrac14+\tfrac1{24}=\tfrac7{24},
    \]
    and the bound is $\sfrac14$.
    \item \textbf{Optimal.} $\delta^\star=\sfrac23-2\kappa$ by \hyref{Claim}{claim:2b1s-explicit-mallows-solution}.
\end{itemize}

\section{Partnership dissolution}
\label{app:partnership}

\subsection{Environment and participation charges}
We consider a partnership dissolution environment with $n\ge2$ risk-neutral agents. Each partner $i\in N$ owns an initial share $r_i\ge0$ of an asset, with $\sum_{i=1}^n r_i=1$. Each partner draws a private value $\theta_i\in[0,1]$ independently from a distribution $F_i$ with a continuous, strictly positive density $f_i$. Write $\mathbf F=(F_i)_{i\in N}$, $\mu_i=\mathbb E[\theta_i]$, and $\overline W=\mathbb E[W(\boldsymbol\theta)]$. Unless a coupling is specified, expectations are taken under the independent prior $\prod_{i\in N}F_i$.

The efficient allocation assigns the asset entirely to a partner with the highest value, yielding ex post welfare
$W(\boldsymbol\theta)=\max_{i\in N}\theta_i$. As in \hyref{Section}{sec:model}, efficient, truthful mechanisms have
Groves transfers
\[
    T_i(\boldsymbol\theta)
    =h_i(\boldsymbol\theta_{-i})-\bigl[W(\boldsymbol\theta)-\theta_iQ_i(\boldsymbol\theta)\bigr].
\]
Each partner's payoff is $W(\boldsymbol\theta)-h_i(\boldsymbol\theta_{-i})$,
and the designer's deficit is
\[
    \mathrm{Deficit}(\boldsymbol\theta;\mathcal M)
    =(n-1)W(\boldsymbol\theta)-\sum_{i=1}^n h_i(\boldsymbol\theta_{-i}).
\]

To ensure voluntary dissolution, the mechanism must give partners at least the values of their initial share. This places an upper bound on the expected participation charge. Under interim participation, the bound comes from the partner's worst-off type; under ex ante participation, it is imposed only in expectation. These bounds are given by:
\begin{align*}
    H_i^{\mathrm{IIR}}
    &=\min_{\theta_i\in[0,1]}\left\{\mathbb E\bigl[W(\theta_i,\boldsymbol\theta_{-i})\bigr]-r_i\theta_i\right\},\\
    H_i^{\mathrm{AIR}}&=\overline W-r_i\mu_i.
\end{align*}
Thus, under participation regime $x\in\{\mathrm{IIR},\mathrm{AIR}\}$, voluntary participation is equivalent to
$\mathbb E[h_i(\boldsymbol\theta_{-i})]\le H_i^x$ for every $i\in N$. Unlike in the trade environment, a partner's worst-off type need not be an endpoint of the type space.

\subsection{The deficit guarantee and its dual}
The designer's problem has the form:
\[
\begin{aligned}
    \textbf{(DG)}\qquad
    \delta^\star={}&\inf_{\mathbf h,\delta}\;\delta\\
    \text{s.t.}\qquad
    &(n-1)W(\boldsymbol\theta)
       -\sum_{i=1}^n h_i(\boldsymbol\theta_{-i})\le\delta
       &&\forall\boldsymbol\theta\in[0,1]^n,\\
    &\mathbb E[h_i(\boldsymbol\theta_{-i})]\le H_i^x
       &&\forall i\in N.
\end{aligned}
\]
As in \hyref{Section}{sec:duality-nm}, we formulate the problem as optimal selection of participation charges and write its value as an infimum. 

Let $\Lambda(\mathbf F)$ be the set of distributions on $[0,1]^n$ whose leave-one-out marginals satisfy
\[
    \lambda_{-i}=\prod_{k\in N\setminus\{i\}}F_k
    \qquad\forall i\in N.
\]
Assigning a nonnegative measure $\lambda$ to the pointwise constraint and multipliers $\alpha_i\ge0$ to the participation constraints gives the same dual restrictions as in \hyref{Theorem}{thm:duality_nm}. The coefficient of $\delta$ requires $\lambda$ to have unit mass; the coefficients of the charges require $\lambda_{-i}=\alpha_i\prod_{k\ne i}F_k$, and therefore $\alpha_i=1$. Nature's adversarial surplus problem is consequently
\[
    \textbf{(AS)}\qquad
    \max_{\lambda\in\Lambda(\mathbf F)}
        \mathbb E_\lambda\bigl[\max_{i\in N}\theta_i\bigr].
\]
These are leave-one-out constraints, rather than merely restrictions on individual marginals. The problem is therefore a multistochastic Monge--Kantorovich problem, just as in the two-sided trade environment.

\begin{proposition}[Strong duality]
\label{prop:partnership-duality}
Under participation regime $x\in\{\mathrm{IIR},\mathrm{AIR}\}$, the optimal deficit ceiling is
\[
    \delta^\star
    =(n-1)\max_{\lambda\in\Lambda(\mathbf F)}
        \mathbb E_\lambda[W]
      -\sum_{i=1}^n H_i^x.
\]
Equivalently, the optimal worst-case revenue guarantee is $r^\star=-\delta^\star$.
\end{proposition}

This duality defines the fiscal limit of dissolving a partnership. The designer can collect at most $\sum_i H_i^x$ in expected participation charges, but must cover the Groves liability $(n-1)W(\boldsymbol\theta)$ at every profile.
Nature chooses dependence invisible in every leave-one-out economy to make this liability as large as possible.

Optimal participation charges follow from conservative welfare contributions, exactly as in \hyref{Theorem}{thm:optimal-mechanism-from-potentials}. Consider the welfare imputation problem
\[
    \inf_{(\phi_i)_{i\in N}}
        \sum_{i=1}^n\mathbb E[\phi_i]
    \quad\text{subject to}\quad
        \sum_{i=1}^n\phi_i(\boldsymbol\theta_{-i})
        \ge W(\boldsymbol\theta)
        \quad\forall\boldsymbol\theta.
\]
\begin{proposition}[Optimal charges]
\label{prop:partnership-optimal-charges}
If integrable potentials $(\phi_i)_{i\in N}$ solve the welfare imputation problem, then the charges
\[
    h_i(\boldsymbol\theta_{-i})
    =H_i^x+(n-1)
       \bigl(\phi_i(\boldsymbol\theta_{-i})-\mathbb E[\phi_i]\bigr)
    \qquad\forall i\in N
\]
attain the optimal deficit ceiling.
\end{proposition}

The existence and regularity of optimal potentials are subject to the qualifications in \hyref{Section}{sec:existence-of-potentials-nm}; strong duality alone does not establish bounded optimal charges.

\subsection{Ex ante participation with a common distribution}
Suppose all partners share a distribution $F$ with mean $\mu$. Then $H_i^{\mathrm{AIR}}=\overline W-r_i\mu$, and
$\sum_iH_i^{\mathrm{AIR}}=n\overline W-\mu$. The optimal revenue guarantee therefore becomes
\[
    r^\star
    =n\overline W-\mu
     -(n-1)\max_{\lambda\in\Lambda(\mathbf F)}
                 \mathbb E_\lambda[W],
    \qquad \mathbf F=(F,\ldots,F).
\]
We can use this dual formulation to recover the possibility result of \citet[Theorem~2]{JagadeesanTordjman2025}. The key is to bound Nature's expected maximum using the independent distribution of every leave-one-out profile.

\begin{proposition}[Efficient Dissolution of Partnership]
\label{thm:ex_ante_partnership}
If all partners share a distribution $F$, efficient, truthful dissolution can be implemented without a deficit under ex ante participation, regardless of the initial ownership shares. In particular, $\delta^{\star,\mathrm{AIR}}\le0$.
\end{proposition}

\begin{proof}
Fix $\lambda\in\Lambda(\mathbf F)$.
For each $i$, Jensen's inequality and leave-one-out restriction give
\[
    \overline W
    \ge\mathbb E\bigl[\max\{\mu,\max_{k\ne i}\theta_k\}\bigr]
    =\mathbb E_\lambda\bigl[\max\{\mu,\max_{k\ne i}\theta_k\}\bigr].
\]
For every realized profile,
\[
    \sum_{i=1}^n\max\{\mu,\max_{k\ne i}\theta_k\}
    \ge(n-1)\max_{k\in N}\theta_k+\mu.
\]
Indeed, all terms except that of a highest-value partner are at least the highest value, and the remaining term is at least $\mu$. Summing the preceding expectations therefore gives $n\overline W\ge(n-1)\mathbb E_\lambda[W]+\mu$. Since this holds for every admissible $\lambda$, \hyref{Proposition}{prop:partnership-duality} implies $\delta^{\star,\mathrm{AIR}}\le0$.

To obtain an actual implementation, consider the following participation charges
\[
\begin{aligned}
    h_i(\boldsymbol\theta_{-i})
    ={}\overline W-r_i\mu
       +\max\{\mu,\max_{k\ne i}\theta_k\}
      -\mathbb E\bigl[\max\{\mu,\max_{k\ne i}\theta_k\}\bigr].
\end{aligned}
\]
Their expectations equal $H_i^{\mathrm{AIR}}$, so ex ante participation binds. Jensen's inequality makes each centering adjustment $\overline W-\mathbb E[\max\{\mu,\max_{k\ne i}\theta_k\}]$ nonnegative. Total share $\sum_i r_i=1$ then gives $\sum_i h_i\ge(n-1)W$ pointwise, so the mechanism never runs a deficit.
\end{proof}

\section{Public good provision}
\label{app:publicgood}

\subsection{Environment and participation charges}
We consider a public good provision environment with $n\ge2$ risk-neutral users. Each user $i\in N$ derives private value $\theta_i\in[0,1]$ from the public good, drawn independently from a distribution $F_i$ with a continuous, strictly positive density $f_i$ and mean $\mu_i$. Write $\mathbf F=(F_i)_{i\in N}$. The cost of provision is $K\in(0,n)$. The efficient allocation provides the public good if and only if aggregate value exceeds the cost:
\[
    Q(\boldsymbol\theta)
    =\mathds{1}\left\{\sum_{i=1}^n\theta_i\ge K\right\},
    \qquad
    W(\boldsymbol\theta)
    =\left(\sum_{i=1}^n\theta_i-K\right)^+.
\]
Let $\overline W=\mathbb E[W(\boldsymbol\theta)]$, with expectations
taken under the independent prior unless otherwise specified. To ensure truthfulness, the mechanism must belong to the family of Groves schemes. The transfer extracted from user $i$ takes the form
\[
    T_i(\boldsymbol\theta)
    =h_i(\boldsymbol\theta_{-i})
     -\left(\sum_{j\ne i}\theta_j-K\right)Q(\boldsymbol\theta),
\]
where $h_i$ does not depend on user $i$'s own report.
User $i$'s payoff is therefore
$W(\boldsymbol\theta)-h_i(\boldsymbol\theta_{-i})$.
Since the designer must pay the provision cost, the net deficit is
\[
\begin{aligned}
    \mathrm{Deficit}(\boldsymbol\theta;\mathcal M)
    &=KQ(\boldsymbol\theta)-\sum_{i=1}^nT_i(\boldsymbol\theta)
    &=(n-1)W(\boldsymbol\theta)
      -\sum_{i=1}^n h_i(\boldsymbol\theta_{-i}).
\end{aligned}
\]
Thus efficient public good provision creates the same fiscal structure
as partnership dissolution: Groves transfers generate the liability
$(n-1)W$, and the remaining design variables are participation charges.

Voluntary participation places an upper bound on each expected charge.
Under interim participation, the worst type is $\theta_i=0$;
under ex ante participation, the constraint is imposed only in expectation.
The respective bounds are
\[
    H_i^{\mathrm{IIR}}
    =\mathbb E\left[\left(\sum_{j\ne i}\theta_j-K\right)^+\right]
    =\overline W_{-i},
    \qquad
    H_i^{\mathrm{AIR}}=\overline W,
\]
where the provision cost is unchanged when a user is removed.
Participation under regime $x\in\{\mathrm{IIR},\mathrm{AIR}\}$ is
equivalent to $\mathbb E[h_i]\le H_i^x$ for every user $i$.

\subsection{The deficit guarantee and its dual}
The designer's problem is
\[
\begin{aligned}
    \textbf{(DG)}\qquad
    \delta^\star={}&\inf_{\mathbf h,\delta}\;\delta\\
    \text{s.t.}\qquad
    &(n-1)W(\boldsymbol\theta)
       -\sum_{i=1}^n h_i(\boldsymbol\theta_{-i})\le\delta
       &&\forall\boldsymbol\theta\in[0,1]^n,\\
    &\mathbb E[h_i(\boldsymbol\theta_{-i})]\le H_i^x
       &&\forall i\in N.
\end{aligned}
\]
As in the main text, the designer choices are participation charges and auxiliary ceiling $\delta$; the same infimum is obtained over bounded, measurable charges. Let $\Lambda(\mathbf F)$ consist of the joint distributions satisfying
\(
    \lambda_{-i}=\prod_{k\in N\setminus\{i\}}F_k, \, \forall i\in N.
\)
The dual restrictions are the same as in
\hyref{Theorem}{thm:duality_nm} and
\hyref{Appendix}{app:partnership}: Nature preserves the distribution of every leave-one-out economy while maximizing expected welfare:
\[
    \textbf{(AS)}\qquad
    \max_{\lambda\in\Lambda(\mathbf F)}
        \mathbb E_\lambda\left[
            \left(\sum_{i=1}^n\theta_i-K\right)^+\right].
\]

\begin{proposition}[Strong duality]
\label{thm:publicgood-duality}
Under participation regime $x\in\{\mathrm{IIR},\mathrm{AIR}\}$, the optimal deficit ceiling is
\[
    \delta^\star
    =(n-1)\max_{\lambda\in\Lambda(\mathbf F)}
                \mathbb E_\lambda[W]
      -\sum_{i=1}^n H_i^x.
\]
Equivalently, the optimal worst-case net revenue guarantee is
$r^\star=-\delta^\star$.
\end{proposition}

\begin{proof}
The argument is identical to that of \hyref{Proposition}{prop:partnership-duality}, with $W(\boldsymbol\theta)=(\sum_i\theta_i-K)^+$ and the participation bounds given above. In particular, the provision cost has already been incorporated into welfare and the net deficit formula.
\end{proof}

This duality defines the fiscal limit of efficient public good provision. The designer can collect at most $\sum_iH_i^x$ in expected participation charges, but must cover the Groves liability $(n-1)W$ at every profile. The adversarial coupling identifies dependence invisible in every leave-one-out marginal that makes this liability as large as possible.

The mechanism construction also carries over unchanged. As in \hyref{Theorem}{thm:optimal-mechanism-from-potentials},
let $(\phi_i)_{i\in N}$ solve the welfare imputation problem
\[
    \inf_{(\phi_i)_{i\in N}}
        \sum_{i=1}^n\mathbb E[\phi_i]
    \quad\text{subject to}\quad
        \sum_{i=1}^n\phi_i(\boldsymbol\theta_{-i})
        \ge W(\boldsymbol\theta)
        \quad\forall\boldsymbol\theta.
\]
\begin{proposition}[Optimal participation charges]
\label{prop:publicgood-optimal-charges}
If integrable optimal potentials exist and solve the welfare imputation problem,  then the charges
\[
    h_i(\boldsymbol\theta_{-i}) =H_i^x+(n-1)
    \bigl(\phi_i(\boldsymbol\theta_{-i})-\mathbb E[\phi_i]\bigr)
    \qquad\forall i\in N
\]
attain the optimal deficit ceiling.
\end{proposition}
The same qualifications on attainment and regularity apply as in \hyref{Section}{sec:existence-of-potentials-nm}.

\subsection{Ex ante participation}
Under ex ante participation, every user's participation bound is $H_i^{\mathrm{AIR}}=\overline W$. The optimal deficit ceiling becomes
\[
    \delta^{\star,\mathrm{AIR}}
    =(n-1)\max_{\lambda\in\Lambda(\mathbf F)}
                  \mathbb E_\lambda[W]-n\overline W.
\]
The following proposition gives a simple sufficient condition for efficient provision without ever running a deficit.

\begin{proposition}
\label{thm:ex_ante_publicgood}
If $\sum_{i=1}^n\mu_i\ge K$, efficient, truthful public good provision can be implemented without a deficit under ex ante participation. In particular, $\delta^{\star,\mathrm{AIR}}\le0$.
\end{proposition}

\begin{proof}
Fix $\lambda\in\Lambda(\mathbf F)$.
For each $i$, Jensen's inequality and leave-one-out restriction give
\[
    \overline W
    \ge\mathbb E\left[
        \left(\mu_i+\sum_{j\ne i}\theta_j-K\right)^+\right]
    =\mathbb E_\lambda\left[
        \left(\mu_i+\sum_{j\ne i}\theta_j-K\right)^+\right].
\]
Moreover, for every realized profile,
\[
    \sum_{i=1}^n\left(\mu_i+\sum_{j\ne i}\theta_j-K\right)^+
    \ge(n-1)\left(\sum_{i=1}^n\theta_i-K\right)^+.
\]
If aggregate value is below $K$, the right-hand side is zero. Otherwise, the sum of the untruncated terms equals
$(n-1)\sum_i\theta_i+\sum_i\mu_i-nK$, which is at least $(n-1)(\sum_i\theta_i-K)$ because $\sum_i\mu_i\ge K$. Summing the preceding expectations therefore gives $n\overline W\ge(n-1)\mathbb E_\lambda[W]$.
Since this holds for every admissible $\lambda$, \hyref{Proposition}{thm:publicgood-duality} yields $\delta^{\star,\mathrm{AIR}}\le0$.

An explicit implementation is given by the bounded, continuous charges
\[
\begin{aligned}
    h_i(\boldsymbol\theta_{-i})
    ={}&\overline W
      +\left(\mu_i+\sum_{j\ne i}\theta_j-K\right)^+
      -\mathbb E\left[
          \left(\mu_i+\sum_{j\ne i}\theta_j-K\right)^+\right].
\end{aligned}
\]
Their expectations equal $\overline W$, so ex ante participation binds. By Jensen's inequality, the added constants are nonnegative. The deterministic inequality then gives $\sum_i h_i\ge(n-1)W$ at every profile, ensuring no deficit.
\end{proof}

\end{document}